\pdfoutput=1
\newcommand*{\ATLASLATEXPATH}{}
\documentclass[cernpreprint, UKenglish, texlive=2016]{\ATLASLATEXPATH atlasdoc}
\usepackage[subfig]{\ATLASLATEXPATH atlaspackage}
\usepackage{\ATLASLATEXPATH atlasbiblatex}
\usepackage{\ATLASLATEXPATH atlasphysics}
 
\graphicspath{{logos/}{figures/}}
 
\usepackage{ANA-EXOT-2017-32-PAPER-defs}

\AtlasTitle{Constraints on mediator-based dark matter and scalar dark energy models using $\sqrt s = 13$~TeV $pp$ collision data collected by the ATLAS detector}

\AtlasAbstract{ 
Constraints on selected  mediator-based dark matter models and a scalar dark energy model using up to $37~\mathrm{fb}^{-1}$
$\sqrt s = 13$~TeV $pp$ collision data collected by the ATLAS detector at the LHC during 2015--2016 are summarised in this paper.
The results of  experimental searches in a variety of final states
are interpreted in terms of a set of spin-1 and spin-0 single-mediator dark matter simplified models and a second set of
models involving an extended Higgs sector plus an additional vector or pseudo-scalar mediator.
The searches considered in this paper constrain spin-1 leptophobic and leptophilic
mediators, spin-0 colour-neutral and colour-charged mediators and
vector or pseudo-scalar mediators embedded in extended Higgs sector models. In this case, also $\sqrt s = 8$~TeV $pp$ collision data are used for the interpretation of the results.
The results are also interpreted for the first time in terms of light scalar particles that could contribute
to the accelerating expansion of the universe (dark energy).
}
 
\author{The ATLAS Collaboration}

\PreprintIdNumber{CERN-EP-2018-334}

\AtlasJournalRef{\JHEP 1905 (2019) 142}
\AtlasDOI{10.1007/JHEP05(2019)142}

\hypersetup{pdftitle={ATLAS document},pdfauthor={The ATLAS Collaboration}}
 
\usepackage{lscape}

\begin{document}
 
\maketitle
 
{\small
\tableofcontents
}
 
\section{Introduction}
 

The existence of a non-luminous component of matter  and the origin of
the accelerating  expansion of the universe are two major unknowns
in our current understanding of the universe.
 
The existence of dark matter (DM) is supported by a variety of astrophysical
measurements, ranging from the rotational speed of stars in galaxies,
over precision measurements of the cosmic microwave
background~\cite{Hinshaw:2012aka,Akrami:2018vks}, to gravitational lensing
measurements~\cite{Trimble1987,Bertone:2004pz,Feng:2010gw}. However,
the nature and properties of the DM remain largely unknown.
Searches for particle DM are performed using different complementary
approaches: the measurement of elastic scattering of DM by nuclei and electrons in a
detector~\cite{Angloher:2015ewa,Akerib:2016vxi,Amole:2017dex,Amaudruz:2017ekt,Cui:2017nnn,Aprile:2018dbl,Aprile:2017iyp,Agnese:2017jvy,Agnese:2017njq,},
the detection of Standard Model (SM) particles produced in the
annihilations or decays of DM in the universe
\cite{Choi:2015ara,Aartsen:2018mxl,hess2,hess1,TheFermi-LAT:2017vmf},
the production of DM particles at
colliders \cite{EXOT-2016-23,EXOT-2016-32,HIGG-2016-18,EXOT-2016-25,HIGG-2016-28,SUSY-2016-18,EXOT-2016-27,SUSY-2016-16,SUSY-2016-15,HIGG-2015-03,EXOT-2017-16,CMS-EXO-16-037,CMS-EXO-16-012,CMS-EXO-16-055,CMS-EXO-16-005,CMS-EXO-16-039,CMS-SUS-17-001,CMS-EXO-16-052,CMS-HIG-17-031},
and the study of the effect of  DM interactions on astrophysical
systems~\cite{Roos2012,Arrenberg:2013rzp}.
Another major unknown in the physics of our universe, beside the nature of DM, is
the origin of its accelerating
expansion~\cite{Riess:1998cb,Perlmutter:1998np}. In the context of a
homogeneous and isotropic universe, this implies the existence of a
repulsive force, which causes the universe to expand at an accelerating
rate~\cite{1917SPAW.......142E}.
Assuming general relativity, one of the simplest explanation for this
repulsive force is a new type of matter which
mimics a constant energy density, thus dubbed dark energy (DE). The
effect of DE on cosmological scales is studied by measuring the
redshift--distance relation using supernovae, baryon acoustic
oscillations, the matter power spectrum and the cosmic microwave
background, as well as gravitational
lensing~\cite{Weinberg:2013raj}. On microscopic scales, DE is probed
by laboratory experiments searching for additional gravitational
forces that would lead to deviations from the $1/r^2$
law~\cite{Kapner:2006si,Hamilton:2015zga,Anastassopoulos:2015yda,
Steffen:2010ze,Brax:2016wjk,
Burrage:2016rkv,Lemmel:2015kwa,Elder:2016yxm,Brax:2013cfa}.
Multi-messenger astronomical
observations also provide important information for understanding the nature of
DE~\cite{TheLIGOScientific:2016src,TheLIGOScientific:2017qsa,GBM:2017lvd}.
 
The work reported in this paper considers the hypothesis that the DM is composed of a
weakly interacting massive particle
(WIMP)~\cite{Steigman:1984ac}.
WIMPs can
account for the relic density of non-relativistic matter in the early
universe~\cite{Kolb:1990vq} measured in data from the
Planck~\cite{Akrami:2018vks} and WMAP~\cite{Hinshaw:2012aka}
experiments.
For benchmarking purposes it is assumed
that WIMPs are Dirac fermions in all models evaluated in this paper.
Theories such as  R-parity-conserving
supersymmetry~\cite{Martin:1997ns,Farrar:1978xj,Goldberg:1983nd,Ellis:1983ew}
can also provide such WIMP DM candidates. These models are examined
using a wide range of experimental
signatures~\cite{SUSY-2014-05,SUSY-2014-06,SUSY-2014-07,SUSY-2014-08,SUSY-2015-12,SUSY-2016-23,SUSY-2016-24,SUSY-2016-25,CMS-SUS-17-004,Sirunyan:2017pjw,CMS-SUS-16-003}
in searches performed by the ATLAS and CMS Collaborations. These searches are not included in this paper.
 
For most of the models in this paper, WIMPs are potentially pair-produced in $pp$ collisions at the Large Hadron Collider (LHC)~\cite{Evans:2008zzb}. These
particles, denoted by the symbol $\chi$ throughout this paper,  are
stable over cosmological scales and do not interact with the
detector. To identify events with DM,  additional particle(s),
$X=\textrm{jet}, \gamma, W, Z, h,(t)\bar{t}, (b)\bar{b}$, need to be produced in
association with DM in a $pp$ collision, in order to tag the event and
detect the recoiling WIMPs as missing transverse momentum (with magnitude \met).
If the DM candidates can be produced at the LHC via an $s$-channel
exchange of a new particle, then this mediator could also decay back
into SM final states: resonance searches can therefore also be used to
constrain DM models. The interplay of resonance and
$X+\met$ searches depends on the specific model choice and is
further outlined in this paper.
In the models under study, some
of which are new with respect to previous ATLAS publications,  one or
more new particles mediate the interaction of DM with the SM
particles.
The first category considers simplified models mediated by a vector, axial-vector, scalar or
pseudo-scalar mediator. In the case of simplified vector and scalar
mediators, different types of interactions are explored
(baryon-charged, neutral-flavour-changing and coloured interactions).
The second category considers less simplified models involving an
extended Higgs sector plus an additional mediator, either a vector or
a pseudo-scalar particle.
The assumptions and choices of the models closely follow the work of
the DM Forum/LHC DM Working Group~\cite{Abercrombie:2015wmb,Boveia:2016mrp,Albert:2017onk,2HDMWGproxi}.
Analyses focusing on signatures compatible with (unstable) long-lived particles decaying in the detector volume are also not considered in this paper~\cite{Lee:2018pag}.
 
Results from particle physics experiments may be used to  elucidate
the microscopic nature of DE~\cite{Kunz:2006ca,Kunz:2007rk}. Hadron
collider data considering $X+\met$ final states are used
to constrain Horndeski models of DE~\cite{Brax:2016did} in an
effective field theory (EFT) framework~\cite{Georgi:1994qn}.

This paper aims to provide an overview of the experimental programme
of ATLAS searches~\cite{PERF-2007-01} for mediator-based DM production performed to date using
13~TeV proton--proton collisions delivered by the LHC in 2015 and
2016.
The studies presented in this paper use public ATLAS results.
Since no significant excess over the expected SM background was
found in any of these analyses,  the results are used to constrain the
available phase space for DM models. Furthermore, DE models are also constrained using these analyses.
 
The paper is structured as follows. The DM and DE models evaluated in this
paper are outlined in Section~\ref{sub:DMmodels},
while the data and simulation
samples are described in Section~\ref{sec:montecarlo}. The data analyses
for each different signature are briefly described in
Section~\ref{sec:analyses}, where the complementarity of
different final states is also discussed. Finally, the dominant systematic
uncertainties affecting the modelling of the signal samples are highlighted in
Section~\ref{sec:syst}
and the results are presented in
Section~\ref{sec:result}, followed by the conclusions (Section~\ref{sec:conclusion}).

\section{Theoretical framework}
\label{sub:DMmodels}
 
All DM results presented in this paper are interpreted in the framework
of simplified DM models~\cite{Fox:2008kb,Cassel:2009pu,Bai:2010hh,Abdallah:2015ter,Abercrombie:2015wmb,2HDMWGproxi},
where a new particle (or set of particles) mediates
the
interaction of DM with the SM particles. These DM
simplified models, which overcome some of the shortcomings of previous EFT-based DM models~\cite{Beltran:2010ww,Goodman:2010yf,Bai:2010hh,Goodman:2010ku,Rajaraman:2011wf,Fox:2011pm},
can be classified according to content and properties of
the particles
that mediate the interaction between DM and SM particles (mediator sector),  giving rise to collider signatures with different
kinematic characteristics and topologies.
 
Two classes of models are taken into account: the case where the mediator sector is composed of a
single particle, either of spin-1 (Section~\ref{sub:spin1theory}) or of
spin-0 (Section~\ref{sub:spin0theory}), and the case where the mediator sector
is composed of an extended Higgs sector plus an additional mediator,
either a spin-1 (Section~\ref{sub:2HDMspin1theory}) or
spin-0 (Section~\ref{sub:2HDMspin0theory}) particle.
 
Finally, a Horndeski model of DE~\cite{Horndeski:1974wa} is studied within an
EFT framework and is used to interpret the
results (Section~\ref{sub:DEtheory}).
 
All models described in this section
are summarised
in Table~\ref{t:ModelSummaryNotation}.
For all models, the width of the mediator is always assumed to be the smallest
width that can be calculated from all other parameters~\cite{Abercrombie:2015wmb} (minimal width assumption).
Furthermore this paper assumes DM to be a Dirac fermion.\footnote{The alternative assumption that DM is a Majorana fermion changes not only the set of allowed interactions, but also the total cross-section for the ones that are allowed. Aside from these, changing the choice of Dirac fermions,
Majorana fermions, or scalars is expected to produce minor changes in the kinematic distributions of the visible particles in the final state. However, these assumptions have not been evaluated further in terms of simplified DM models~\cite{Abercrombie:2015wmb}.}

\begin{table}
\caption{Summary of the mediator-based simplified models considered in
this paper, along with the associated model acronym ($2^{\mathrm{nd}}$ column, defined in the text) and mediator
symbol ($3^{\mathrm{rd}}$ column) used throughout.
The $4^{\mathrm{th}}$ and $5^{\mathrm{th}}$ columns indicate the quantum numbers of the mediator.
The '$\times$' indicates the cases where no other charge than the new mediator's interaction is present.
The $6^{\mathrm{th}}$ column indicates the final-state signatures (details in Sec.~\ref{sec:analyses}) and the $7^{\mathrm{th}}$ column gives the reference to the interpretation.}
\label{t:ModelSummaryNotation}
\centering
\scalebox{1}{
\begin{tabular}{p{37mm}lllp{24mm}p{28mm}l}
\toprule
Short description & Acronym & Symbol & $J^{P}$ & Charge & Signatures & Results \\
& & & & & (Sec.~\ref{sec:analyses}) & Section: \\[1ex]\midrule
\raggedright Vector/axial-vector mediator & V/AV & \vvec/\avec & $1^\mp$                  &\raggedright  $\times$ &\raggedright jet/$\gamma$/$W$/$Z$+\met, difermion resonance& \ref{ssub:spin1res1}\\[1ex]\hline
\raggedright Vector baryon-number-charged mediator  & VBC & \bvec & $1^-$                          &\raggedright  baryon-number &\raggedright \monoH & \ref{ssub:spin1res2}\\[1ex]\hline
\raggedright Vector flavour-changing mediator  & VFC & \nvec & $1^-$                       &\raggedright  flavour  &\raggedright $tt$, \monotop   & \ref{ssub:spin1res3}\\[1ex]\hline
\raggedright Scalar/pseudo-scalar mediator  & S/PS & \scal/\pscal & $0^\pm$ &\raggedright  $\times$ &\raggedright \monojet, $\ttbar/\bbbar+$\met& \ref{ssub:spin0res1}\\[1ex]\hline
\raggedright Scalar colour-charged mediator  & \scg$_{q/b/t}$ & $\eta_{q/b/t}$ & $0^+$            &\raggedright  colour, 2/3 electric-charge &\raggedright \monojet, \monob, \monotop & \ref{ssub:spin0res2}\\[1ex]\hline
\raggedright Two-Higgs-doublet  plus vector mediator & \small \thdmZ  & \vvec & $1^-$             &\raggedright  $\times$  &\raggedright \monoH & \ref{sub:2HDMspin1res}\\[1ex]\hline
\raggedright Two-Higgs-doublet  plus pseudo-scalar mediator & \small \thdmS  & \pscal & $0^-$      &\raggedright  $\times$ &\raggedright  $W/Z/h+\met$,  $\ttbar/\bbbar+$\met, \hinv,  \fourtop& \ref{sub:2HDMspin0res}\\[1ex]\hline
\raggedright Dark energy  & DE & $\phiDE$  & $0^+$                                 &\raggedright  $\times$ &\raggedright \monojet, \dmtt& \ref{sub:DEres}\\[1ex]
\bottomrule
\end{tabular}
}
\end{table}

\subsection{Vector or axial-vector dark matter models}
\label{sub:spin1theory}
 
The first category of models under study in this paper consists of a
set of simplified models with a single spin-1 particle that acts as the mediator.
This category of models that assume the existence
of new gauge symmetries is among the most commonly studied extensions of the
SM~\cite{Alves:2015pea} and provides a convenient framework to describe the interaction between
the SM and DM.
Three types of simplified models involving a single spin-1 particle
are investigated:
a neutral mediator~\cite{Gondolo:2011eq,Frandsen:2011cg,Fox:2011pm,An:2012va,Alves:2013tqa,Lebedev:2014bba,Papucci:2014iwa,Buchmueller:2014yoa},
a baryon-number
charged mediator~\cite{Carone:1994aa,Agashe:2004ci,FileviezPerez:2010gw,Carpenter:2013xra} and
a flavour-changing neutral-current mediator~\cite{Andrea:2011ws,Agram:2013wda,Boucheneb:2014wza}.
 
\subsubsection{Neutral interaction}
\label{ssub:spin1theory1}
 
\begin{figure}
\centering
\subfloat[]{\includegraphics[width=.25\textwidth]{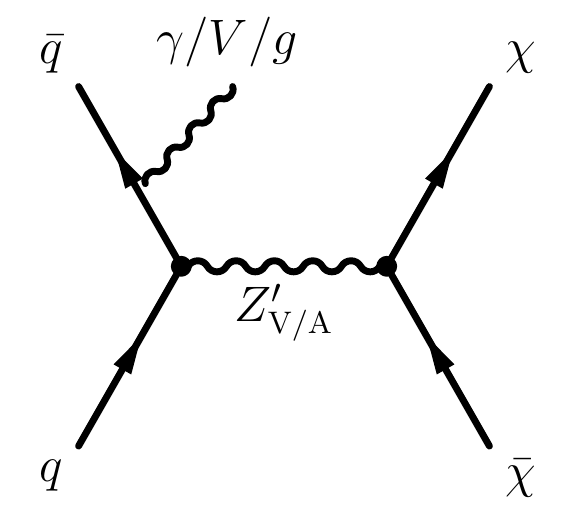}
\label{g:Diag-AVVneut1}
}
\subfloat[]{\includegraphics[width=.25\textwidth]{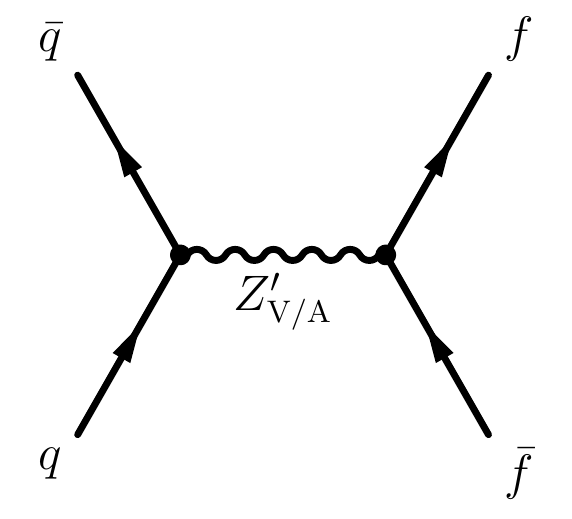}
}
\caption{Schematic representation of the dominant production and decay
modes for the V/AV model. }
\label{g:Diag-AVVneut}
\end{figure}

One vector or axial-vector simplified model (V/AV)~\cite{Abercrombie:2015wmb}
consists of a simple extension of the SM with an
additional U(1) gauge symmetry under which the DM particles are charged.
The new mediator ($Z'$) is either a vector (\vvec) or an
axial-vector (\avec) boson. The model has five parameters~\cite{Abercrombie:2015wmb}:
the masses of the mediator and the DM particle ($m_{Z'_{\text{V/A}}}$ and $m_\chi$, respectively), the flavour-universal
coupling of the $Z'$ boson to all flavour quarks, $g_q$; the coupling to
all lepton flavours, $g_\ell$; and the coupling to DM, $g_\chi$.
Representative diagrams for this model are shown in
Figure~\ref{g:Diag-AVVneut}. The $Z'$ mediator can decay into a pair of quarks, a pair of leptons, or a pair
of DM particles. In the latter case, an additional visible object has to
be produced in association with the mediator as initial-state
radiation (ISR), as shown in Figure~\ref{g:Diag-AVVneut1}. The visible
object can either be a jet, a photon or a $W$ or $Z$ boson.
In order to
highlight the complementarity of dedicated searches based  on different
final states~\cite{Albert:2017onk},
two coupling scenarios, a leptophobic and a leptophilic $Z'$ mediator, respectively,
are considered for the interpretation of these models (see Section~\ref{ssub:spin1res1}).
 
\begin{figure}
\centering
\subfloat[]{\includegraphics[width=.25\textwidth]{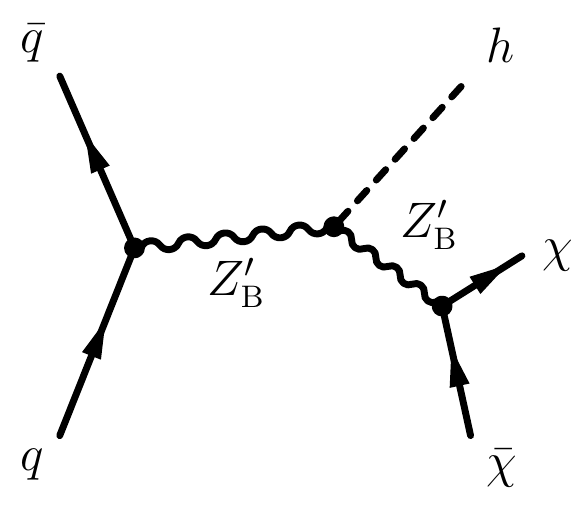}
\label{g:diag-AVVbaryonic}
}
\subfloat[]{\includegraphics[width=.25\textwidth]{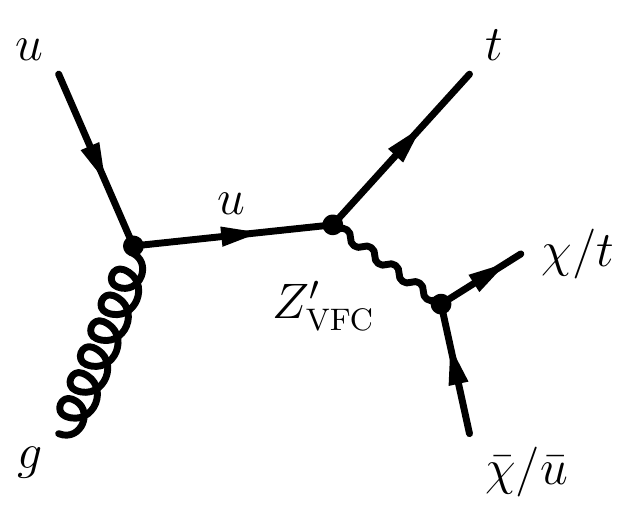}
\label{g:Diag-AVVFCN1}
}
\subfloat[]{\includegraphics[width=.25\textwidth]{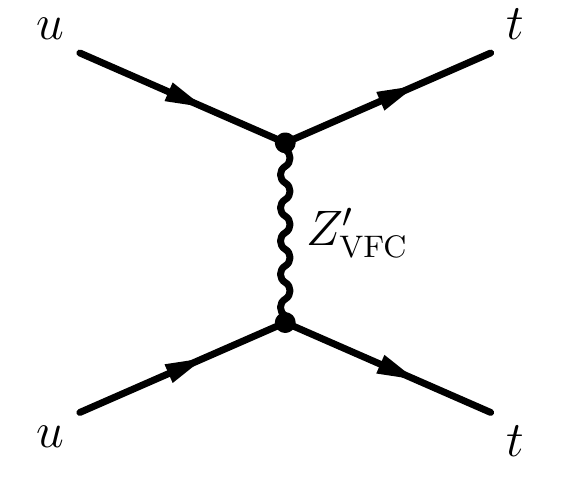}
\label{g:Diag-AVVFCN2}
}
\subfloat[]{\includegraphics[width=.25\textwidth]{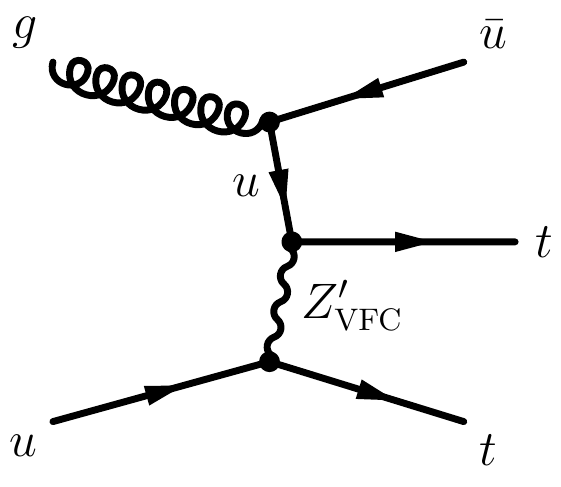}
\label{g:Diag-AVVFCN3}
}
\caption{Schematic representation of the dominant production and decay
modes for the (a) \vbc\ model and (b,c,d) \vfc\ model.}
\end{figure}
 
\subsubsection{Baryon-number-charged interaction}

The baryon-number-charged mediator simplified model~\cite{Carpenter:2013xra,Abercrombie:2015wmb}
(\bcm) considers a spin-1 vector mediator.
It also assumes that
the charge of the U(1) symmetry coincides with the baryon
number and it is spontaneously broken by a baryonic Higgs scalar.
The DM candidate in this model is a stable baryonic
state and it is neutral under the SM gauge symmetry.
While the model can provide an ISR signature through
$s$-channel \bvec-mediator production subsequently decaying into
a pair of DM candidates as for the V/AV models described
in the previous section (Figure~\ref{g:Diag-AVVneut1}), it can
also exhibit a distinctive \monoH\
signature~\cite{Carpenter:2013xra}, as shown in Figure~\ref{g:diag-AVVbaryonic}.
The model has five
parameters~\cite{Carpenter:2013xra}, whose values are chosen to enhance the cross-section for \monoH\ final states relative to
traditional ISR signatures.
The mixing angle between the baryonic and the SM Higgs bosons,
$\theta$, is fixed to $\sin\theta = 0.3$ in order to comply with the current Higgs boson coupling measurements.
The coupling of the
mediator $\bvec$ with the quarks, $g_q$, and the DM, $g_\chi$, are set to $1/3$
and $1$, respectively. The coupling of the mediator with the Higgs boson, $g_{\bvec}$, is set to
the ratio of the mediator mass to the vacuum expectation value (VEV) of the baryonic Higgs boson: $m_{\bvec} / v_\text{B}$.
The mediator is naturally leptophobic, thus evading the current constraints coming from the
dilepton resonance searches.
Different mediator and DM masses are
investigated.

\subsubsection{Flavour-changing interaction}
\label{ssub:spin1theory3}

The flavour-changing vector mediator model (\vfc)~\cite{Boucheneb:2014wza} permits the interaction of the
DM candidate with the top quark. A spin-1 colour-neutral mediator \nvec\
enables a flavour-changing interaction of the DM
with ordinary matter, for instance between the top quark and
the up quark.
For simplicity, the mediator is allowed to couple only to the right-handed component of the top-quark
field~\cite{Boucheneb:2014wza}.
This model predicts flavour-changing neutral current (FCNC)
processes which are suppressed in the SM.
A representative diagram of the on-shell production of the new
mediator \nvec\ is shown in Figure~\ref{g:Diag-AVVFCN1}. The mediator can
either decay invisibly, leading to a final state involving a single
top quark and large missing transverse momentum, or decay visibly,
producing a distinctive final state containing two top quarks with the same
electric charge ($tt/\bar{t}\bar{t}$). A similar topology arises from the
$t$-channel exchange of the \nvec\ mediator, as depicted in Figs.~\ref{g:Diag-AVVFCN2}~and~\ref{g:Diag-AVVFCN3}.

The model is fully predictive once the four main parameters are
specified~\cite{EXOT-2016-16}: the mass of the mediator $m_{\nvec}$, the DM mass $m_\chi$, and the couplings
of the mediator to the DM particles and to the quarks, \gDM\ and \gSM, respectively. In the context of the analyses described in this paper,
the mass of the DM candidate $m_{\chi}$ has negligible impact on
the kinematics, provided that $m_{\nvec} > 2 m_{\chi}$, and it is fixed to $1\;\GeV$.
This reduces the number of dimensions of the parameter space to three.
The sensitivity of the experimental analyses to this model is
explored in three scenarios that investigate different parameter planes
as a function of $m_{\nvec}$, \gSM and the invisible branching ratio of the \nvec mediator, $\mathcal{B}(\chi\bar{\chi})$.

\subsection{Scalar or pseudo-scalar dark matter models}
\label{sub:spin0theory}
 
The second category of models under study in this paper consists of a
set of simplified models with a single spin-0 particle that composes
the mediator sector. In simplified models
the mediator couples to SM
fermions proportionally to the Higgs Yukawa couplings. These models can therefore
be easily included in the extended Higgs boson sectors of ultraviolet-complete (UV-complete) theories.
The various models can be grouped in
two broad categories:
colour-neutral~\cite{Cheung:2010zf,Haisch:2013fla,Buckley:2014fba,Buckley:2015ctj,Haisch:2015ioa,Backovic:2015soa,Arina:2016cqj,Haisch:2016gry,Banerjee:2017wxi,Haisch:2018hbm}
or colour-charged mediators~\cite{Andrea:2011ws,Agram:2013wda,Boucheneb:2014wza,Batell:2011tc,Agrawal:2011ze,Cheung:2011zza,Kile:2013ola,An:2013xka,Bai:2013iqa,DiFranzo:2013vra,Batell:2013zwa,Chang:2013oia,Agrawal:2014aoa,Agrawal:2014una,Gomez:2014lva,Kilic:2015vka,Blanke:2017tnb,Blanke:2017fum}.
The latter category is divided into three further models with different final states.

\begin{figure}[h]
\centering
\subfloat[]{\includegraphics[width=.3\textwidth]{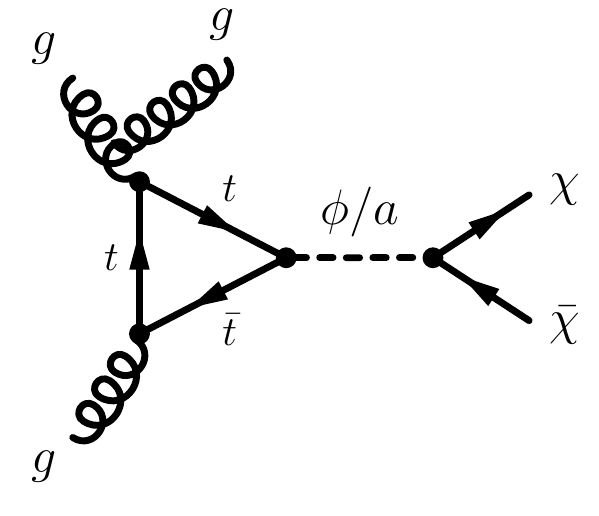}
}
\subfloat[]{\includegraphics[width=.26\textwidth]{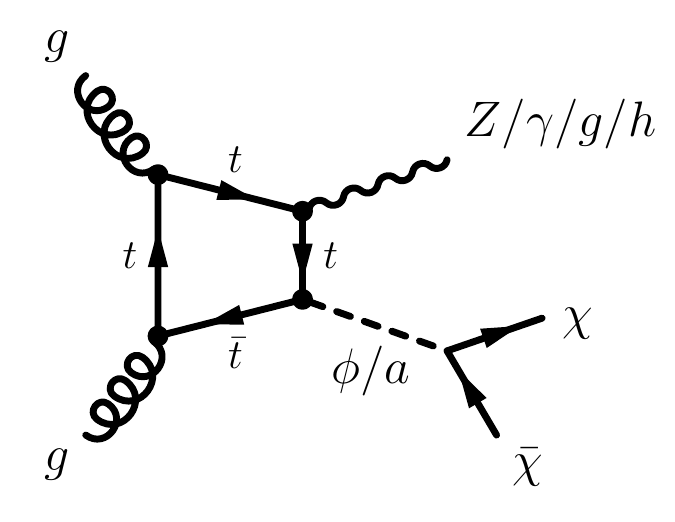}
}
\subfloat[]{\includegraphics[width=.26\textwidth]{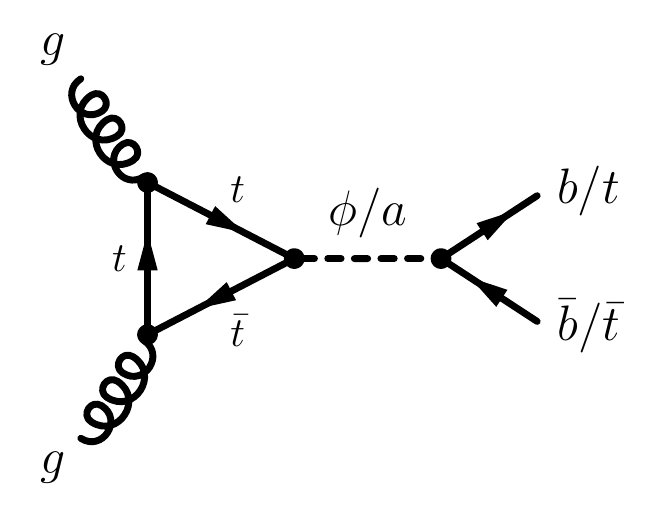}
} \\
\subfloat[]{\includegraphics[width=.26\textwidth]{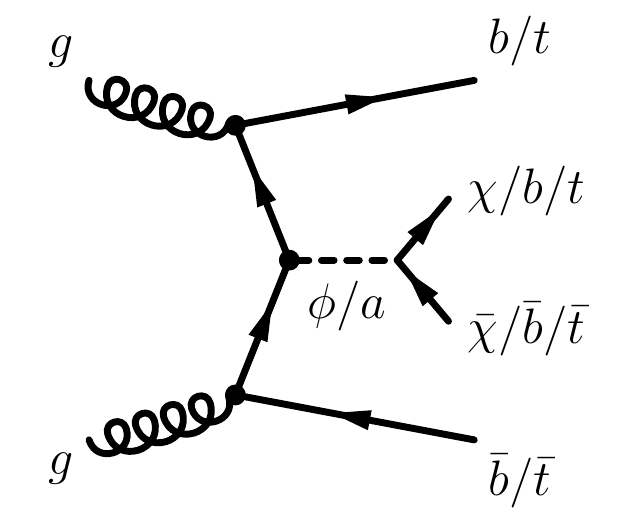}
}
\subfloat[]{\includegraphics[width=.26\textwidth]{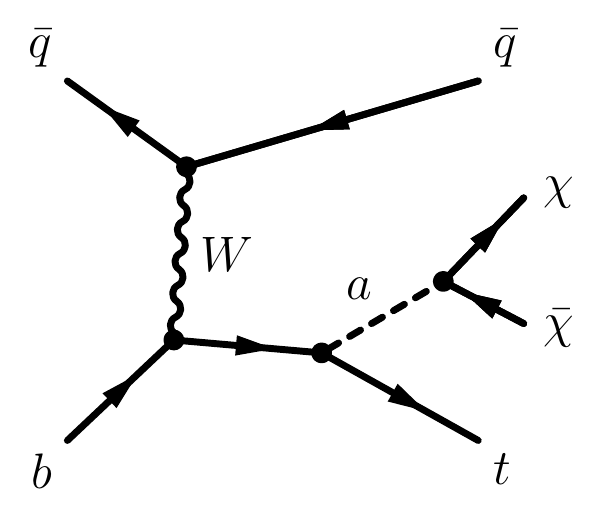}
}
\subfloat[]{\includegraphics[width=.26\textwidth]{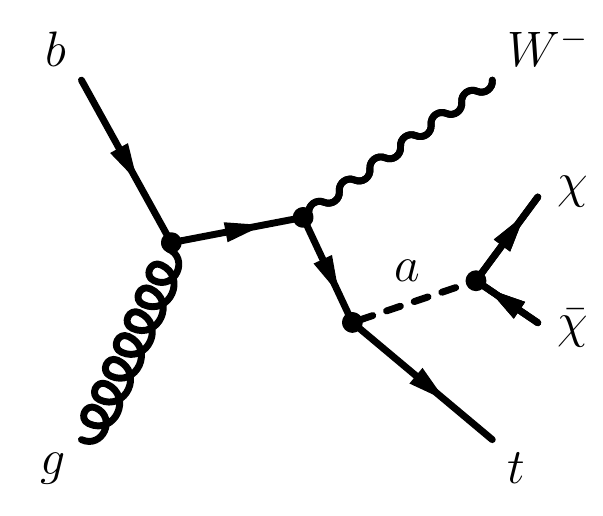}
} \\
\caption{Schematic representation of the dominant production and decay
modes for the S/PS models.}
\label{g:diagSpin0}
\end{figure}

\subsubsection{Colour-neutral interaction}
 
In the scalar or pseudo-scalar simplified models (S/PS) a new spin-0 gauge particle mediates the interaction,
at tree level, with a  DM particle~\cite{Buckley:2014fba,Abercrombie:2015wmb}.
The mediator is considered to be either a scalar (\scal) or a
pseudo-scalar (\pscal). This model has four parameters~\cite{SUSY-2016-18}: the mass of the
mediator $m_{\scal}$ or $m_{\pscal}$; the DM mass; the DM-mediator coupling,
$g_\chi$; and the coupling of the mediator with the
SM fermions. The latter is composed of a flavour-universal term,
$g_q$, which is a free parameter of the model and multiplies the
SM-Yukawa coupling for each of the fermions~\cite{Buckley:2014fba}.
This particular form of interaction, common to all models with
spin-0 mediators evaluated in this paper, is typically referred to as the
minimal flavour violation (MFV) ansatz and by construction, it relaxes the
severe restrictions on the coupling of new spin-0
colour-neutral particles to the SM fermions imposed by flavour
measurements~\cite{DAmbrosio:2002vsn,Charles:2004jd,PhysRevD.98.030001}. Furthermore, it implies that these  mediators are sizeably produced through
loop-induced gluon fusion or in association with heavy-flavour quarks
(see Figure~\ref{g:diagSpin0}).
According to whether the mediator decays into a pair of DM or SM
particles, different final states are sensitive to these
models. Due to the Yukawa-like structure of the couplings, visible
final states with two or four top quarks are particularly important
signatures. Final states involving a single top quark and \met\ may also
play an important role in constraining these models~\cite{Pinna:2017tay,Pani:2017qyd,Plehn:2017bys,Brooijmans:2018xbu,ATL-PHYS-PUB-2018-036,Haisch:2018bby}.
Despite the absence of a
dedicated parameter that regulates the relative importance of up-type
and down-type quark couplings (otherwise present in UV completions
of these models as in Section~\ref{sub:2HDMspin0theory}), it
is also important to study final states
involving bottom quarks separately, since these become a relevant signature if the
up-type couplings are suppressed.

\subsubsection{Colour-charged interaction}

The scalar colour-charged interaction model (\scg) assumes that the scalar mediator couples
to
left- or right-handed quarks and it is a colour triplet. The DM particle(s) is
produced via a
$t$-channel exchange of the mediator which leads to a
different phenomenology from that of colour-neutral interactions.
These models have a strong connection with the
minimal supersymmetric Standard Model (MSSM)~\cite{Fayet:1976et,Fayet:1977yc}
with a neutralino DM and
first- and second-generation squarks with universal masses. They share with it the same
cross-sections and phenomenology when the mediator is pair-produced via strong interaction. Nevertheless, additional production
diagrams are taken into account in this scenario, since
values assumed for the couplings of the mediator to quarks and DM differ from those of  the MSSM.
 
As in the case of the MSSM, it is reasonable to decouple the first two
generations from the third, considering the different mass scales. For this purpose, three different models are
considered:\footnote{These three scenarios provide benchmarks for each signature considered and
do not aim to be an exhaustive set of models involving colour-charged interactions}
 
\begin{figure}
\centering
\subfloat[]{\includegraphics[width=.24\textwidth]{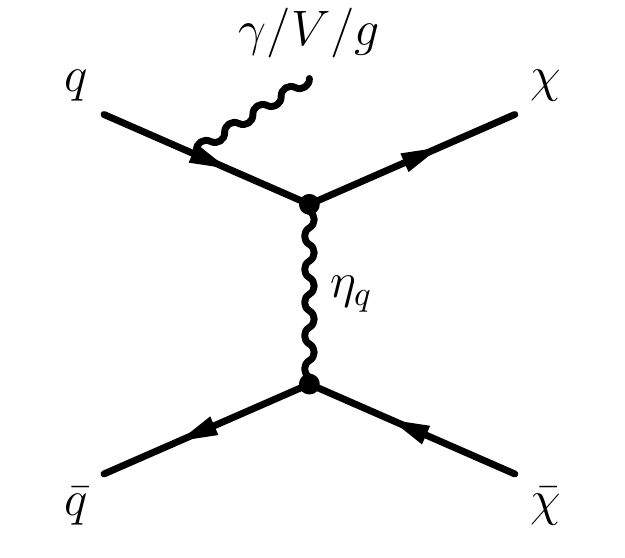}
\label{g:diagSpin0C1}
}
\subfloat[]{\includegraphics[width=.24\textwidth]{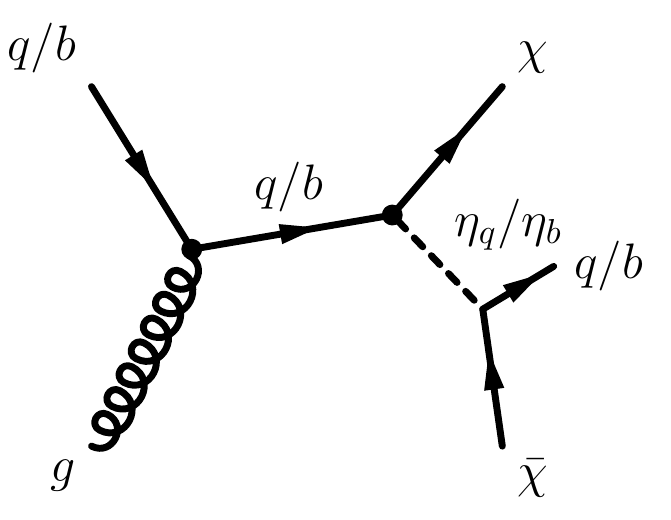}
\label{g:diagSpin0C2}
}
\subfloat[]{\includegraphics[width=.24\textwidth]{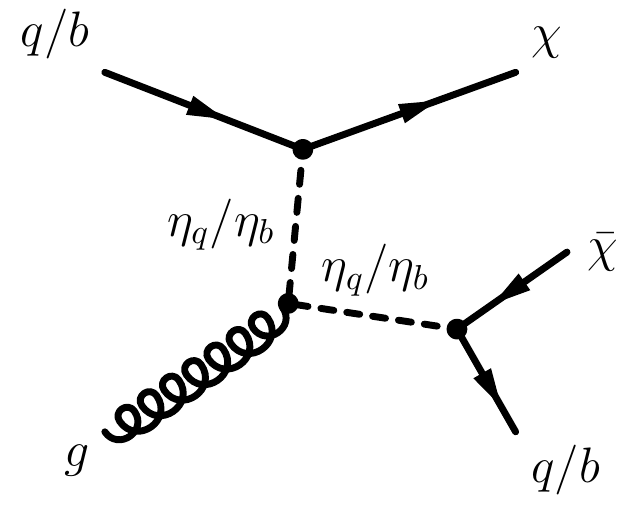}
\label{g:diagSpin0C3}
}
\subfloat[]{\includegraphics[width=.24\textwidth]{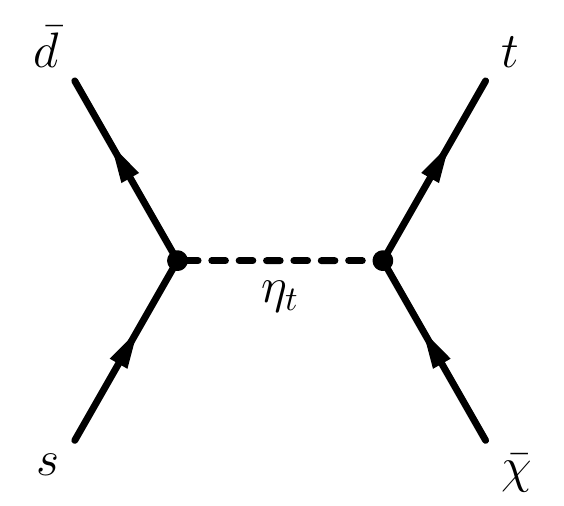}
\label{g:diagSpin0C4}
}
\caption{Schematic representation of the dominant production and decay
modes for the \scg\  models.}
\end{figure}

\begin{enumerate}
\item In the \scg$_{q}$ model, the mediator, $\eta_{q}$, couples to the left-handed quarks of
the first and second generations and is a SU(2) singlet under the
SM. The mediator decays into a quark--DM pair, so that the strongest
sensitivity for these models is provided by searches involving jets
and missing transverse momentum.
The three model parameters are the mediator mass, the DM mass, and
the flavour-universal coupling to quarks and DM, $\lambda_q$. This
model is described in detail in Refs.~\cite{Papucci:2014iwa,EXOT-2016-27} and
representative diagrams are shown in Figures.~\ref{g:diagSpin0C1},~\ref{g:diagSpin0C2}~and~\ref{g:diagSpin0C3}.
 
\item In the \scg$_{b}$ model, the mediator, $\eta_{b}$, couples to the right-handed bottom
quark. Following previous publications~\cite{EXOT-2014-01,SUSY-2016-18}, the
specific realisation of this model is obtained within the framework of
``flavoured'' DM, where the DM candidate is the lightest component of a
flavour triplet~\cite{Agrawal:2014una}. With these
assumptions, the mediator always decays into a $b$-quark--DM pair.
Of the three parameters
of the model, the mediator and DM masses and the coupling,
$\lambda_b$, only the first two are varied, while the last one is
set to the value predicting a DM relic density compatible with
astrophysical observations~\cite{DAmbrosio:2002vsn}. Representative
diagrams for these models are presented in
Figures.~\ref{g:diagSpin0C2}~and~\ref{g:diagSpin0C3}.
 
\item In the \scg$_{t}$ model, the mediator, $\eta_{t}$, consists of a SU(2)$_\text{L}$-singlet field
that couples to right-handed quarks, and is produced by down-type
quark--anti-quark fusion, and it decays into a top quark and a DM particle.  The representative diagram is
shown in Figure~\ref{g:diagSpin0C4}. This
specific realisation of the model~\cite{Boucheneb:2014wza}, which gives rise to a
characteristic signature composed of a single top quark and an
invisible particle, can be related to the MSSM if an additional R-parity violating interaction
of the top squark with the down-type quarks is assumed.
The coupling strength of the mediator to DM and top quarks,
denoted by $\lambda_{t}$, and the coupling strength to light-flavour down-type quarks, $g_{ds}$,
are free parameters of the model.

\end{enumerate}

\subsection{Extended Higgs sector dark matter models}
 
The third category of models aims to extend the simplified DM mediator models by
involving an extended two-Higgs-doublet sector (2HDM)~\cite{Ipek:2014gua,No:2015xqa,Goncalves:2016iyg,Bell:2016ekl,Bauer:2017ota,Bauer:2017fsw,Haisch:2017gql,Arcadi:2017wqi,Bell:2017rgi},
together with an additional mediator to DM, either a vector or a pseudo-scalar.
This embeds the simplified models in
a UV-complete
and renormalisable framework and allows the investigation of a broad phenomenology predicted by these
types of models. In both models, the 2HDM sector has
a CP-conserving potential and a softly broken $\mathbb{Z}_2$ symmetry~\cite{Gunion:2002zf},
and the alignment limit is assumed, so that the lightest CP-even
state, $h$, of the Higgs sector can be identified with the SM Higgs boson.

\subsubsection{Two-Higgs-doublet models with a vector mediator}
\label{sub:2HDMspin1theory}
 
\begin{figure}
\centering
\subfloat[]{\includegraphics[width=.26\textwidth]{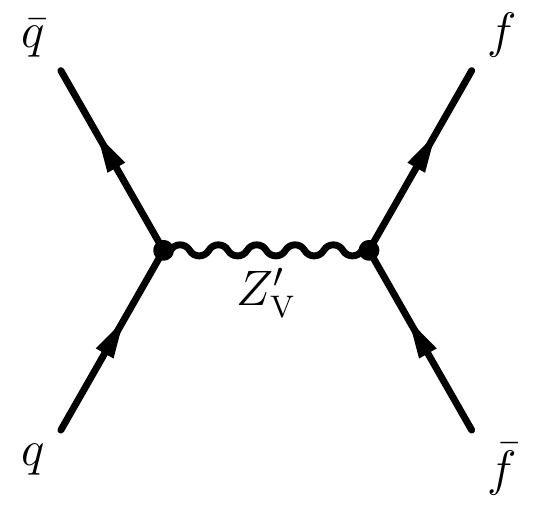}
}
\subfloat[]{\includegraphics[width=.26\textwidth]{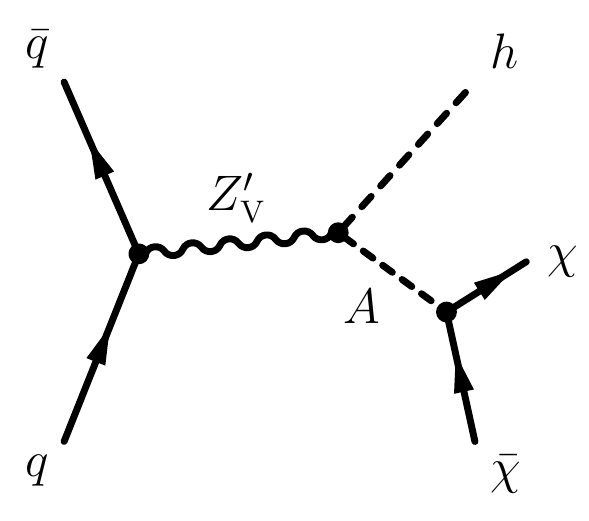}
}
\caption{Schematic representation of the dominant production and decay
modes for the \thdmZ\ model.}
\label{diag2HDMZ}
\end{figure}

The first two-Higgs-doublet model~\cite{Berlin:2014cfa}, denoted for brevity \thdmZ\ in
the following, is based on a type-II 2HDM~\cite{Gunion:2002zf,Branco:2011iw} with an
additional U(1) gauge symmetry, which gives rise to a new massive $\vvec$
gauge boson state. The $\vvec$ boson, which can mix with the \Zboson\ boson,  couples only to right-handed quarks and
only to the Higgs doublet that couples to the up-type fermions.
The CP-odd scalar mass eigenstate, $A$,  from the extended Higgs sector
couples to DM particles and complies with  electroweak precision
measurement constraints.
The phenomenology of this model is extended with respect to the simplified case
due to the presence of a new decay mode $\vvec \rightarrow hA$, as shown in
Figure~\ref{diag2HDMZ}, with the $A$ boson decaying into a pair of DM particles with a large branching ratio
(when kinematically possible),
as long as the decay into a pair of top quarks is kinematically forbidden~\cite{CMS-EXO-16-012}.
Additional signatures involving decays of the \vvec\ boson into SM particles or the $H$ and $H^\pm$ bosons
are possible in the model. However, the model parameters are chosen in order to be consistent with
the constraints from
searches for heavy-boson resonances on this model~\cite{EXOT-2014-20}, and therefore these signatures are not
considered further in the context of this interpretation.
The model has six parameters~\cite{EXOT-2014-20}:
$\tan\beta$, the ratio of the vacuum expectation values of the
two Higgs doublets, is set to unity;    $m_\chi$, the DM mass, is set to $100\;\GeV$; and $g_Z$,
the coupling of the new $\vvec$  U(1) gauge symmetry, is set to 0.8.
The masses $m_h$ and $m_H = m_{H^\pm}$ of the two CP-even and charged Higgs bosons are set to $125\;\GeV$,
and $300\;\GeV$, respectively, while $m_A$, the mass of the CP-odd Higgs partner and $m_{\vvec}$ are
free parameters and varied in the interpretation.
 
\subsubsection{Two-Higgs-doublet models with a pseudo-scalar mediator}
\label{sub:2HDMspin0theory}

The second 2HDM model~\cite{Bauer:2017ota},
\thdmS,
includes an additional pseudo-scalar
mediator, $\pscal$. In this case also, the 2HDM coupling structure is chosen to be of type-II,
although many of the interpretations in this paper hold for a type-I case too.
The additional pseudo-scalar mediator of the
model couples the DM particles to the SM and mixes with the
pseudo-scalar partner of the SM Higgs boson.
The physics of the model is fully captured by 14
parameters: the masses of the CP-even ($h$ and $H$), CP-odd ($\pscal$ and $A$)
and charged ($H^\pm$) bosons; the mass of the DM particle (\mchi);
the three quartic couplings between the scalar doublets and the $\pscal$ boson ($\lambda_\text{P1}, \lambda_\text{P2}$ and $\lambda_3$);
and the coupling between the $\pscal$ boson and the DM, $y_\chi$;
the electroweak VEV, $v$; the ratio of the VEVs of the
two Higgs doublets, $\tan\beta$; and the mixing angles of the
CP-even and CP-odd weak eigenstates, $\alpha$ and $\theta$, respectively.
The coupling $y_\chi = 1$ is chosen, having a negligible effect on the kinematics in the final states of interest.
The alignment and
decoupling limit ($\cos(\beta-\alpha) = 0$) is assumed,
thus $h$ is the SM Higgs boson and $v = 246\;\GeV$.
The quartic
coupling $\lambda_3 = 3$ is chosen to ensure the stability of the Higgs
potential for our choice of the masses of the heavy Higgs bosons which are
themselves fixed to the same value ($m_A = m_{H^\pm} = m_H$) to simplify the phenomenology and evade the
constraints from electroweak precision measurements~\cite{Bauer:2017ota}.
The other quartic couplings are also set to $3$ in
order to maximise the trilinear couplings between the CP-odd and the
CP-even neutral states.

\begin{figure}[h!]
\centering
\subfloat[]{\includegraphics[width=0.26\textwidth]{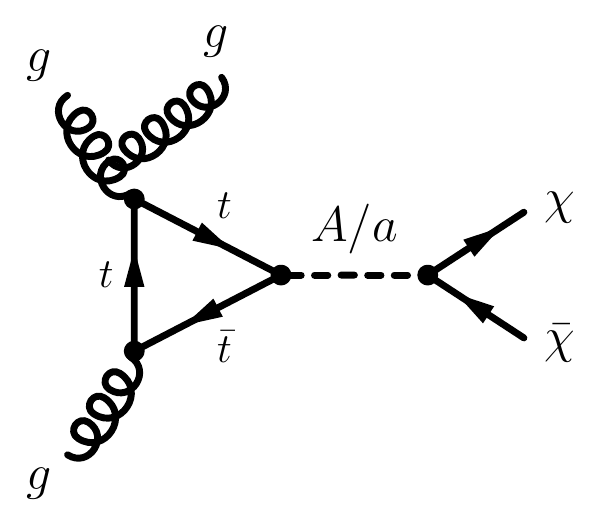}
\label{fig:THDMpa_feyna}
}
\subfloat[]{\includegraphics[width=0.26\textwidth]{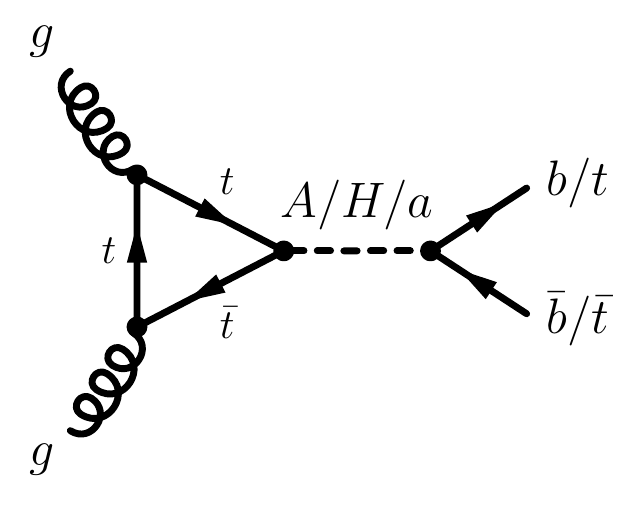}
}
\subfloat[]{\includegraphics[width=0.26\textwidth]{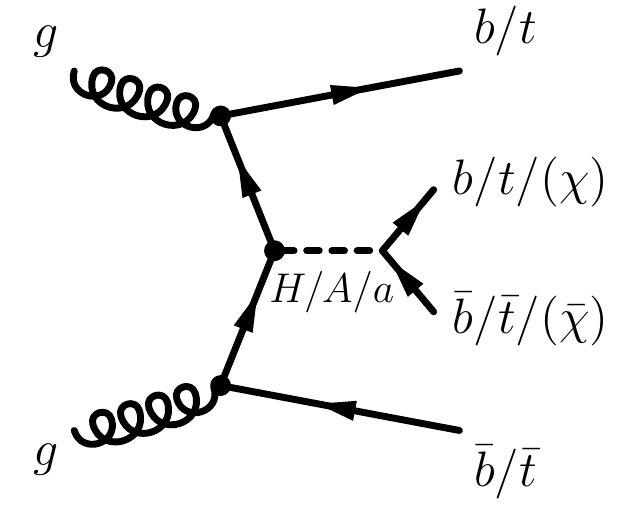}
\label{fig:THDMpa_feync}
} \\
\subfloat[]{\includegraphics[width=0.26\textwidth]{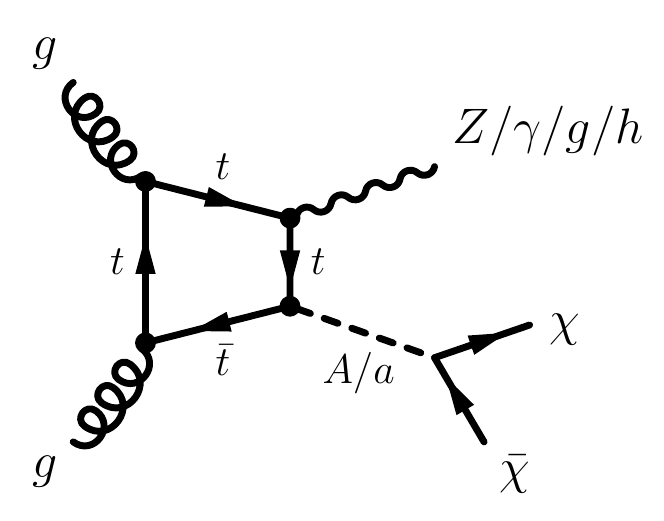}
\label{fig:THDMpa_feynd}
}
\subfloat[]{\includegraphics[width=0.26\textwidth]{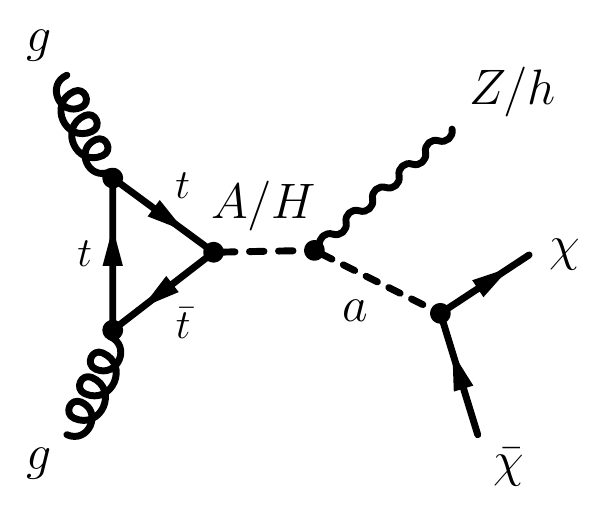}
\label{fig:THDMpa_feyne}
}
\subfloat[]{\includegraphics[width=0.26\textwidth]{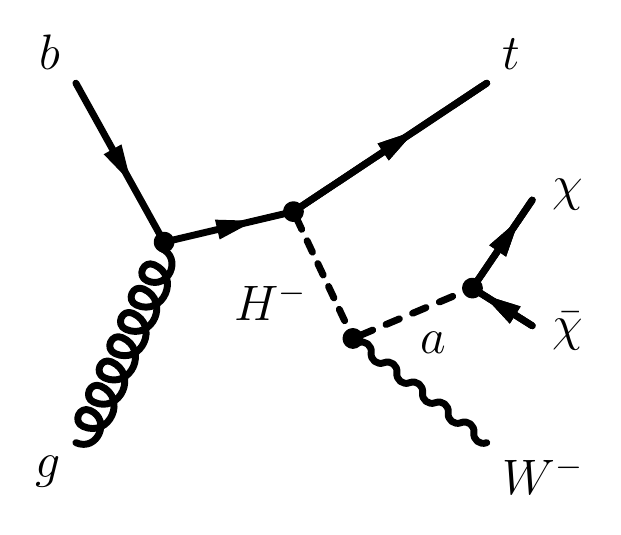}
\label{fig:THDMpa_feynf}
}\\
\caption{Schematic representation of the dominant production and decay
modes for the \thdmS\ model.}
\end{figure}

This model is characterised by a rich phenomenology.
The production of the lightest pseudo-scalar
is dominated by loop-induced gluon fusion,
followed by associated production with heavy-flavour quarks or associated
production with a Higgs or $Z$ boson (Figures~\ref{fig:THDMpa_feyna}-\ref{fig:THDMpa_feync}). Furthermore, according to
the Higgs sector's mass hierarchy,  Higgs and
$Z$ bosons can be produced in the resonant decay of the heavier bosons into the
lightest pseudo-scalar (see for example Figures~\ref{fig:THDMpa_feynd}-\ref{fig:THDMpa_feynf}). The pseudo-scalar mediator can
subsequently decay into either a pair of DM particles or a pair of SM particles
(mostly top quarks if kinematically allowed), giving
rise to very diverse  signatures.
The four-top-quark signature~\cite{ATL-PHYS-PUB-2018-027} is particularly interesting in this model if the neutral Higgs partner
masses are kept above the \ttbar\ decay threshold, since, when kinematically allowed, all
heavy neutral bosons can contribute to this final state, as depicted in the diagram of Figure~\ref{fig:THDMpa_feync}.
Four benchmark scenarios~\cite{2HDMWGproxi} that are consistent
with bounds from electroweak precision, flavour and Higgs observables
are chosen to
investigate the
sensitivity to this model as a function of relevant parameters:
$m_{\pscal},m_A, \tan\beta, \sin\theta$ and \mchi.

\subsection{EFT model of scalar dark energy}
\label{sub:DEtheory}

The Horndeski theories~\cite{Horndeski:1974wa} introduce a dark energy
scalar which couples to gravity and provide a useful framework for
constraining the cosmological constant problem
and the source of the acceleration of the expansion of the universe.
The model considered in this paper is an EFT implementation of these
theories~\cite{Brax:2016did}. In this model, the dark energy field is
assumed to couple to matter universally.
The model
contains two classes of effective operators: operators which are invariant under
shift-symmetry $\phiDE\rightarrow\phiDE + \mathrm{constant}$, where $\phiDE$ denotes the DE scalar
field, and operators which break this symmetry. Shift-symmetric operators
contain derivative interactions of $\phiDE$ with the SM particles, while
operators that break the shift-symmetry contain direct interactions of $\phiDE$
with the SM.
In the former case the DE scalar is pair-produced and does not decay
within the volume of collider experiments, thereby resulting in \met\ in the
detector, while the latter case includes Yukawa-type interactions
$\phiDE\bar{\psi}\psi$, which allow the scalar to decay
into SM fermions, thereby changing the expected signatures. The
interactions
arising from the shift-symmetry breaking operators
are tightly constrained~\cite{Joyce:2014kja} and are not
evaluated here.
 
\begin{figure}[h!]
\centering
\subfloat[]{\includegraphics[width=0.26\textwidth]{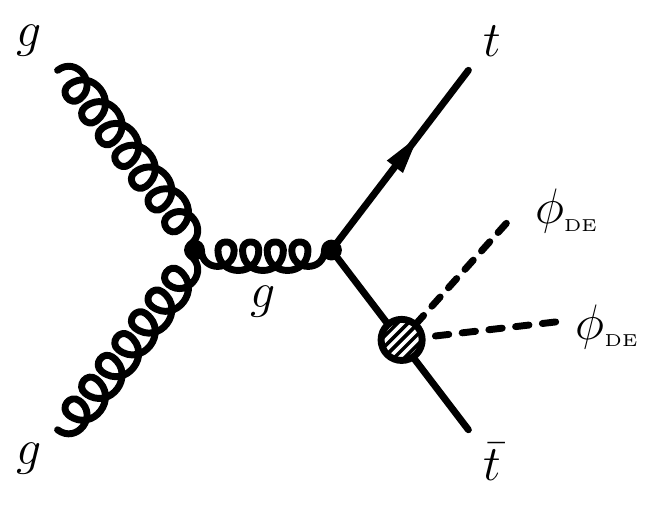}
}
\subfloat[]{\includegraphics[width=0.26\textwidth]{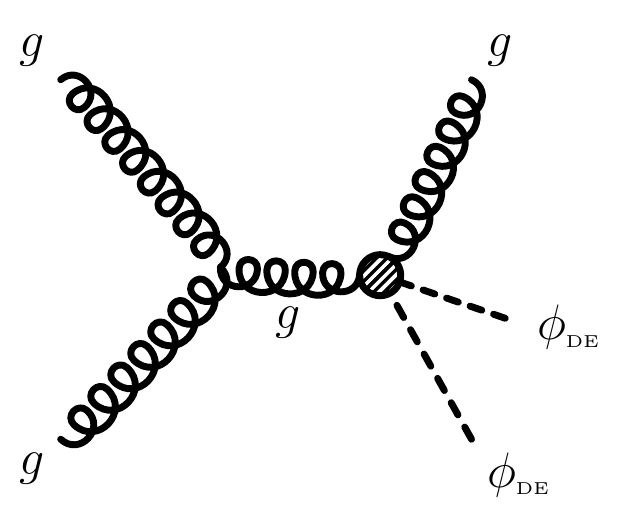}
}
\caption{    Schematic representation of representative production
modes for the DE model for the Lagrangian effective operators
(a) $\mathcal{L}_1$ and (b) $\mathcal{L}_2$.}
\label{fig:de_feynman_diagrams}
\end{figure}

There are nine shift-symmetric Lagrangian effective operators in the model, each
suppressed by powers of a characteristic energy scale $M$ according to the
operator's dimensionality:
\begin{equation*}
\mathcal{L}=\mathcal{L}_{\mathrm{SM}}+\sum_{i=1}^9 c_i\mathcal{L}_i=\mathcal{L}_{\mathrm{SM}}+\sum_{i=1}^9 \frac{c_i}{M_i^{d-4}}\mathcal{O}^{(d)}_i,
\end{equation*}
where $d$ is the operator's dimension and $c_i$ are the Wilson coefficients.
Operators $\mathcal{L}_1$--$\mathcal{L}_5$ correspond to interactions of the
DE field with SM fields.
The leading, i.e. least suppressed, operators of dimension eight are
\begin{eqnarray*}
\mathcal{L}_1&=&\frac{\partial_{\mu}\phiDE\partial^{\mu}\phiDE}{M_1^4}T^{\nu}_{\nu}\\
\mathcal{L}_2&=&\frac{\partial_{\mu}\phiDE\partial_{\nu}\phiDE}{M_2^4}T^{\mu\nu},
\end{eqnarray*}
where $T^{\mu\nu}$ is the energy-momentum tensor corresponding to the SM
Lagrangian. The $\mathcal{L}_1$ operator corresponds to a derivative coupling of the DE field to the conformal
anomaly, $T^{\nu}_{\nu}$ ($=m\bar{\psi}\psi$ for a Dirac field), and
is therefore proportional to
the mass of the SM fermions to which DE couples. Signatures which probe DE production in association with $t\bar{t}$ are therefore the most sensitive to this type of coupling and are used here. The $\mathcal{L}_2$ operator
involves derivatives of the SM fields and is therefore proportional to their
momenta. Final states involving large momentum transfers, such as the \monojet\ signature,  offer the
highest sensitivity to this type of coupling. The $\mathcal{L}_1$ and $\mathcal{L}_2$ operators are referred to as (kinetically dependent) conformal~\cite{Brax:2016kin} and
disformal, respectively.
Operators $\mathcal{L}_3$--$\mathcal{L}_5$ correspond to higher-order versions of $\mathcal{L}_1$ and $\mathcal{L}_2$.
The operator $\mathcal{L}_6$ corresponds to a generalised kinetic term for the DE scalar and operators $\mathcal{L}_7$--$\mathcal{L}_9$ correspond to the non-trivial Galilean terms~\cite{Nicolis:2008in}.
In this paper, only $\mathcal{L}_1$ and $\mathcal{L}_2$ are considered. Due to the absence
of terms allowing the decay of the DE scalars into SM particles, the DE particles ($\phiDE$) are
stable and they escape the detector producing a missing-momentum signature.
 
The validity of the EFT approach in the context of
collider data~\cite{Busoni:2013lha,Busoni:2014sya,Busoni:2014haa} is assessed with the
procedure described in Ref.~\cite{Abercrombie:2015wmb},
imposing the condition $\sqrt{\hat{s}}<g_{*}M$,
where $\sqrt{\hat{s}}$ is the centre-of-mass energy of the hard interaction and
$g_{*}$ is the effective coupling associated with the UV completion of the EFT.
 
Representative Feynman diagrams corresponding to the $\mathcal{L}_1$ and
$\mathcal{L}_2$ operators for the $t\bar{t}+\met$ and mono-jet signatures are shown in Figure~\ref{fig:de_feynman_diagrams}.


\section{Dataset and Signal simulation}
\label{sec:montecarlo}
 
This paper interprets analyses of $pp$ collision data recorded at a centre-of-mass energy of $\sqrt{s} = 13~\TeV$  by the ATLAS detector during 2015 and 2016. Unless otherwise
specified, the integrated luminosity of the dataset, after requiring that all detector subsystems were operational during data recording, amounts to $36.1 \pm 0.8$~\ifb.

Monte Carlo (MC) simulated event samples were used to aid in the
estimation of the background from SM processes and to model the DM and
DE signals.
Simulated events were processed either through a detector
simulation~\cite{SOFT-2010-01} based on {\textsc
Geant4}~\cite{Agostinelli:2002hh} or  through a fast
simulation~\cite{SOFT-2010-01} with a parameterisation of the
calorimeter response and {\textsc Geant4} for the other parts of the
detector~\cite{ATL-PHYS-PUB-2010-013}. Either of these ATLAS detector simulations were used
for background processes (details in the specific analysis references)
and most of the signal processes, as detailed in the following.
 
Two sets of samples were used for the modelling of the signal processes
considered in this paper. One set of samples is based on
signal events processed through the ATLAS detector
simulation, referred to as ``reconstructed'' samples. The second set of
samples consists of  signal events composed of
particle-level objects, defined according to the guiding principles
outlined in Ref.~\cite{ATL-PHYS-PUB-2015-013}, and not including
any resolution effect due to the ATLAS detector. These are referred to as ``particle-level''
samples.
Particle-level samples were used to define
a rescaling procedure specifically designed to
broaden the range of signal models and parameter choices
considered in the interpretation of the results.
The procedure allows the use of less extensive  computational resources that would be needed
to provide a full detector simulation for the large set of considered signals,
while providing a complete picture of the current  experimental coverage for these models.
The rescaling procedure calculated a set of correction weights for a reference model as
the ratio of the acceptance for a baseline signal sample
to the acceptance of the signal sample of interest.
Both of these acceptances are derived in a particle-level simulation.
These weights were then applied to the reconstructed baseline signal sample of the reference model,
assuming similar detector effects for the two models.
The same procedure was used in some cases to rescale
between signal samples of the same reference model but for different parameter
choices which affect the kinematics of the final state. Closure-tests
were performed to determine the reliability of this procedure and assign specific
systematic uncertainties when needed.
Further details about the rescaling used in the V/AV, VFC and the \thdmS\ signal samples
are given in Appendix~\ref{app:rescaling}.

The generation settings for signal models considering a spin-1 mediator are
summarised in Table~\ref{tab:MCvectorsignal}. For each model the table indicates the
Universal FeynRules Output (UFO)
model~\cite{Degrande:2011ua} implementation, the matrix element (ME)
generator, the parton shower (PS), and the cross-section normalisation,
at QCD leading-order or next-to-leading order accuracy (LO and NLO,
respectively).   Following the notation of
the previous section, the simplified models are indicated with $Z'_\text{V/A}$,
while the baryon-charged and flavour-changing interactions are
indicated as \bvec\ and \nvec, respectively. The 2HDM model with an additional
vector mediator is indicated as \thdmZ. When relevant for the
generations settings, each separate final state considered in
this paper is indicated for each model.
 
The generation settings for signal models considering a spin-0 mediator are
summarised in Table~\ref{tab:MCscalarsignal}. Following the notation of
the previous section, the colour-neutral (colour-charged) simplified models are indicated with $\phi/a$
($\eta$).
The 2HDM with additional
pseudo-scalar mediator is indicated as \thdmS.

The model implementations, settings and parameter scans follow the
prescriptions of the DM Forum/LHC DM Working Group
\cite{Abercrombie:2015wmb,Boveia:2016mrp,Albert:2017onk,2HDMWGproxi}.
 
Finally, the generation settings for the DE model are also indicated  in Table~\ref{tab:MCdesignal}.

 
\section{Experimental signatures}
\label{sec:analyses}
 
Dark matter searches are an important component of the ATLAS physics
programme. Several final-state signatures are targeted to maximise the
discovery potential.
This section presents summaries of the different searches for DM and is not intended to be exhaustive.
More details are available in the reference papers.
Table~\ref{tab:invisible} summarises the DM searches for invisible final states, while
Table~\ref{tab:visible} summarises the searches for visible final states. These tables also provide an overview of the models (Table~\ref{t:ModelSummaryNotation})
which are constrained by each of these signatures and which of these intepretations have not been presented elsewhere.

Electrons, muons, photons and jets are reconstructed by combining the
signals from the different components of the ATLAS
detector\footnote{\AtlasCoordFootnote\ The rapidity is defined at $y = 1/2 \ln [(E+p_z) /  (E-p_z)]$,
where $E$ is the energy and $p_z$ is the component of its momentum along the $z$-axis. The rapidity
difference between two jets is defined as $y^* = (y_1 - y_2)/2$. }~\cite{PERF-2016-01,ATLAS-CONF-2016-024,PERF-2015-10,PERF-2017-02,ATL-PHYS-PUB-2015-036}.
Leptons ($\ell$) in the following refers to electrons or muons. In several
analyses, events with identified leptons are rejected from the signal
region selection. This is referred to as a lepton veto. The analyses
may implement different lepton and photon selection criteria for
particle identification \cite{PERF-2017-02,ATL-PHYS-PUB-2015-041,PERF-2015-10,PERF-2016-01},
isolation \cite{PERF-2017-03,PERF-2015-10,PERF-2016-01}, and kinematic requirements
($\pt,\eta$).
Small-$R$ and large-$R$ jets are reconstructed from energy deposits in the calorimeters using
the anti-$k_t$ jet
algorithm~\cite{paper:antikt,Cacciari:2011ma} and using a radius parameter
of $R = 0.4$ and $R=1.0$, respectively.
Reclustered large-$R$ jets are reconstructed
from small-$R$ jets using a radius parameter of either
$R=0.8$ or $R=1.2$.
Multivariate algorithms  are used to
identify
small-$R$ jets with $\pt > 20\;\GeV$ containing $b$-hadrons
($b$-jets)~\cite{PERF-2012-04,ATL-PHYS-PUB-2016-012}.
This is referred to as $b$-tagging.
For large-$R$
jets, $b$-tagging is applied to their associated track-jets, which
are constructed from tracks reconstructed in the inner detector
using the anti-$k_t$ jet algorithm with $R=0.2$.
The missing transverse momentum $\ptmiss$ (with
magnitude $\met$) is calculated from the negative vector sum of transverse
momenta ($\pt$) of electrons, muons
and jet candidates and an additional soft term~\cite{PERF-2016-07}
which includes activity in the tracking system
originating from the primary vertex but not
associated with any reconstructed particle.
Some analyses may also consider photons in the \met\ reconstructions.

\subsection{Searches for invisible final states}
Searches for WIMP candidates at the LHC are characterised by the requirement of large \met\
since WIMPs escape detection. Therefore, final states with additional
visible particles are considered in the selection of the events. These
additional particles may come from initial-state radiation or from
associated production. Several signatures  that are listed in
the following are exploited and optimised to enhance the sensitivity to different DM models.
 
\paragraph{Jet $+ \met$}
The \monojet\ analysis~\cite{EXOT-2016-27}, commonly referred to as the mono-jet analysis, is characterised by the presence of an energetic jet and large \met.
The analysis selects events with
$\met>250\;\GeV$,
at least one jet with $\pt > 250\;\GeV$,
and at most three additional jets with  $\pt > 30\;\GeV$.  Events are required to pass a lepton veto.
To reduce the contribution from multi-jet background where large \met can originate from jet energy under-measurement, a minimum separation in the azimuthal angle between each selected jet and the \met direction is also required: $\Delta\phi(\textrm{jet},\ptmiss)>0.4$.
The $W$+jets, $Z$+jets, and top-quark-related backgrounds are constrained using MC event samples normalised to data in selected control regions containing leptons.
In the case of $W$+jets and $Z$+jets events, MC predictions are reweighted to account for higher-order QCD and electroweak corrections as a function of the vector-boson $\pt$~\cite{Lindert:2017olm}.
The normalisation factors for these backgrounds are extracted
simultaneously using a binned likelihood fit of the \met\ distributions in all control and signal regions that includes systematic uncertainties. 
The remaining SM backgrounds from diboson processes are determined using MC simulated samples, while the multi-jet background contribution is extracted from data.
 
\paragraph{\hinv}
Searches for invisible Higgs boson decays have been performed using
several production and decay channels at a centre-of-mass energy of
$\sqrt{s} = 8\;\TeV$~\cite{HIGG-2015-03}. Results of searches in the vector-boson
fusion (VBF) production channel and in
associated production of a Higgs boson with a $W/Z$
boson are statistically combined with the
measured production and decay rates of the Higgs boson in the $\gamma\gamma$, $ZZ$, $WW$, $Z\gamma$, $bb$, $\tau\tau$, and $\mu\mu$ channels
to set an upper limit on the Higgs boson's invisible branching ratio of
0.23 at 95\% confidence level (CL).
This combined limit is used in the results quoted in Section~\ref{sec:result}.
Among the direct searches, the VBF production of Higgs bosons decaying into invisible particles~\cite{HIGG-2013-16} is the most sensitive one, setting an upper limit on the invisible branching ratio of 0.28.
The VBF+\met\ analysis requires $\met > 150\;\GeV$ and two jets with $\pt>35\;\GeV$. Three orthogonal signal regions are defined by varying the threshold imposed on the leading jet $\pt$ and the invariant mass of the two jets. Additional requirements on the angular separation of the two jets are applied to enhance the sensitivity to VBF production.
In particular, the two leading jets are required to be well separated in pseudorapidity.
Lepton and $b$-jet vetoes are applied to reduce contamination from
$W+$jets and top-quark backgrounds, respectively.
Dedicated control regions with one and
two leptons in the final state are used to constrain the contributions
from
dominant $Z/W$+jets backgrounds, through a simultaneous fit together
with the signal region.
The multi-jet background is estimated using a data-driven technique.
Searches for $Zh$(inv)~and $Vh$(inv)~\cite{HIGG-2016-28,EXOT-2016-23,Aaboud:2019rtt} have been performed at centre-of-mass energy
$\sqrt{s}=13\;\TeV$. Constraints using a VBF+\met\ analysis are also
available using $\sqrt{s}=13\;\TeV$ $pp$ collision data~\cite{EXOT-2016-37,Sirunyan:2018owy}. However, the $8\;\TeV$ combination gives more stringent limits, thus it is used here.
 
\paragraph{\monophoton}
Events in the \monophoton\ analysis~\cite{EXOT-2016-32} are required to pass the lepton veto and to have a photon with $\ET > 150\;\GeV$.
Events with more than one jet ($\pt > 30\;\GeV$) or with a jet fulfilling $\Delta\phi(\textrm{jet},\ptmiss)<0.4$ are rejected.
Three exclusive signal regions with \met  ranges between $150\;\GeV$, $225\;\GeV$, $300\;\GeV$ and above are defined.
The $W\gamma$, $Z\gamma$, and $\gamma$+jets backgrounds are normalised in control regions using a
simultaneous likelihood fit of all \met regions, but with independent normalisation factors for each region.
The backgrounds due to photons from the misidentification of electrons or jets in processes such as $W/Z$+jets, diboson, and multi-jet events are estimated using data-driven techniques.
 
\paragraph{\monoZll}
The event selection criteria in this analysis~\cite{HIGG-2016-28} require large \met and a pair of high-$\pt$ leptons. Two opposite-sign, same-flavour leptons satisfying $\pt > 30\;\GeV$ and $\pt > 20\;\GeV$ are required.
The lepton pair is required to have an invariant mass between $76\;\GeV$ and $106\;\GeV$ to be consistent with originating from a $Z$ boson.
Events with an additional lepton with $\pt>7\;\GeV$ or a $b$-jet with $\pt > 20\;\GeV$ are vetoed.
To target events consistent with a boosted $Z$ boson produced in the direction opposite to $\ptmiss$,
additional requirements are imposed on the azimuthal angle between the dilepton system and $\ptmiss$ and on the angular distance between leptons.
A single inclusive \met signal region is defined with $\met>90\;\GeV$ for each of the $ee$ and $\mu\mu$ channels.
The dominant background in this analysis, $ZZ$ production, is estimated from MC simulation.
The $WZ$ background is normalised to data in a three-lepton control region.
The contributions from $Z$+jets and non-resonant-$\ell\ell$ backgrounds are estimated using data-driven techniques.
A statistical combination of the two decay channels is used for the final results.
 
\paragraph{\ensuremath{W(qq')/Z(q\bar{q})+\met}}
This analysis~\cite{EXOT-2016-23} selects events with $\met > 150\;\GeV$ and a hadronically decaying $W$ or $Z$ boson candidate. The vector-boson candidate is defined with one large-$R$ jet with $\pt > 250\;\GeV$ in a boosted topology ($\met > 250\;\GeV$) or with two small-$R$ jets with $\pt >20\;\GeV$ in a resolved topology. In both cases, a lepton veto is applied. Additional requirements are applied to the invariant mass of the boson candidate. Several signal regions are defined according to the $b$-jet multiplicity. Similarly, several control regions are defined according to lepton and $b$-jet multiplicity. The normalisations of the $\ttbar$ and $W/Z$+jets background processes are constrained using a simultaneous fit of all control and signal regions of the $\met$ distribution.
The subdominant contribution from diboson and single-top-quark production is obtained from simulation. Multi-jet contributions are estimated with a data-driven technique.
 
\paragraph{\monoHbb}
The \monoHbb\ analysis~\cite{EXOT-2016-25} is defined by the requirement of $\met > 150\;\GeV$, a lepton veto, and the presence of a Higgs boson candidate decaying to two $b$-jets with suitable invariant mass.
Events with mis-measured \met are rejected by imposing constraints on $\Delta\phi(\textrm{jet},\ptmiss)$, between the missing momentum direction and the direction of any selected jet in the event.
Two sets of signal regions are defined targeting moderate-momentum (resolved) and high-momentum (boosted) Higgs boson candidates. In each case, the regions are further split according to whether there are one or two $b$-jets.
The resolved regime, defined in three exclusive \met regions between $150\;\GeV$ and $500\;\GeV$, selects a Higgs boson candidate reconstructed from the two leading $b$-tagged small-$R$  jets (or from a $b$-tagged and a non-$b$-tagged small-$R$ jet) with $\pt>20\;\GeV$.
In the boosted regime, defined by $\met>500\;\GeV$, the leading large-$R$ jet with $\pt>200\;\GeV$ is the Higgs boson candidate. The $b$-jet multiplicity is defined by the number of $b$-tagged track-jets associated with the large-$R$ jet.
Backgrounds involving the production of $W/Z$ bosons in association with heavy-flavour quarks or top-quark pairs
are normalised in dedicated control regions distinct from the signal regions by requiring one or two leptons.
A simultaneous binned likelihood fit to the
invariant mass of the Higgs boson candidate is performed in all signal and control regions. The multi-jet background is obtained with a data-driven technique. Other subdominant backgrounds are estimated from simulation.
 
\paragraph{\monoHgg}
The \monoHgg\ events in this analysis~\cite{HIGG-2016-18} are selected by requiring at least two photons with $\pt>25\;\GeV$. The two leading photons
are chosen to reconstruct the Higgs candidate, which is required to satisfy $105\;\GeV < m_{\gamma\gamma} < 160\;\GeV$.
The leading (sub-leading) photon is also required to have $\ET^\gamma / m_{\gamma\gamma}>0.35$~$(0.25)$. Events with leptons are vetoed.
Events with $\pt(\gamma\gamma)>90\;\GeV$ and $\met / \sqrt{\sumET} > 7\;\GeV^{1/2}$ in Ref.~\cite{HIGG-2016-18} are used for the interpretation of DM models, where $\sumET$ is the scalar sum of the transverse momentum of all reconstructed objects in the event. The backgrounds are extracted by fitting an analytic function to the diphoton invariant mass distribution.
In the case of the non-resonant background, the normalisation and shape are obtained by fitting the invariant mass distribution in data to an exponential function.
The SM Higgs boson background shape is modelled with a double-sided Crystal Ball function and fitted to the MC simulation.

\paragraph{\monotop}
The \monotop\ analysis~\cite{EXOT-2017-16} searches for events with one top quark and relatively large \met. Two signal regions are defined depending on the decay channel of the top quark. The leptonic channel selects events with a positively charged lepton with $\pt>30\;\GeV$, $\met > 50\;\GeV$, and transverse mass of the lepton and the \met, $\mT^W$, larger than $260\;\GeV$. One $b$-jet with $\pt > 30\;\GeV$ is additionally required. The hadronic channel is optimised to select events with a top quark produced with a large boost. Events are selected with $\met > 200\;\GeV$ and one large-$R$ jet with $\pt > 250\;\GeV$ with one $b$-tagged track-jet associated with it. Events failing the lepton veto are rejected.
Dedicated control regions are defined to constrain the $\ttbar$ and $W/Z$+jets backgrounds from data. The multi-jet background is estimated from data, whereas other remaining backgrounds are taken from simulation. All signal and control regions for the two decay channels are fitted simultaneously to extract the background normalisation. In the case of the hadronic channel, the transverse mass of the large-$R$ jet  and the \met  are the discriminating variables, while for the leptonic channel, the \met distribution is used to discriminate signal from background.

\paragraph{$b(\bar{b}) + \met$}
The $b + \met$ analysis~\cite{SUSY-2016-18} selects events with two
energetic jets ($\pt > 160\;\GeV$), at least one of which is \btagged,
$\met > 650\;\GeV$, and additional total hadronic energy restricted to be less than
$100\;\GeV$.
This last requirement rejects top-quark background.
The dominant background for this analysis, $Z$+jets events, is
constrained from data in a dedicated control region, which is fitted
together with the signal region.
The \dmbb\ analysis~\cite{SUSY-2016-18} instead exploits a selection
with at least two $b$-jets
and a considerably lower \met\ requirement, $\met > 180\;\GeV$.
The azimuthal separations between the $b$-jets and
$\ptmiss$  are exploited
to enhance the separation between
the signal and the irreducible background in this channel
($Z(\nu\bar{\nu})$+$b\bar{b}$), which is constrained from data in a dedicated control region.
The results are extracted by fitting an observable that relies on the pseudorapidity difference between the two $b$-jets:
$\cos\theta^*_{bb} = \left|
\tanh\left(\Delta\eta_{bb}/2\right)\right|$.

\paragraph{$\ttbar + \met$}
The \dmtt\ analysis~\cite{SUSY-2016-15,SUSY-2016-18,SUSY-2016-16}
is split into three channels according to the decays of the $W$ bosons
from the top-quark decays: 0-lepton, where both $W$ bosons decay
hadronically, 1-lepton, where one of the two $W$ bosons decays leptonically
and 2-leptons where both $W$ bosons decay leptonically.
The analyses targeting the 0-lepton channel exploit two
sets of signal topologies: spin-0 DM
models \cite{SUSY-2016-18}, used for the DM interpretations
presented in this paper,  and top-squark decays into a top quark and a neutralino
\cite{SUSY-2016-15}, used for the DE interpretation in this paper.
Both analyses are characterised by a
set of signal regions which select events with at least four energetic
jets, at least two of which are \btagged, and relatively
high \met.
Requirements on the invariant mass of reclustered large-$R$ jets are imposed
to identify events where a $W$ boson or a top quark are boosted.
The dominant backgrounds ($Z$+jets, top-quark processes and $\ttbar+Z$)
are constrained in dedicated control regions. The three signal regions
used for the DE interpretation
are statistically combined, while the two signal regions
in the DM analysis are not.
The analysis targeting the 1-lepton channel selects events with
at least four energetic jets, at least one of which is \btagged, one isolated lepton and high \met.
The events are also required to have at least one hadronic top candidate with invariant mass
loosely compatible with the mass of the top quark.
Requirements on the transverse and asymmetric stransverse masses \cite{SUSY-2016-16}
are used to suppress semileptonic and dileptonic \ttbar\ events, respectively.
Requirements on the azimuthal angle between the lepton and $\ptmiss$ and
on $\Delta\phi(\textrm{jets},\ptmiss)$ are also exploited to further
suppress the background contamination of the signal regions. All top-quark background processes are estimated in dedicated control regions.
Finally, the analysis targeting the 2-lepton channel selects events
with two opposite-sign leptons which are inconsistent
with being produced in the decay of a $Z$ boson. At least
one $b$-jet is also required in the selections.
The \met and the
stransverse mass ($m_{\mathrm T2}^{\ell\ell}$ \cite{SUSY-2016-18}) requirements are such that
$m_{\mathrm T2}^{\ell\ell}+0.2\cdot\met > 170\;\GeV$. The dominant backgrounds in
this channel ($\ttbar$ and $\ttbar+Z$) are estimated in dedicated
control regions.

None of these analyses shows a significant deviation from the expected SM background, and thus exclusion limits can be set for the relevant models. These limits are discussed in Section~\ref{sec:result}.
The observed \met distributions compared with the background predictions are shown in Figure~\ref{fig:Bkg2HDMps} for the \monoHbb\ and \monoZll\ analyses, with representative \thdmS\  signal distributions shown in each case.
These two analyses have the strongest sensitivity for this model, as discussed in Section~\ref{sub:2HDMspin0res}.
The observed $m_{\mathrm{T}2}^{\chi^{2}}$ and \met\  distributions compared with the background predictions are shown in Figure~\ref{fig:BkgDE} for the \dmtt\ (0-lepton channel) and \monojet\ analyses, respectively, with representative DE signal distributions shown in each case. Figures~\ref{fig:Bkg2HDMps}~and~\ref{fig:BkgDE} show background predictions after the corresponding fit in each analysis.

\begin{table}
\caption{Summary of searches for invisible final states used to constrain the  different DM models defined in Section~\ref{sub:DMmodels}. The $(*)$ indicates models which were presented in the original publication, all others are either new or updated.}
\label{tab:invisible}
\centering
\resizebox{\textwidth}{!}{
\begin{tabular}{m{40mm}m{30mm}m{55mm}m{50mm}m{30mm}}
\toprule
Analysis  & Models targeted & Final-state signature & Key Characteristics & Results \\\midrule
Jet $+ \met$~\cite{EXOT-2016-27} & V/AV$(*)$, S/PS$(*)$, SCC$_{q}$$(*)$, DE
& 1--4 jets, \met, 0 $\ell$. 
& Binned likelihood fit of \met.
& Section~\ref{ssub:spin1res1}, \ref{ssub:spin0res1}, \ref{ssub:spin0res2}, \ref{sub:DEres}\\[0.5ex]
\midrule
\hinv~\cite{HIGG-2013-16,HIGG-2015-03} & \thdmS
& 2 jets, \met, $m_{jj}$, $\Delta\eta_{jj}$. 
& Single-bin likelihood fit.
& Section~\ref{sub:2HDMspin0res}\\[0.5ex]
\midrule
\monophoton~\cite{EXOT-2016-32} & V/AV$(*)$
& 1 photon, 0--1 jets, \met, 0 $\ell$. 
& Binned likelihood fit of \met.
& Section~\ref{ssub:spin1res1}\\[0.5ex]
\midrule
\monoZll~\cite{HIGG-2016-28} & V/AV, \thdmS
& 2 $\ell$, \met, $m_{\ell\ell} \sim m_Z$. 
& Binned likelihood fit of \met
& Section~\ref{ssub:spin1res1}, \ref{sub:2HDMspin0res}\\[0.5ex]
\midrule
\monoVqq~\cite{EXOT-2016-23} & V/AV, \thdmS
& \met, $W/Z$ candidate (resolved and boosted topologies). 
& Binned likelihood fit of \met.
& Section~\ref{ssub:spin1res1}, \ref{sub:2HDMspin0res}\\[0.5ex]
\midrule
\monoHbb~\cite{EXOT-2016-25} & VBC, \thdmZ$(*)$, \thdmS
&  \met, $h$ candidate (resolved and boosted topologies). 
& Binned likelihood fit of $m_h$ in bins of \met.
& Section~\ref{ssub:spin1res2}, \ref{sub:2HDMspin1res}, \ref{sub:2HDMspin0res}\\[0.5ex]
\midrule
$\monoHgg$~\cite{HIGG-2016-18} & VBC, \thdmZ$(*)$, \thdmS
& 2 photons, $m_{\gamma\gamma} \sim m_h$, \met. 
& Analytic function fit of $m_{\gamma\gamma}$.
& Section~\ref{ssub:spin1res2}, \ref{sub:2HDMspin1res}, \ref{sub:2HDMspin0res}\\[0.5ex]
\midrule
$\monotop$~\cite{EXOT-2017-16} & \vfc
& \met, $t$ candidate (all decay channels). 
& Binned likelihood fit of $\met$ ($\mT(\met, \textrm{large-}R\textrm{ jet})$) in the leptonic (hadronic) channel.
& Section~\ref{ssub:spin1res3}\\[0.5ex]
\midrule
$b(\bar{b}) + \met$~\cite{SUSY-2016-18}  & S/PS$(*)$, SCC$_{b}$$(*)$, \thdmS
& 1--2 $b$-jets, \met, 0 $\ell$. 
& Binned likelihood fit of $\cos\theta^*_{bb}$.
& Section~\ref{ssub:spin0res1}, \ref{ssub:spin0res2}, \ref{sub:2HDMspin0res}\\[0.5ex]
\midrule
$t\bar{t} + \met$~\cite{SUSY-2016-18,SUSY-2016-18,SUSY-2016-16}  & S/PS$(*)$, SCC$_{t}$$(*)$, \thdmS, DE
& 0--2$\ell$, 1--2 $b$-jets, $\geq $1--4 jets, \met, $m_{\mathrm T2}^{\ell\ell}$.
& Binned likelihood fit.
& Section~\ref{ssub:spin0res1}, \ref{ssub:spin0res2}, \ref{sub:2HDMspin0res}, \ref{sub:DEres}\\[0.5ex]
\bottomrule
\end{tabular}
}
\end{table}
 
\begin{figure}[htbp]
\centering
\subfloat[]{\includegraphics[width=0.49\linewidth]{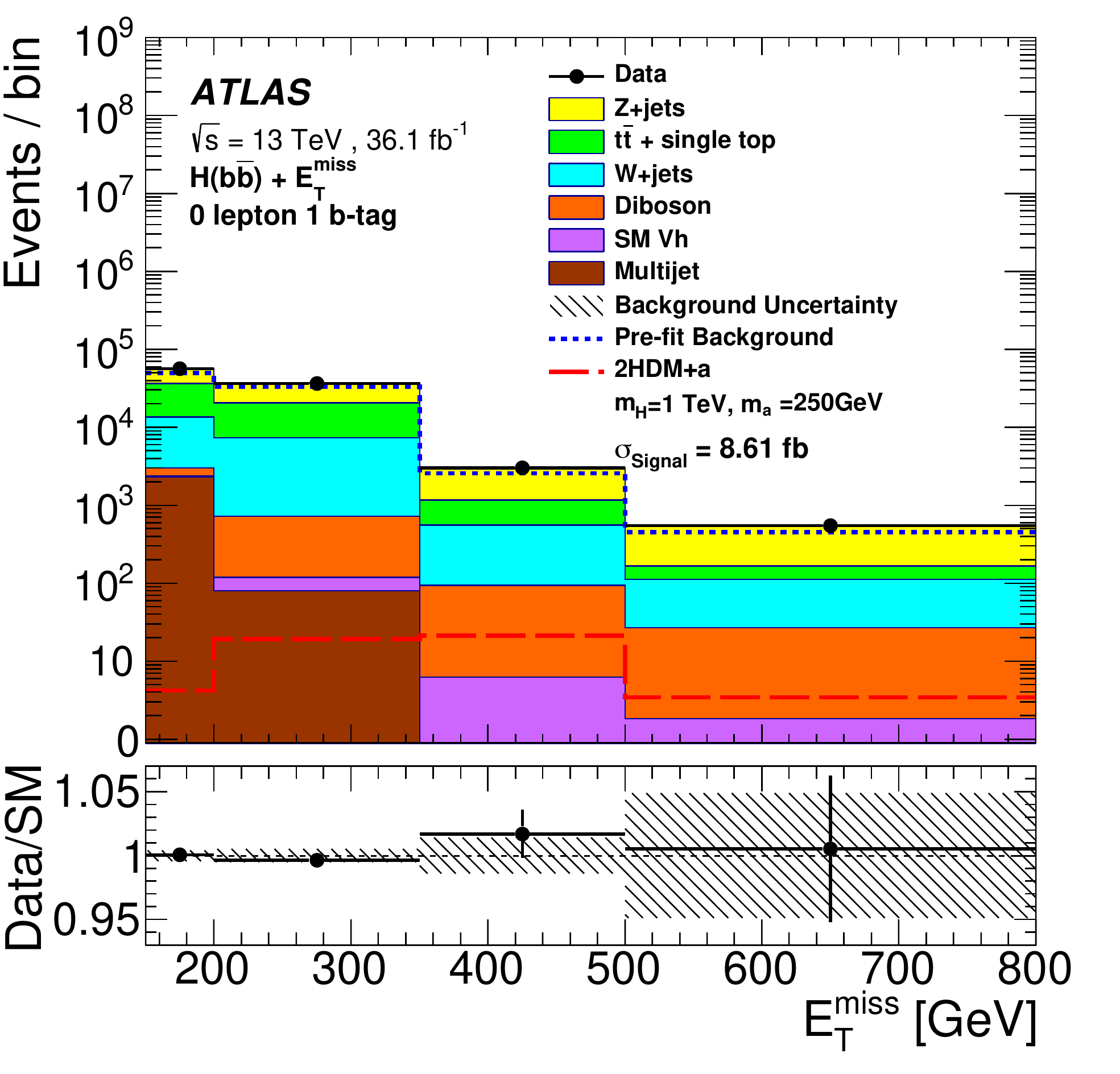}
}
\subfloat[]{\includegraphics[width=0.49\linewidth]{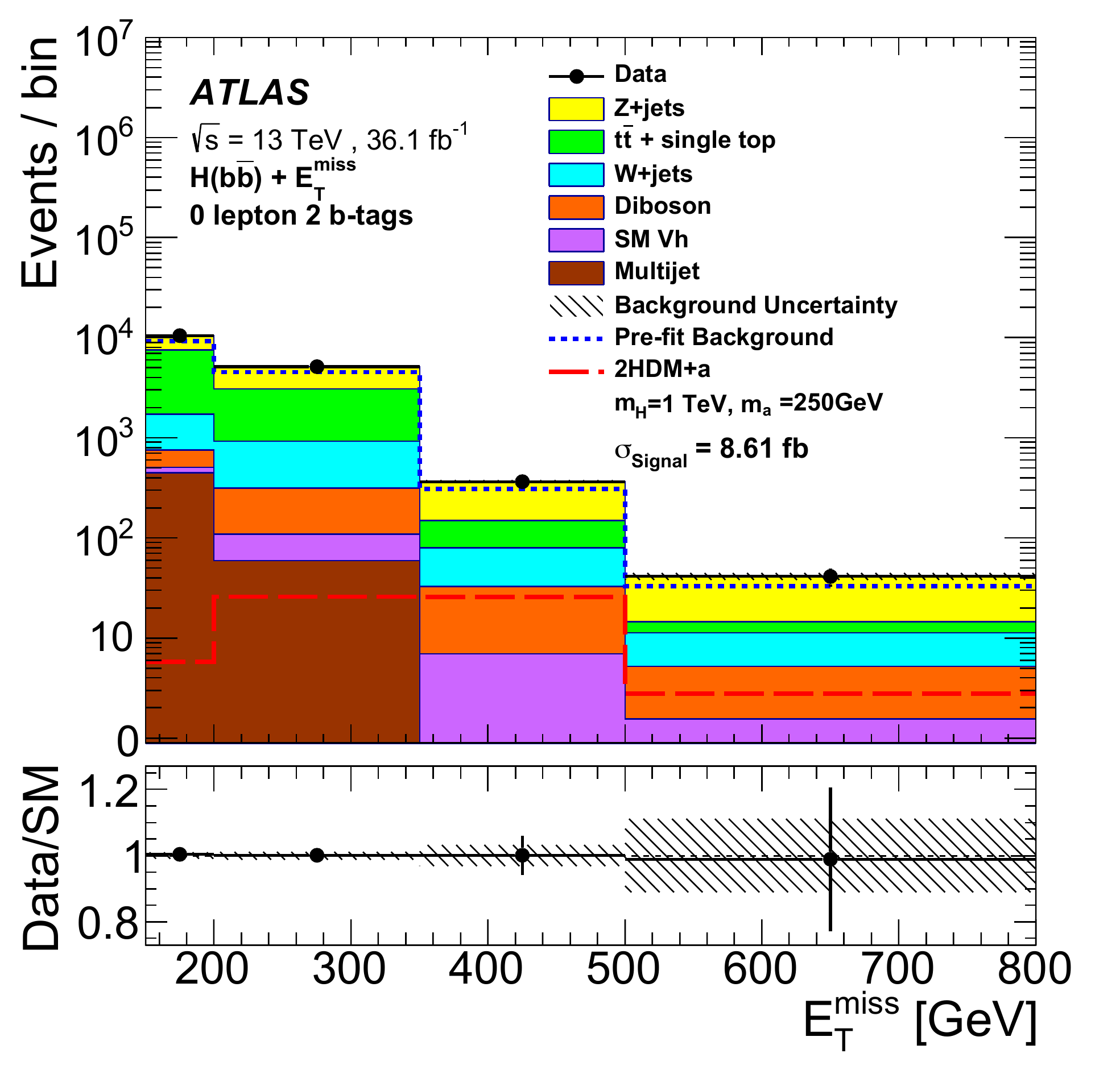}
} \\
\subfloat[]{\includegraphics[width=0.49\linewidth]{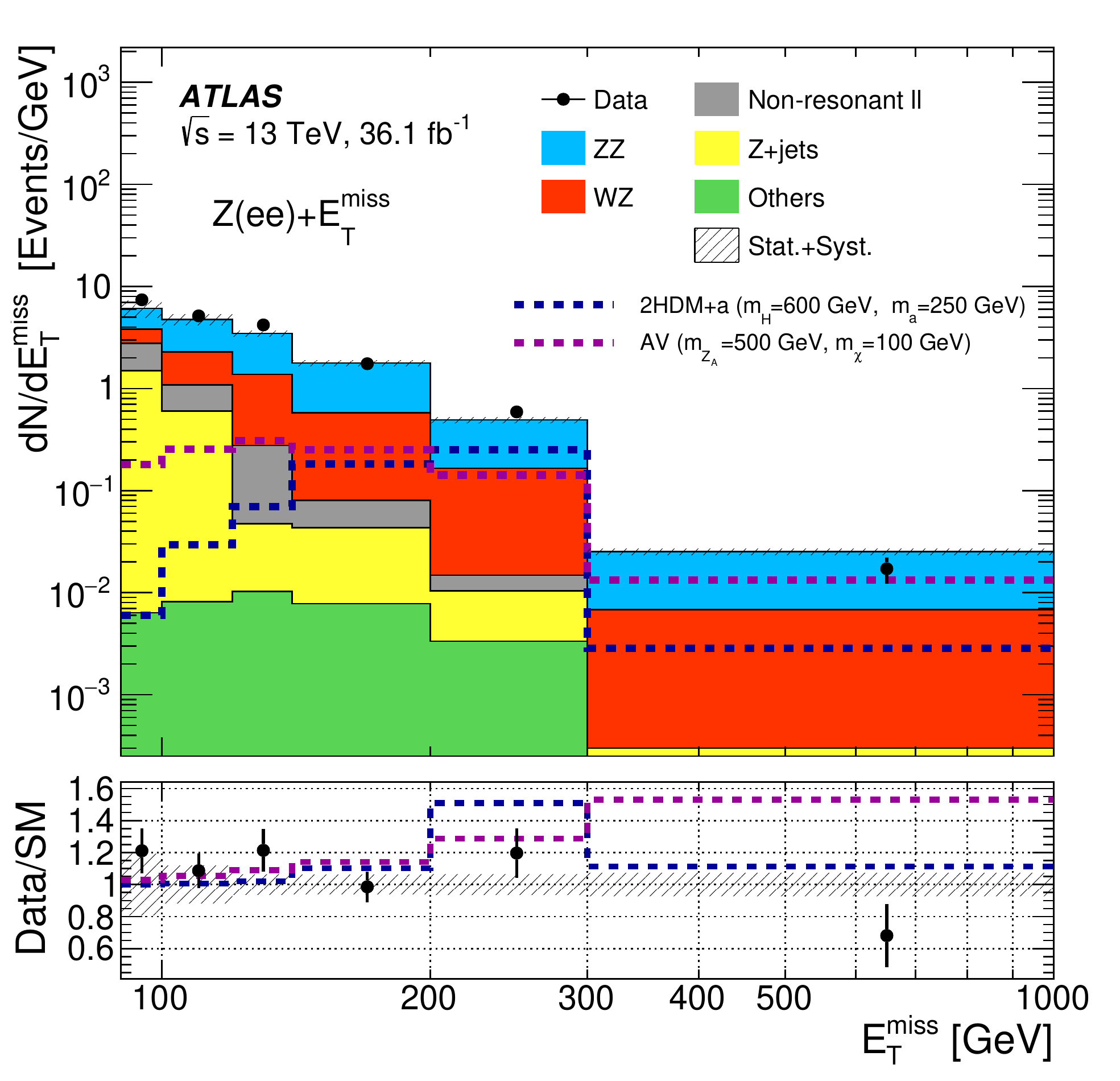}
}
\subfloat[]{\includegraphics[width=0.49\linewidth]{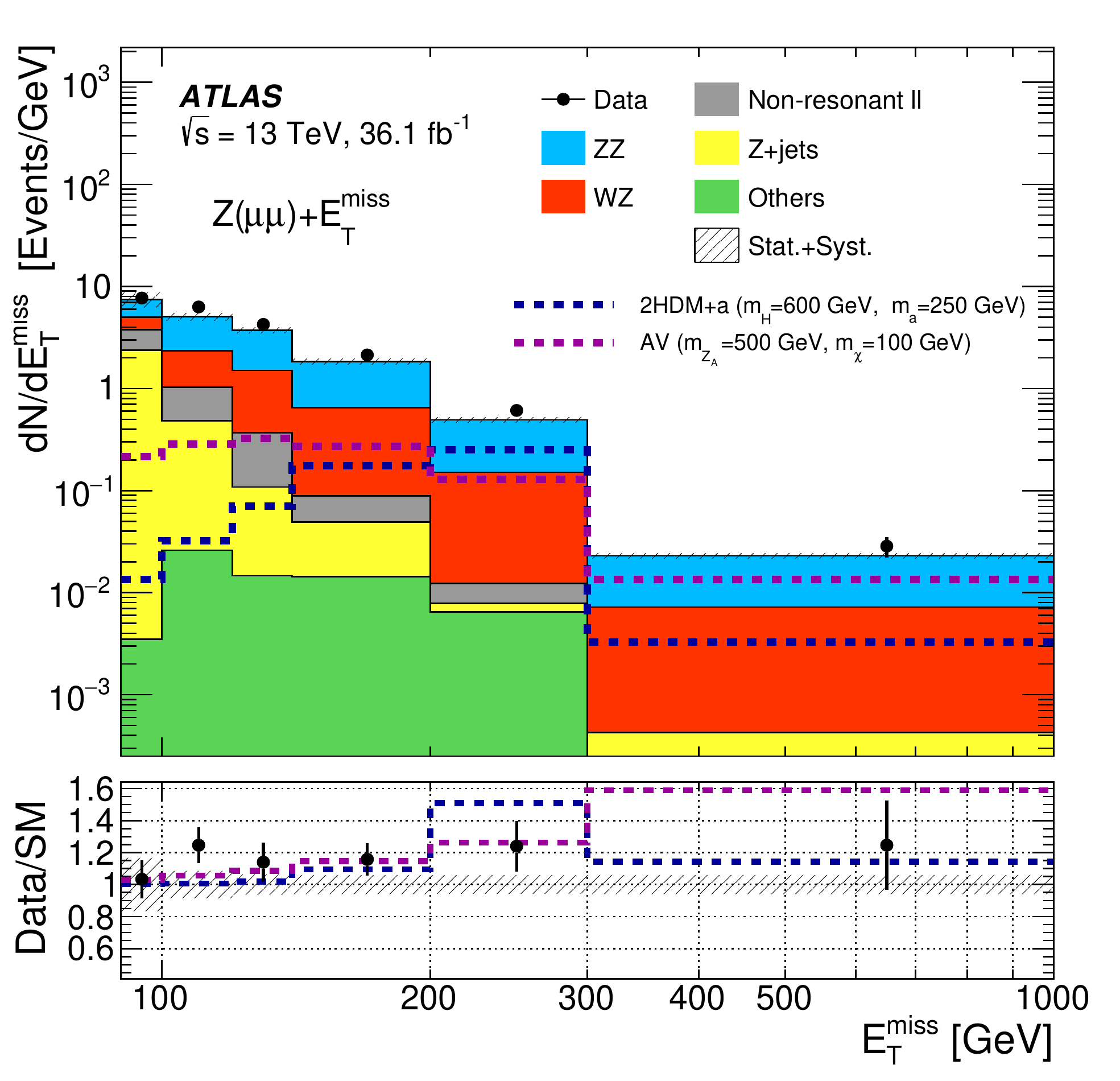}
}
\caption{Observed \met\ distribution in the \monoHbb\ analysis in the (a) 1-$b$-tag and (b) 2-$b$-tag signal regions compared with the background predictions.
The error bands show the total statistical and systematic uncertainties of the background predictions. The expected \met distribution for a representative  signal model is also shown. It corresponds to a \thdmS\ signal with $m_{\pscal} = 250\;\GeV$, $m_H=m_{H^{\pm}}=m_A = 1000\;\GeV$, $\tan\beta = 1.0$, $\sin\theta = 0.35$, $g_\chi = 1.0$ and $m_\chi = 10\;\GeV$.
Observed \met\ distribution in the \monoZll\ analysis in the (c) $ee$ and (d) $\mu\mu$ signal regions compared with the background predictions. The expected \met\ distribution for representative  signal models are also shown. They correspond to a \thdmS\ signal with $m_{\pscal} = 250\;\GeV$, $m_H=m_{H^{\pm}}=m_A = 600\;\GeV$, $\tan\beta = 1.0$, $\sin\theta = 0.35$, $g_\chi = 1.0$ and $m_\chi = 10\;\GeV$, and an AV signal with $m_{\avec}=500\;\GeV$, $m_\chi=100\;\GeV$, $g_q =0.25$, $g_\ell =0$, and $g_\chi = 1.0$.
The background predictions are after the corresponding fit in each analysis.
}
\label{fig:Bkg2HDMps}
\end{figure}
 
\begin{figure}[htbp]
\centering
\subfloat[]{\includegraphics[width=0.49\linewidth]{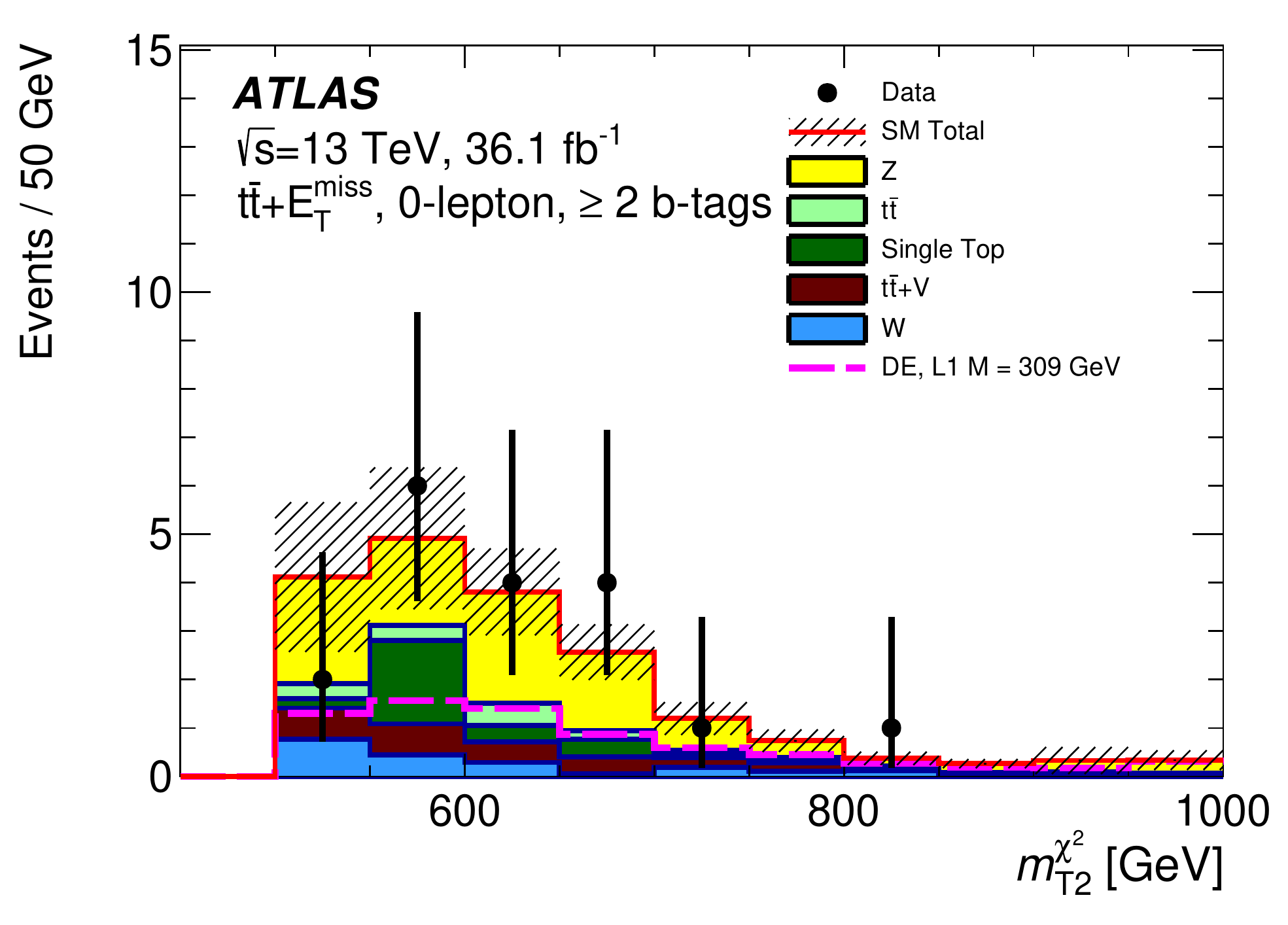}
}
\subfloat[]{\includegraphics[width=0.49\linewidth]{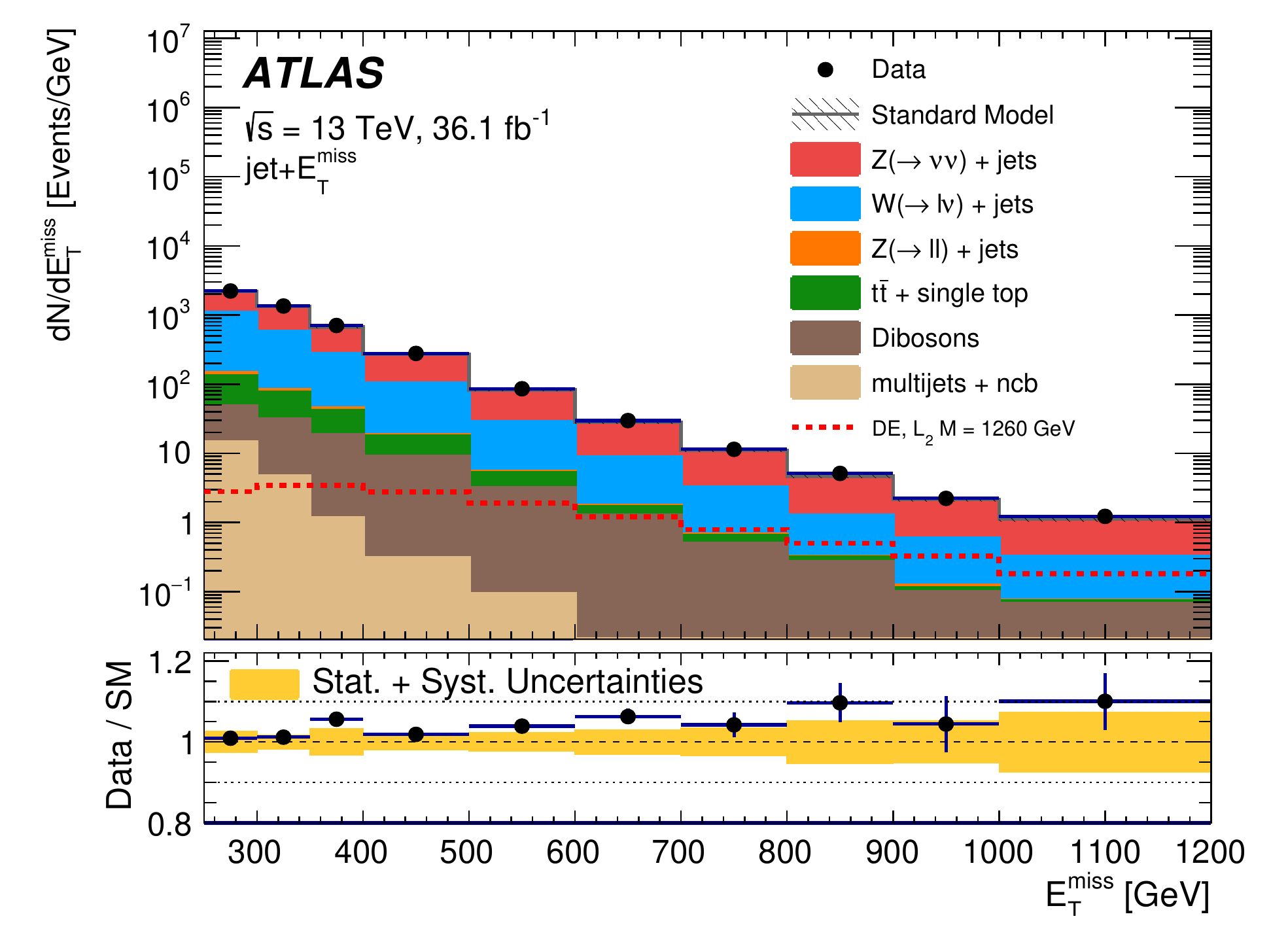}
}
\caption{Observed $m_{\mathrm{T}2}^{\chi^{2}}$ and \met\ distributions in the (a) $\ttbar (0L) + \met$ and (b) \monojet analyses, respectively, compared with a representative DE signal and the post-fit background predictions. The error bands show the total statistical and systematic uncertainties of the background predictions. Representative DE signal distributions are shown for $\mathcal{L}_1$ and $\mathcal{L}_2$ operators in (a) and (b), respectively.
}
\label{fig:BkgDE}
\end{figure}
 
\subsection{Searches for visible final states}
\label{ssec:visible}
Several searches for narrow resonances are interpreted in terms of the DM models described in Section~\ref{sub:DMmodels}. These searches explore
several final-state signatures by selecting different visible particles,
thus requiring the presence of reconstructed objects such as jets or leptons,
covering a variety of kinematic regions.
In some of the analyses described below, further identification techniques are employed to select final states with top quarks.

\paragraph{Dijet}
For this analysis~\cite{EXOT-2016-21} events with at least two small-$R$ jets are selected if  the $\pt$ of the leading (sub-leading) jet is greater than $440$~$(60)\;\GeV$.
The dijet selection requires a rapidity difference $|y^*| < 0.6$
and the invariant mass of the dijet system to be $m_{jj} > 1.1\;\TeV$.
The background estimation is obtained by fitting the falling $m_{jj}$ distribution. Bin widths are chosen to approximate the $m_{jj}$ resolution, and thus are wider for higher masses.
A sliding-window fitting technique is used, where restricted regions of the spectrum are fitted with a functional form.
The background is constructed bin-by-bin by performing a likelihood fit to the data in each window and using the fit value
in the central bin
for the background estimate. The values from the full set of windows are then combined to create the background estimate for the full mass range.
Model-independent limits on the visible cross-section for a hypothetical signal that produces a Gaussian contribution to the $m_{jj}$ distribution (for several signal widths) are provided for this analysis (see Appendix A of Ref.~\cite{EXOT-2013-11}).
This analysis was performed in data corresponding to an integrated luminosity of $37.0\;\ifb$.

\paragraph{Dijet angular}
A dijet selection can also be exploited to search for deviations from the SM expectation in angular distributions, characteristic of wider resonances
where the nominal dijet search would lose sensitivity. A dijet angular analysis~\cite{EXOT-2016-21} is performed in events with two jets following the $\pt$ requirements of the dijet search, but relaxing the $|y^*|$ requirement to 1.7. Due to
different kinematics
in this loosened selection, the mass of the dijet pair is required to be $m_{jj}>2.5\;\TeV$.
The analysis makes use of the variable
$\chi _{jj} = \text{e}^{2|y^*|} \sim (1+\cos\theta^*)/(1-\cos\theta^*)$\footnote{$\theta^*$ is defined as the polar angle with respect to the direction of the initial partons in the dijet centre-of-mass frame.},
constructed so that, in the limit of massless parton scattering and when only the $t$-channel scattering contributes to the partonic cross-section, the angular distribution $\text{d}N/\text{d}\chi_{jj}$ is approximately independent of $\chi_{jj}$.
MC events from multi-jet production are modelled at LO in QCD, and reweighted to NLO predictions from  NLOJET++~\cite{Nagy:2001fj,Nagy:2003tz} using mass- and angle-dependent correction factors. Additional electroweak mass- and angle-dependent  correction factors are applied.
The data are compared with a SM template in different  $m_{jj}$ ranges, and different $\chi_{jj}$ bins.

\paragraph{Trigger-object-level dijet}
For the dijet analysis described before, the high $\pt$ threshold imposed on the leading jet is limited by the  trigger selection
driven by the bandwidth available for single-jet triggers, thus it only targets $m_{jj} > 1.5\;\TeV$.
The limitation from the high-level trigger selection
is overcome by recording only high-level trigger jet information, rather than the full detector readout, to a dedicated data stream,
reducing the storage 
needs per event.
This strategy allows to record all events passing the single-jet level-one (L1) trigger (with lower threshold than in the high-level trigger) with minimal bandwidth increase.
The dataset collected corresponds to an integrated luminosity of $29.3\;\ifb$.
This trigger-object-level dijet analysis (TLA dijet)~\cite{EXOT-2016-20} selects events with at least two trigger-level jets with $\pt > 85\;\GeV$.
Two selection criteria are used: $|y^*| < 0.6$ in the mass range $700\;\GeV < m_{jj} < 1.8\;\TeV$ and $|y^*| < 0.3$ for $450\;\GeV < m_{jj} < 700\;\GeV$.
The leading trigger-level jet is required to have $\pt>185\;\GeV$ and $\pt>220\;\GeV$ for the $|y^*| < 0.3$ and $|y^*| < 0.6$ selections, respectively, to ensure full efficiency for the L1 triggers.
The search is then interpreted in terms of resonances with a mass between $450\;\GeV$ and $1.8\;\TeV$.
The background strategy used in the dijet search is also used here.

\paragraph{Resolved dijet + ISR}
Another alternative strategy to search for low-mass dijet resonances is to select events with
a pair of jets recoiling against a photon or an additional jet from ISR.
The resolved dijet + ISR analyses~\cite{ATLAS-CONF-2016-070} select events with a high-$\pt$ ISR object ($\gamma$ or jet), used to trigger the event, and a relatively low mass dijet resonance.
Dijet+$\gamma$ events contain at least one photon with $p_\gamma >150\;\GeV$ and at least two jets with $\pt > 25\;\GeV$.
The two leading jets must satisfy $|y^*| < 0.8$, which allows to probe of dijet invariant masses between $170\;\GeV$ and $1.5\;\TeV$.
The three-jet selection requires at least one jet with $\pt > 430\;\GeV$ as well as two additional jets with $\pt > 25\;\GeV$.
The leading jet is chosen as the ISR candidate, and the second- and third-highest-$\pt$ jets are required to satisfy $|y^*| < 0.6$. This selection probes a mass range between about $300\;\GeV$ and $600\;\GeV$.
The background contribution is estimated by fitting the $m_{jj}$ distribution.
This analysis was performed in $13\;\TeV$ collision data corresponding to an integrated luminosity of $15.5\;\ifb$.

\paragraph{Boosted dijet + ISR}
In the case of a dijet+ISR selection,
if the associated ISR photon or jet has large transverse momentum, the dijet resonance candidate is reconstructed as a large-$R$ jet~\cite{EXOT-2017-01} of radius 1.0 with mass $m$.
To enhance the sensitivity to quark pair decays, jet substructure techniques are used to discriminate between a two-particle jet from a decay of a boosted resonance and a single-particle jet~\cite{Thaler:2010tr}.
Events are required to have a large-$R$ jet, the resonance candidate, and at least one ISR object candidate. The azimuthal angular separation between the resonance candidate and the ISR object is required to satisfy $\Delta\phi > \pi/2$.
A $\pt > 2 m$ requirement ensures sufficient collimation of the resonance candidate.
In the ISR jet (photon) channel, the large-$R$ jet satisfies $\pt > 450$~$(200)\;\GeV$ and the ISR jet (photon) has $\pt > 450$~$(155) \;\GeV$.
A data-driven technique is used to model the expected background in the signal region via a transfer factor that extrapolates from a control region with inverted jet substructure requirements.

\paragraph{Dibjet}
The \dibjet\ search~\cite{EXOT-2016-33} targets dijet resonances with one or two jets identified as $b$-jets.
Two different analyses cover both the low and high invariant mass regions.
Events in the high invariant mass region  are selected with at least two jets, one of which has $\pt > 430\;\GeV$ and passes an inclusive jet trigger. The rapidity difference is required to be $|y^*| < 0.8$.
This analysis covers the region with $m_{jj} > 1.2\;\TeV$.
The low invariant mass region uses a trigger targeting events with two jets containing $b$-hadrons,
which provides access to lower dibjet invariant masses ($m_{jj}$) compared to the single jet trigger: $570\;\GeV < m_{jj} < 1.5\;\TeV$. The rapidity difference requirement is tightened to $|y^*| < 0.6$.
In this case, only the two-$b$-jets selection is considered.
Because the double $b$-jet trigger was not available during the full data-taking period, the total integrated luminosity used for the low-mass analysis corresponds to $24.3\;\ifb$ of  $13\;\TeV$ collision data.
A background estimation strategy similar to that of the dijet analysis is used in these analyses.
 
\paragraph{Dilepton}
The dilepton analysis~\cite{EXOT-2016-05} selects events with at least two same-flavour leptons. The pair of electrons (muons) with highest $\ET$ ($\pt$) are chosen as the candidate decay products  of the resonance.
Only the muon channel candidates are required to have opposite charge, due to higher charge misidentification for high-$\ET$ electrons
and the $\pt$ misreconstruction associated with wrongly measured charge in muons.
Background processes with two prompt leptons are modelled using MC samples.
The $Z/\gamma^*\rightarrow\ell\ell$ background is smoothed for $120\;\GeV < m_{\ell\ell} < 1\;\TeV$.
This is done by fitting the MC spectrum and the resulting fitted function is used to set the expected event yields in that mass range.
A data-driven method is employed to estimate backgrounds with at least one misidentified lepton.
The $m_{\ell\ell}$ distribution is explored between $80\;\GeV$ and $6\;\TeV$.
 
\paragraph{Same-sign $tt$}
Events in the same-sign $tt$ analysis~\cite{EXOT-2016-16} are selected with exactly two leptons with positive charge and at least one $b$-jet. Events are required to satisfy $\HT>750\;\GeV$, where $\HT$ is defined as the scalar sum of the \pt\ of all selected objects, including jets. Additionally, requirements on \met\ and the azimuthal separation between the two leptons are imposed.
Signal regions for the different lepton flavours ($ee$, $e\mu$ and $\mu\mu$) are treated separately.
Irreducible SM backgrounds are determined using MC simulation samples. Backgrounds from fake leptons are estimated using data-driven techniques.

\paragraph{\ttbar resonance}
The $\ttbar$ resonance analysis~\cite{EXOT-2015-04} selects events with two top-quark candidates.
Events are required to have a leptonic top-quark decay, selected by requiring a charged lepton and \met\ consistent with a leptonic decay of a $W$ boson, and a small-$R$ jet close by.
Events are classified as boosted or resolved depending on their hadronic activity.
In the boosted selection, events contain one large-$R$ jet passing top-tagging requirements. In the resolved selection, events  have at least four small-$R$ jets and fail the boosted selection.
The $\ttbar$ invariant mass, $m_{\ttbar}$, is reconstructed from the decay products of the two top-quark candidates in the event.
The $b$-jet multiplicity is used for further categorisation.
The SM $\ttbar$ production is estimated using MC samples and fixed-order theory calculations.
The multi-jet and $W$+jets background contribution is estimated using data-driven techniques.
 
\paragraph{\fourtop}
The \fourtop\ analysis~\cite{SUSY-2016-11} searches for events characterised by a single lepton and high jet multiplicity.
Events are categorised by their jet multiplicity, which is defined using three \pt\ thresholds: $40\;\GeV$, $60\;\GeV$, and $80\;\GeV$. Events are further separated into five bins corresponding to the $b$-jet multiplicity.
The $\ttbar$+jets and $W/Z$+jets background production is estimated using a combined fit to these jet and $b$-jet multiplicity bins.
The normalisation of these backgrounds is extrapolated from lower to higher jet multiplicity, while the $b$-jet multiplicity shape is taken from a parameterised extrapolation from data (simulation) for the $\ttbar$ ($W/Z$+jets) background.

As in the case of the searches for invisible final states, these
analyses found no significant deviation from the expected SM
backgrounds.
Therefore, exclusion limits are placed on the allowed phase space of the
corresponding
signal models, as discussed in Section~\ref{sec:result}.
 
\begin{table}
\caption{Summary of searches for visible final states used to constrain the  different DM models defined in Section~\ref{sub:DMmodels}. The $(*)$ indicates models which were  presented in the original publication, all others are either new or updated.}
\label{tab:visible}
\centering
\resizebox{\textwidth}{!}{
\begin{tabular}{m{40mm}m{35mm}m{60mm}m{50mm}m{30mm}}
\toprule
Analysis  & Models targeted & Final-state signature & Key Characteristics & Results \\\midrule
Dijet~\cite{EXOT-2016-21} & V/AV
& 2 jets, $m_{jj}$, $y^*$. 
& Sliding-window fit of the $m_{jj}$ distribution.
& Section~\ref{ssub:spin1res1}\\[0.5ex]
\midrule
Dijet angular~\cite{EXOT-2016-21} & V/AV
& 2 jets, $m_{jj}$, $y^*$. 
& Binned likelihood fit of $\chi_{jj}$.
& Section~\ref{ssub:spin1res1}\\[0.5ex]
\midrule
TLA dijet~\cite{EXOT-2016-20}  & V/AV
& 2 trigger-level jets, $m_{jj}$, $y^*$. 
&Sliding-window fit of the $m_{jj}$ distribution.
& Section~\ref{ssub:spin1res1} \\[0.5ex]
\midrule
Resolved dijet+ISR~\cite{ATLAS-CONF-2016-070}  & V/AV
& 3 jets (or 2 jets and 1 photon), $m_{jj}$, $y^*$. 
& Parametric function fit of the $m_{jj}$ distribution.
& Section~\ref{ssub:spin1res1} \\[0.5ex]
\midrule
Boosted dijet+ISR~\cite{EXOT-2017-01}  & V/AV$(*)$
& 1 large-$R$ jet, 1 jet or photon, $m_{J}$. 
& Data-driven extrapolation from control region via transfer factor.
& Section~\ref{ssub:spin1res1} \\[0.5ex]
\midrule
Dibjet~\cite{EXOT-2016-33} & V/AV
& 2 jets (1 and 2 $b$-jets), $m_{jj}$, $y^*$. 
& Sliding-window parametric fit of the $m_{jj}$ distribution.
& Section~\ref{ssub:spin1res1} \\[0.5ex]
\midrule
Dilepton~\cite{EXOT-2016-05} & V/AV
& 2 $e$ or 2 $\mu$. 
& $Z/\gamma^*\rightarrow\ell\ell$ from fitted MC spectrum. 
& Section~\ref{ssub:spin1res1}\\[0.5ex]
\midrule
Same-sign $tt$~\cite{EXOT-2016-16} & \vfc
& 2 same-sign $\ell$, 2 $b$-jets, $\HT$, \met. 
& Background with real leptons from MC.
& Section~\ref{ssub:spin1res3}\\[0.5ex]
\midrule
$\ttbar$ resonance~\cite{EXOT-2015-04} & V/AV
& 1 $\ell$, hadronic $t$ candidate (resolved and boosted topologies), \met. 
& $\ttbar$ bkg from MC, binned likelihood fit of $m_{\ttbar}$.
& Section~\ref{ssub:spin1res1}\\[0.5ex]
\midrule
\fourtop~\cite{SUSY-2016-11} & \thdmS
& 1 $\ell$, high jet multiplicity. 
& Parameterised extrapolation from low to high jet multiplicity. Binned likelihood fit of jet and $b$-jet multiplicities.
& Section~\ref{sub:2HDMspin0res}\\[0.5ex]
\bottomrule
\end{tabular}
}
\end{table}

\FloatBarrier
\subsection{Complementarity and combination of signatures}
It can be seen from Tables~\ref{tab:invisible} and~\ref{tab:visible}
that several analyses are often sensitive to the same model.
In cases like the  $X+\met$ final-states searches, $X$ originating
from initial-state radiation or associated production, a variety of
final states are evaluated: $X=\textrm{jet}, \gamma, W, Z, h,
t(\bar{t}), b(\bar{b})$.
Since the mediator couples DM to SM particles, it is also possible to
reinterpret results from  resonance searches targeting the mediator directly. The
complementarity depends on the choice of model as well as coupling
values.
For the V/AV model, this paper considers $\textrm{jet}/\gamma/V + \met$,
dijet, \dibjet\, and \ditop\ final states. All results for this model
are new or have been updated from previous publications, except for
the $\textrm{jet}/\gamma + \met$ interpretations.
For the VBC and \thdmZ\
models, this paper considers only \monoH, which
dominates the sensitivity for the chosen parameter values.
All possible final states (same-sign $tt$
and the $t+\met$) are taken into account for the VFC model. Only invisible final
states, $\ttbar/\bbbar/\textrm{jet}+\met$, are considered
for the S/PS model.
The SCC$_{q}$, SCC$_{b}$ and SCC$_{t}$ models
are each addressed with a specific signature: $\textrm{jet}+\met$,
$b+\met$, $t+\met$, respectively, and  all results were
presented in each specific analysis paper.
Various final states are evaluated in order to place the first constraints
on the \thdmS\ model by ATLAS searches: $Z/h+\met$,
$\ttbar/\bbbar+\met$, \hinv\, and \fourtop.
Finally, the
constraints on DE models are set
using \monojet\ and \dmtt\ final states.
 
Complementarity can also be found when studying different SM decay
channels of a given signature. Two natural candidates from
the analyses discussed here are the $h+\met$ ($\bbbar,\gamma\gamma$)
searches and the $\ttbar + \met$ (fully hadronic, semileptonic and
fully leptonic) searches.
 
The results from the $h+\met$ searches presented in
Section~\ref{sub:2HDMspin1res} correspond to a statistical combination of
the \monoHbb\ and \monoHgg\ searches and not published before elsewhere.
The \monoHbb\ analysis has a larger reach in mediator masses, however  its sensitivity
is limited at lower masses by the threshold requirement of
the \met\ trigger used to record the events for this analysis.
The \monoHgg\ analysis covers a lower mass region owing to its selection based
on a photon trigger.
For the combination, the luminosity, experimental, and signal modelling uncertainties
were taken to be correlated
between the two channels. In the \monoHgg\ analysis the systematics
uncertainties are not significantly constrained by the fit. This is
mainly due to the use of a single signal region and no control regions.
In the case of \monoHbb\, the systematics uncertainties
are constrained due to the use of dedicated control regions. It is
observed, however, that the results from the combination and the
individual \monoHbb\ results are very similar.
While the \monoHbb\ channel dominates the sensitivity, searches in
different decay channels are of interest in probing different kinematic
regions defined by different analyses strategies.

For this paper, the $\ttbar + \met$ exclusion limits discussed in
Section~\ref{sec:result} are combined  based on the best expected
exclusion for each signal model, unless separate contours are shown.
The combination and comparison of this channel is also a novel result of this paper.
 
\section{Systematic uncertainties}
\label{sec:syst}
Systematic uncertainties for both the background and signal models are considered in each of the analyses presented in Section~\ref{sec:analyses}. These uncertainties, as well as statistical uncertainties, depend on the 
event selection, the phase space covered by a given analysis, and its background estimation strategy.
The systematic uncertainties include experimental and theoretical uncertainties.
Experimental uncertainties may include uncertainties in the absolute jet and \met\ energy scales and resolutions, the jet quality requirements, pile-up corrections, $b$-tagging efficiencies, the soft contributions to \met\ and the luminosity.
Uncertainties in lepton identification and reconstruction efficiencies, energy/momentum scale and resolution are included for events with selected or vetoed leptons.
Uncertainties due to the finite statistics of the background MC samples and others related to the modelling of the background processes are also included in the analyses.
 
The signal modelling is  subject to experimental and theoretical uncertainties. The experimental uncertainties are the same as for the background processes.
Theoretical uncertainties affecting the production cross-section (normalisation) and the acceptance are considered separately.
The strategy used to estimate systematic uncertainties for those signal models studied in this paper which are not discussed in previous publications is outlined below.
 
The results for the \thdmS\ and DE signal models include theoretical
systematic uncertainties due to parton distribution functions (PDFs),
evaluated following the PDF4LHC recommendations~\cite{Butterworth:2015oua}.
The choice of different PDFs results in up to 2\% uncertainty in the acceptance and up to 10\% uncertainty in the cross-section.
Uncertainties related to the choice of renormalisation and factorisation scales are derived by
varying independently such scales by a factor of $2.0$ and $0.5$ relative to the nominal values used for the MC generation.
They account for an uncertainty in the acceptance below 5\% for the different analyses.
Uncertainties in initial- and final-state radiation due to the parton shower modelling are estimated by generating MC samples
with alternative underlying event and multi-parton interaction parameter choices
resulting in uncertainties 
between 5\% and 15\% in the signal acceptance, typically increasing at higher mediator masses. In the very large jet multiplicity phase space of the \fourtop\ analysis they reach values of 50\%.
 
In some cases, additional uncertainties are included to
account for non-closure effects of the
rescaling procedure described in
Section~\ref{sec:montecarlo}.
These uncertainties include a contribution from the statistical
uncertainty associated with the acceptance ratios determined from the
baseline signal sample.
For the \monoHgg\ (\monoZll) analysis this translates in up to 7\% (8\%) uncertainty in the final \thdmS\ signal yields.

 
\section{Interpretation of the results}
\label{sec:result}

This section summarises the exclusion limits set by ATLAS published searches
briefly outlined in Section~\ref{sec:analyses},
on the various signal models described in Section~\ref{sub:DMmodels}
(following the notation in
Table~\ref{t:ModelSummaryNotation}). The analyses and corresponding
signal regions are referred to by their analysis labels defined
in Tables~\ref{tab:invisible}~and~\ref{tab:visible}.
The
observed and expected 95\% confidence level (CL) exclusion limits are
obtained from the signal region or combination of regions of each
contributing analysis using the CL$_\mathrm{s}$~\cite{ReadCLs} method.
The signal contamination in CRs across different analyses are kept minimal through dedicated selection requirements and hence are not explicitly taken into account in the following  results.
In Section~\ref{sub:2HDMspin1res} a statistical combination of the
\monoH\ final states is used to derive the results.
 
\subsection{Vector or axial-vector dark matter models}

\FloatBarrier
\subsubsection{Neutral interaction}
\label{ssub:spin1res1}
 
\begin{figure}[b]
\centering
\includegraphics[width=0.8\linewidth]{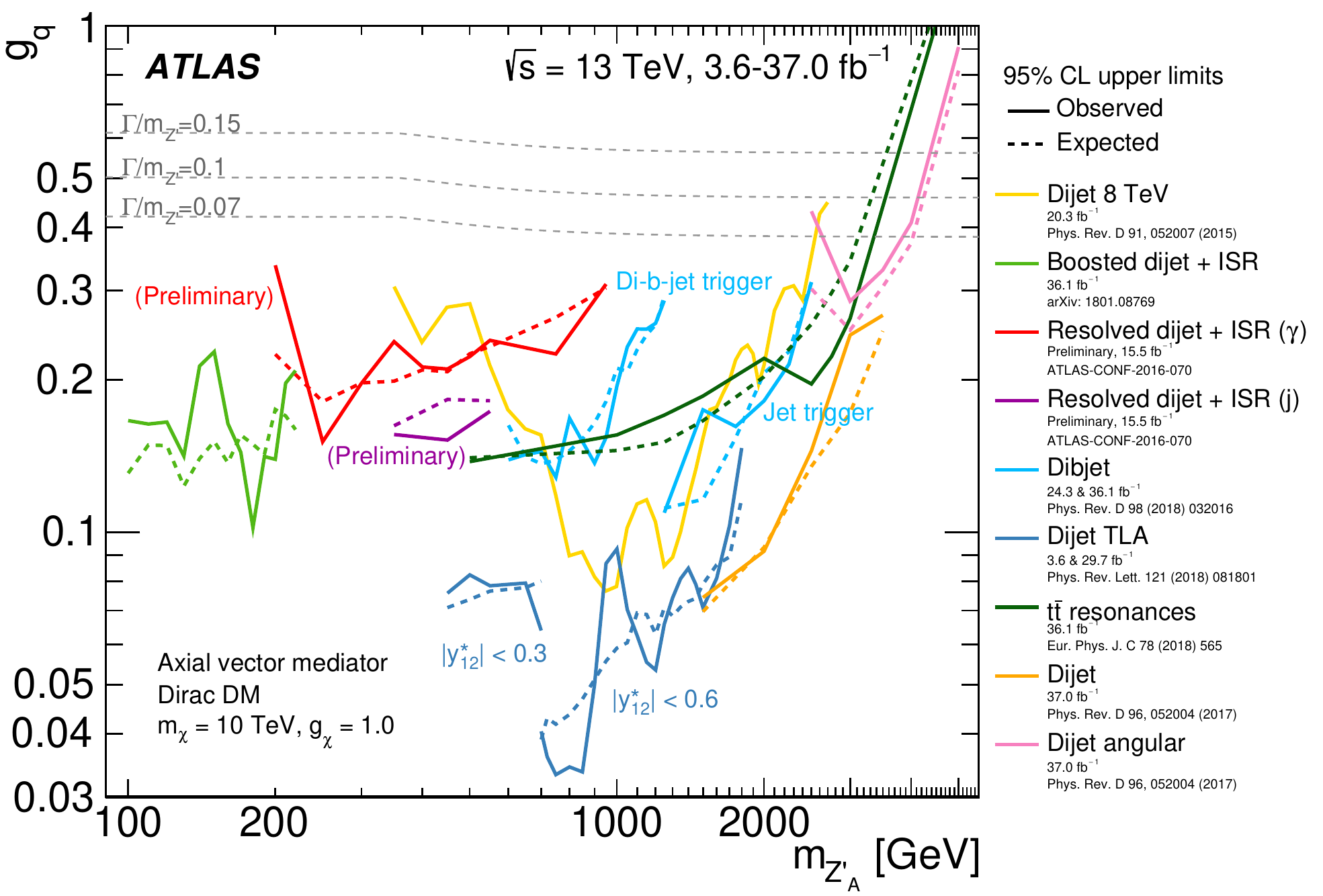}
\caption{
Dijet search contours for 95\% CL upper limits on the coupling $g_q$
as a function of the resonance mass $m_{\avec}$ for the
leptophobic axial-vector \avec\ model.
The expected limits from each search are indicated by dotted
lines. For illustrative purpose, the dijet + ISR Preliminary searches are shown in the plot.
The TLA dijet analysis has two parts,
employing different datasets with different selections in the rapidity
difference $y^*$ as indicated.  The
yellow contour shows the results of the dijet search using $20.3$
fb$^{-1}$ of $8\; \TeV$ data. Coupling values above the solid lines are excluded, as long as
the signals are narrow enough to be detected using these searches.
The TLA dijet search with $|y^*|<0.6$ is sensitive up to $\Gamma/m_{Z'} = 7\%$, the TLA dijet with $|y^*|<0.3$ and dijet + ISR searches are sensitive up to $\Gamma/m_{Z'} =10\%$, and the dijet and \dibjet\ searches are sensitive up to $\Gamma/m_{Z'} = 15\%$. The dijet angular analysis is sensitive up to $\Gamma/m_{Z'} = 50\%$. No limitation in sensitivity arises from large width resonances in the
\ditop\ analysis. Benchmark width lines are indicated in the canvas. The $\Gamma/m_{Z'} = 50\%$ lies beyond the canvas borders.
}
\label{fig:SummaryAV_mZgq}
\end{figure}
 
\begin{figure}[htbp]
\centering
\subfloat[]{\includegraphics[width=0.7\linewidth]{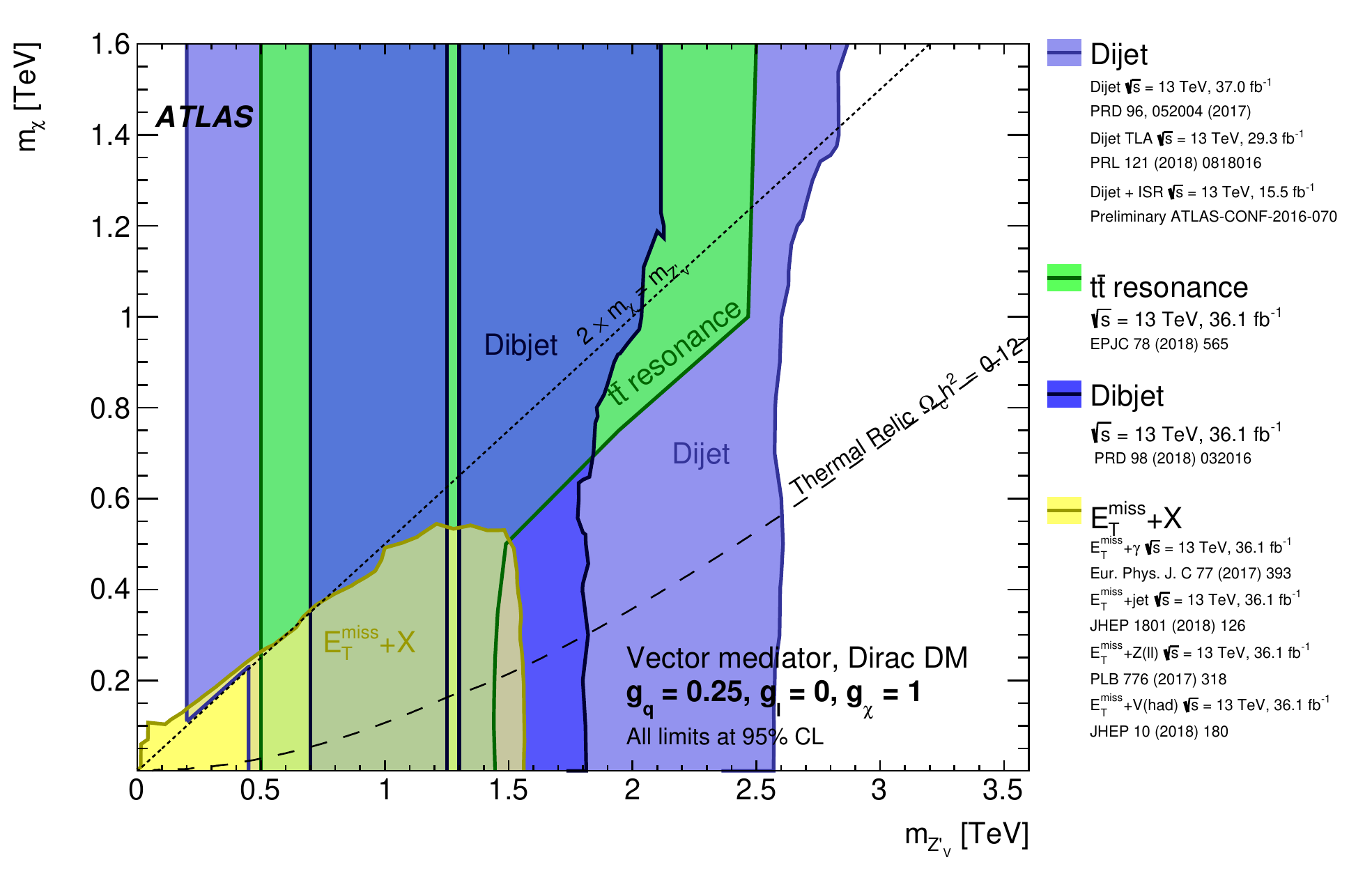}
\label{fig:SummaryAV_V1}
} \\
\subfloat[]{\includegraphics[width=0.7\linewidth]{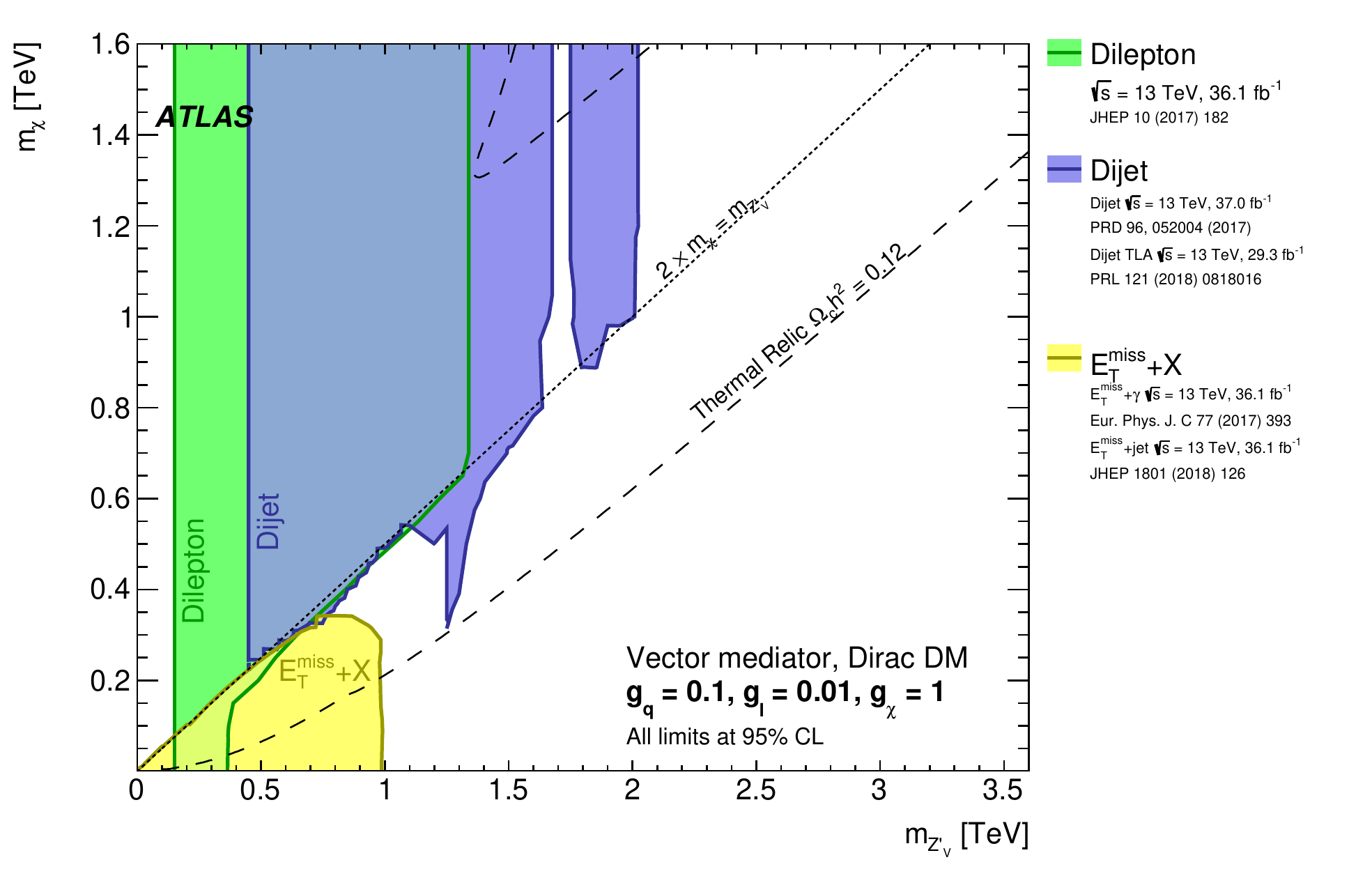}
\label{fig:SummaryAV_V2}
}
\caption{Regions in a (mediator-mass, DM-mass)
plane excluded at
95\% CL by dijet, dilepton and \met+$X$ searches, for
leptophobic (a) or leptophilic (b) vector  mediator simplified models
described in Section~\ref{ssub:spin1theory1}. The exclusions are
computed for a DM coupling \gDM, quark coupling $g_q$,
universal to all flavours, and lepton coupling $g_\ell$ as
indicated in each case.
Dashed curves labelled ``thermal relic'' correspond to combinations of
DM and mediator mass values that are consistent with a DM
density of $\Omega h^2= 0.12$ and a standard thermal history
as computed in MadDM~\cite{Backovic:2015tpt,Albert:2017onk}.
Above the curve in (a)
annihilation processes described by the simplified model deplete
$\Omega h^2$ to below $0.12$.
In (b), this occurs between the two dashed curves.
The dotted line indicates the
kinematic threshold where the mediator can decay on-shell into
DM. }
\label{fig:SummaryV}
\end{figure}
 
\begin{figure}[htbp]
\centering
 
\subfloat[]{\includegraphics[width=0.7\linewidth]{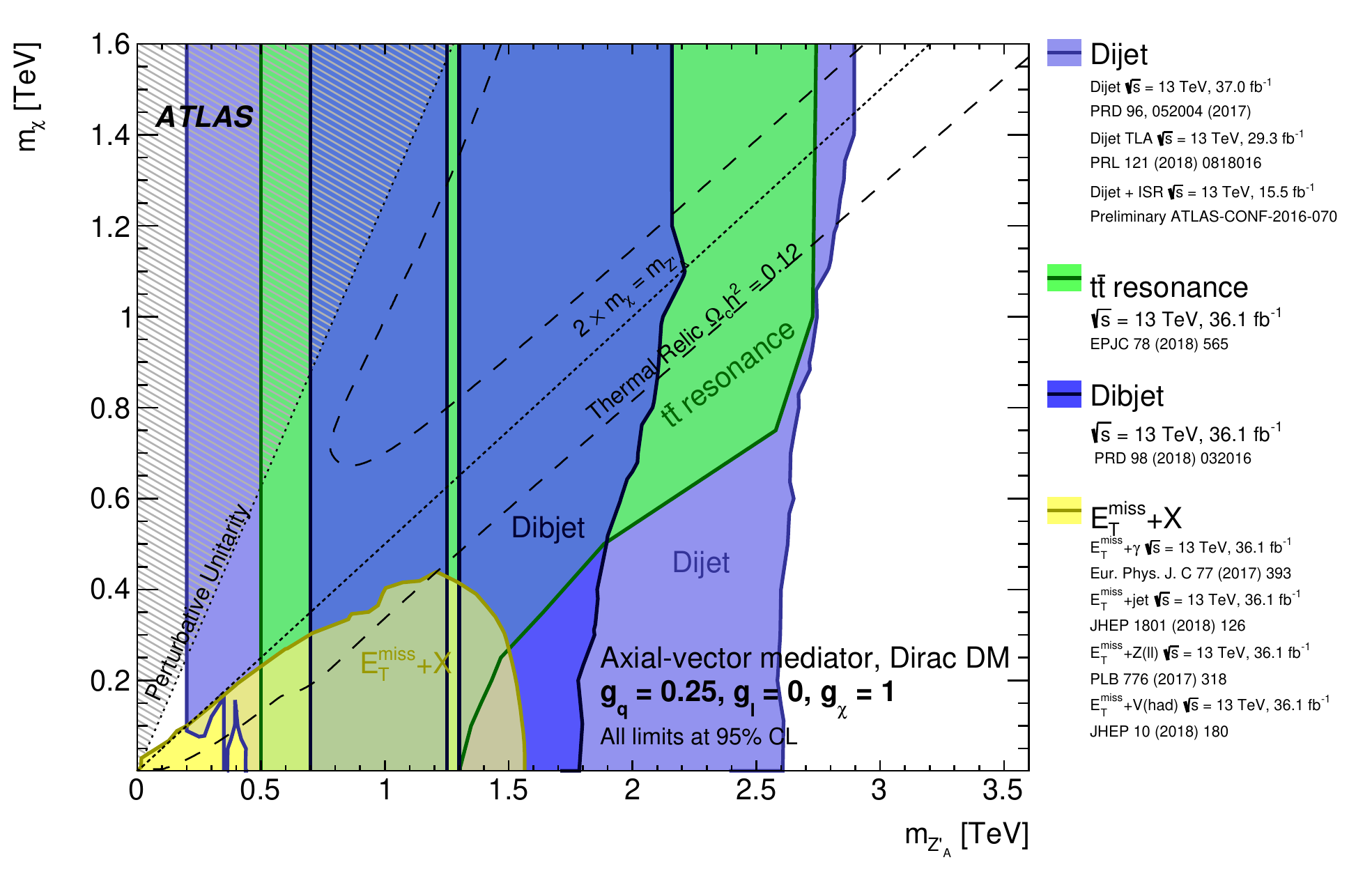}
\label{fig:SummaryAV_A1}
} \\
\subfloat[]{\includegraphics[width=0.7\linewidth]{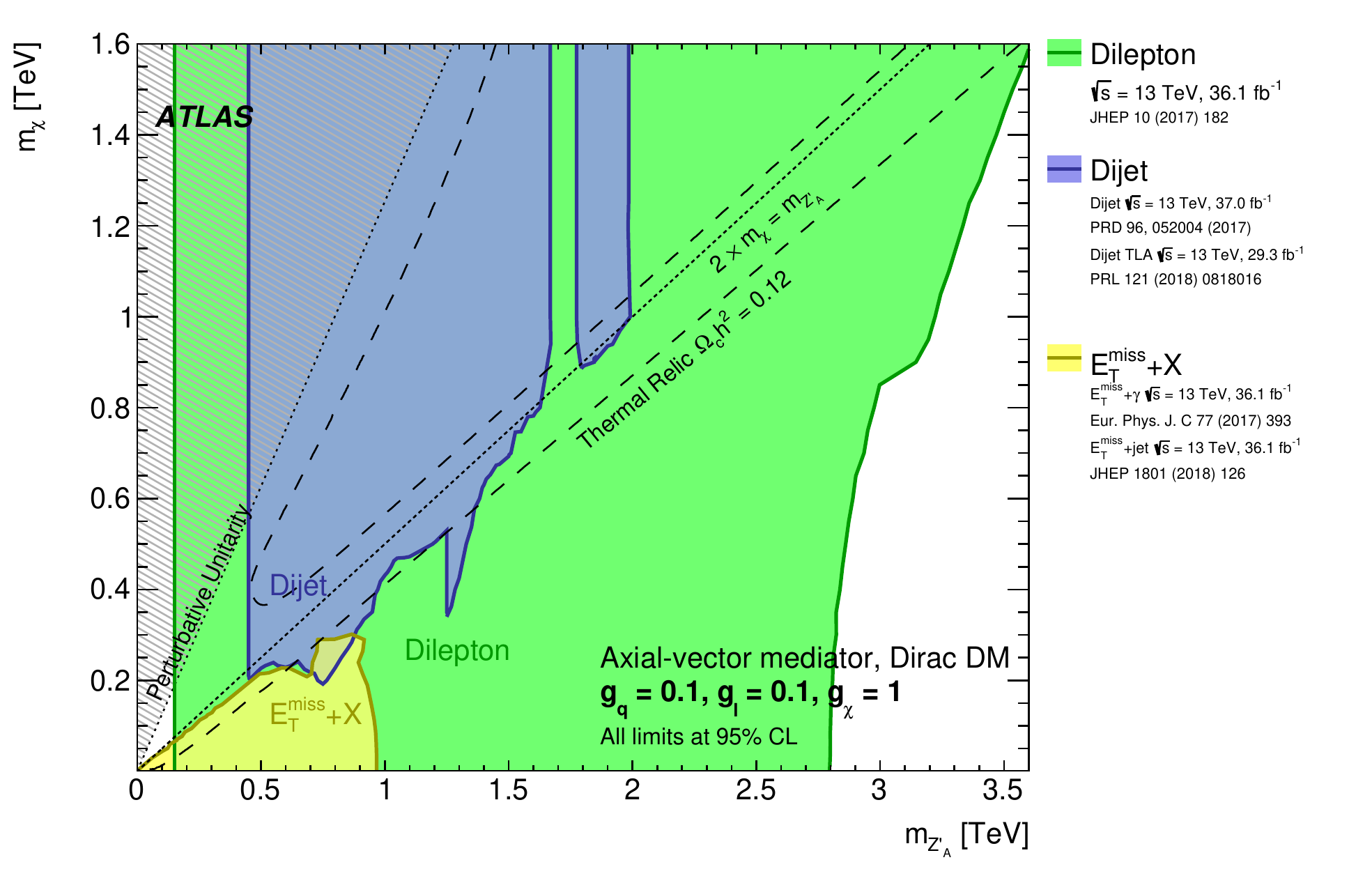}
\label{fig:SummaryAV_A2}
} \\
 
\caption{Regions in a (mediator-mass, DM-mass) plane excluded at
95\% CL by visible and invisible searches, for
leptophobic (a) or leptophilic (b) axial-vector  mediator simplified models
described in Section~\ref{ssub:spin1theory1}. The exclusions are
computed for a DM coupling \gDM, quark coupling $g_q$,
universal to all flavours, and lepton coupling $g_\ell$ as
indicated in each case.
Dashed curves labelled ``thermal relic'' correspond to combinations of
DM and mediator mass values that are consistent with a DM
density of $\Omega h^2= 0.12 $ and a standard thermal history,
as computed in MadDM~\cite{Backovic:2015tpt,Albert:2017onk}. Between the two curves,
annihilation processes described by the simplified model deplete
$\Omega h^2$ to below $0.12 $. A dotted line indicates the
kinematic threshold where the mediator can decay on-shell into
DM. Excluded regions that are in tension with the
perturbative unitary considerations of Ref.~\cite{Kahlhoefer:2015bea} are
indicated by shading in the upper left corner. }
\label{fig:SummaryAV}
\end{figure}

The V/AV simplified model is strongly constrained by
searches for a high-mass resonance decaying into a pair of fermions and searches for
associated production of DM particles with an ISR object.
 
As presented in
Figure~\ref{fig:SummaryAV_mZgq} for the case of an axial-vector
mediator, each resonance search analysis is sensitive to complementary regions of the mass--coupling parameter space.
Couplings above the exclusion line are excluded, as long as the width predicted by the
model is smaller than the maximal ratio of width to mediator mass ($\Gamma/m_{Z'}$) to which the analysis is sensitive.
This limitation arises where the background model is estimated via a sliding-window fit of the $m_{jj}$ distribution.
Specifically, the TLA dijet analysis assuming $|y^*|<0.6$ is sensitive up to around $\Gamma/m_{Z'} = 7\%$, the TLA dijet analysis requiring $|y^*|<0.3$ and the boosted dijet+ISR analysis are
sensitive up to around $\Gamma/m_{Z'} = 10\%$, while the dijet and \dibjet\ analyses are sensitive
up to around $\Gamma/m_{Z'} = 15\%$.
Finally, the dijet angular analysis is sensitive up to $\Gamma/m_{Z'} = 50\%$.
No limitation in sensitivity arises from large width resonances in the
\ditop\ analysis, as the background is constrained
in dedicated control regions.
The different dijet analyses
(see Section~\ref{ssec:visible} for details) are sensitive to different mass regimes as
well as coupling values.
For illustrative purpose, the dijet + ISR Preliminary searches are shown in the plot, as they
constraint a unique portion of the parameter space. At the time of writing, a dijet + ISR analysis based on
$80\;\ifb$ of integrated luminosity was published \cite{Aaboud:2019zxd}, which also probes a similar parameter space.
The boosted dijet+ISR analysis has the best reach for low  masses,
excluding \avec\ mediator masses between $100\;\GeV$ and $220\;\GeV$.
Two new interpretations, for the  \dibjet\ and \ditop\ analyses, are presented for these models.
The \dibjet\ (\ditop) analysis places constraints
on \avec\ mediators with masses between $500\;\GeV$ and $2.5$~$(2)\;\TeV$, in the same region of sensitivity of the
dijet, TLA dijet, and boosted dijet+ISR analyses.
 
To illustrate the complementarity of
the searches~\cite{Abercrombie:2015wmb,Albert:2017onk},
three different coupling
scenarios are also considered in the interpretation of the results:
\begin{description}
\item[Scenario 1] $g_{q}=0.25$, $g_{\ell}=0$, $g_\chi = 1$ (leptophobic \vvec/\avec);
\item[Scenario 2] $g_{q}=0.1$, $g_{\ell}=0.01$, $g_\chi = 1$ (leptophilic \vvec);
\item[Scenario 3] $g_{q}=0.1$, $g_{\ell}=0.1$, $g_\chi = 1$ (leptophilic \avec).
\end{description}
In particular, the lower lepton coupling value is set to highlight the
dilepton search sensitivity even for very small values of this parameter.
 
The exclusions from the resonance searches (dijet, \dibjet, dilepton)
in the $(m_{Z'_{\text{V/A}}}, m_\chi)$ plane
are derived from the limits placed on  resonances reconstructed with a Gaussian shape,
while the limits from the \met+$X$
and \ditop\ analyses are derived using a mixture of simulated signal samples
and rescaling procedures as described in
Section~\ref{sec:montecarlo}. For each scenario in Figures~\ref{fig:SummaryV}~and~\ref{fig:SummaryAV}, dashed curves labelled
``thermal relic'' correspond to combinations of DM and mediator mass values
that are consistent with a DM density of $\Omega h^2= 0.12$
and a standard thermal history, as computed in MadDM~\cite{Albert:2017onk,Backovic:2015tpt}.
Between the two curves, annihilation processes described
by the simplified model deplete the relic density to below the thermal
value (except for Figure~\ref{fig:SummaryAV_V1}, where this occurs above the dashed curve).
A dotted line indicates the kinematic threshold where the
mediator can decay on-shell into DM particles. Excluded regions that
are in tension with the perturbative unitary considerations of
Ref.~\cite{Kahlhoefer:2015bea} are indicated by shading in Figure~\ref{fig:SummaryAV}.
 
The sensitivity reach of the
various experimental signatures for the leptophobic vector-mediator scenario
as a function of the DM and mediator masses is summarised in Figure~\ref{fig:SummaryAV_V1}.
Since the chosen universal quark coupling is relatively high in comparison with other benchmarks, the strongest limits are
obtained from the resonance searches.
These analyses are sensitive to mediator masses between $200\;\GeV$ and $2.5\;\TeV$ with little dependence on the DM mass.
Opening of the $\vvec\rightarrow \chi\bar{\chi}$ decay
channel, significantly reduces the sensitivity observed at high mediator masses and for
$200 < m_{\vvec} < 450\;\GeV$ when $m_{\vvec} > 2\mchi$.
The boosted dijet + ISR search is not reinterpreted here but has sensitivity in this  region.
The
lower limit on the mass is determined by the trigger requirements of the
resolved dijet + ISR analysis.
For $m_{\vvec} < 2\mchi$, masses up to $2.9\;\TeV$ are
excluded by the resolved dijet + ISR, dijet TLA and dijet searches.
Compared to the dijet searches,  the \ditop\ analysis is particularly sensitive
to the reduction in effective cross-section related to changes of the branching ratio,
as can be inferred from the coupling reach of Figure~\ref{fig:SummaryAV_mZgq}.
Conversely, the sensitivity of the \met+$X$ signatures is
highest in the region $m_{\vvec} > 2 \mchi$, up to mediator
masses of $1.5\;\TeV$ and provides unique coverage for masses below
$500\;\GeV$. The sensitivity of these analyses is strongly decreased
for $m_{\vvec} < 2\mchi$, where the DM particles
are produced off-shell, with a consequent strong reduction of the
production cross-section. For this reason, only the \monojet\
and \monophoton\ analyses can probe the off-shell regime for this
benchmark scenario, and only in the case of very low mediator and DM
masses. It is important to highlight that if the value chosen
for $g_q$ were reduced, the relative interplay
between the dijet and \met+$X$ searches would change, as
exemplified by the change of the dijet limit in the different
coupling scenarios described in the following.
 
The experimental limits for
the leptophilic vector-mediator model are summarised in Figure~\ref{fig:SummaryAV_V2}.  In this
case, the mediator decay rates into quarks are reduced in favour of a
higher branching ratio to DM particles, reducing the
sensitivity of dijet searches to this scenario, whereas the leptonic
branching ratio allows dilepton searches to impose
constraints on a wide range of mediator masses. The gap in sensitivity of dijet searches around mediator
masses of about $1.7\;\TeV$ is due to statistical fluctuation in the dijet spectrum. This benchmark
highlights the complementarity among dijet, dilepton, and \met+$X$ final states.
In this case, \dibjet\ and \ditop\ searches are not included in the final result.
The resonance searches exclude mediator masses between
$150\;\GeV$ and $2\;\TeV$ (except for a small gap around $1.7\;\TeV$),
if $m_{\vvec} < 2 \mchi$ and between $150$ and $350\;\GeV$ for
all DM masses. Complementarily, the \met+$X$ searches exclude
mediator masses up to $1\;\TeV$ for $m_{\vvec} > 2 \mchi$.
 
Similar considerations can be made for the axial-vector mediator
models, presented in
Figure~\ref{fig:SummaryAV},
with the exception that in the presence of non-vanishing coupling to leptons
(leptophilic scenario),
the dilepton resonance search becomes by far the most
sensitive analysis for this model, excluding the mass range $150\;\GeV
< m_{\avec} < 2.8\;\TeV$ for any DM mass.
Also in this case, the sensitivity of this
analysis increases when the $\avec\rightarrow \chi\bar{\chi}$ is
kinematically forbidden and becomes independent of the DM mass above
threshold.
For $m_{\avec} < 2 \mchi$, masses up to $3.5\;\TeV$ are excluded.

\begin{figure}[htbp]
\centering
\subfloat[]{\includegraphics[width=0.7\linewidth]{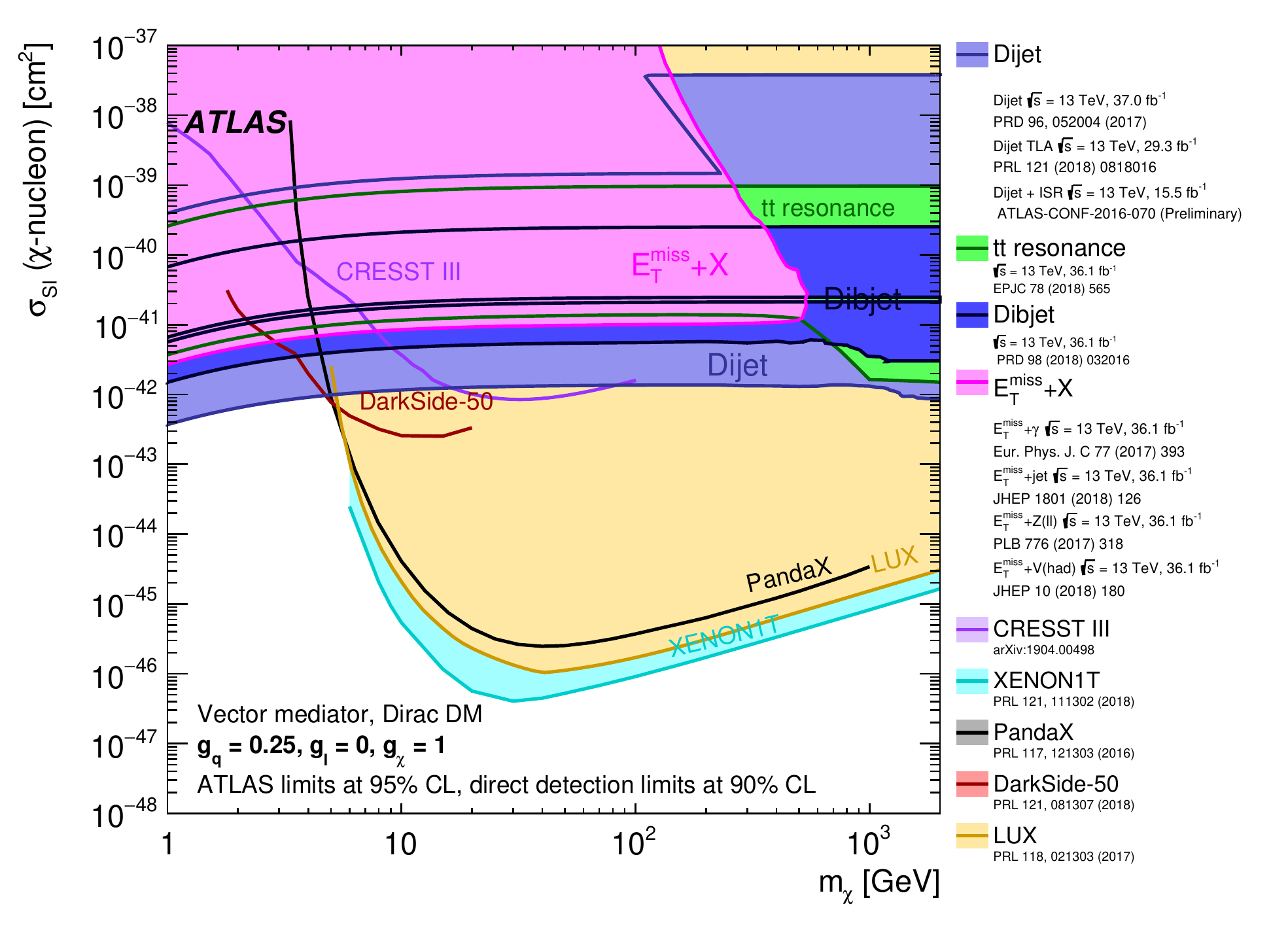}
\label{fig:SummaryAV_DD_V1}
}
\\
\subfloat[]{\includegraphics[width=0.7\linewidth]{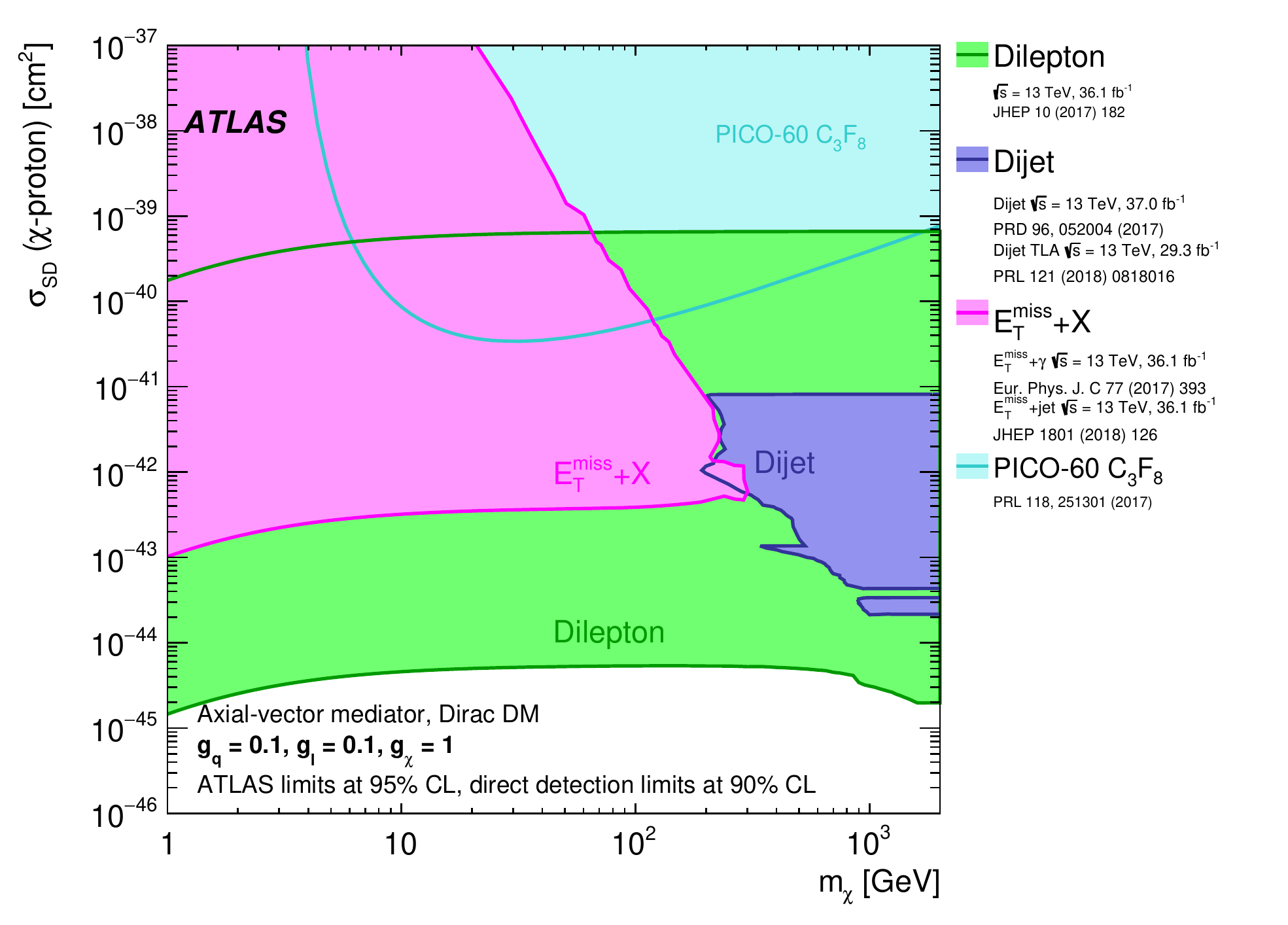}
\label{fig:SummaryAV_DD_V2}
}
\caption{A comparison of the inferred limits with the constraints from
direct detection experiments on (a) the spin-dependent WIMP--proton
scattering cross-section in the context of the vector leptophobic model  and (b) the  spin-independent WIMP--nucleon scattering cross-section in the context of the  axial-vector leptophilic model.
The results from this analysis,
excluding the region inside or to the left of the contour, are compared with
limits from direct detection experiments. ATLAS limits are shown at 95\% CL and
direct detection limits at 90\% CL. ATLAS
searches and direct detection experiments exclude the shaded
areas.
Exclusions beyond the canvas are not implied for the ATLAS results.
The
dijet and \met+$X$ exclusion regions represent the union of
exclusions from all analyses of that type.}
\label{fig:SummaryAV_DD}
\end{figure}
 
Collider experiments provide an approach to DM searches which is complementary to direct and
indirect detection experiments~\cite{Boveia:2016mrp}.
It is therefore interesting and informative, though model-dependent, to compare the V/AV limits with the results from other DM searches.
Figure~\ref{fig:SummaryAV_DD} shows the translation of the V/AV model
limits into limits on the spin-dependent $\chi$--proton and
spin-independent $\chi$--nucleon
scattering cross-sections as a function of the DM mass.
The direct detection experiments dominate the sensitivity by a few orders of magnitude for DM masses
above $10\;\GeV$, thanks to coherence effects, the spin-independent interaction cross-section with heavy
nuclei is enhanced by $A^2$, where $A$ is the number of nucleons in a
nucleus.
However, with the assumed coupling strengths, the analyses presented
in this paper complement direct detection limits in the low DM mass
range where the direct DM search experiments have less sensitivity
due to the very low energy recoils that such low-mass DM
particles would induce. The lower edge of the sensitivity contour for
all analyses in Figure~\ref{fig:SummaryAV_DD_V1} (Figure~\ref{fig:SummaryAV_DD_V2})
is driven by the high-mass reach of each analysis in
Figure~\ref{fig:SummaryAV_V1} (\ref{fig:SummaryAV_A2}),
as the scattering cross-section limit is inversely
proportional to the mediator mass reach (raised to the fourth power).
Conversely, the upper edge of the scattering cross-section contour for the dijet and dilepton analyses is driven
by their low-mass sensitivity limit due to the trigger requirements
employed in these analyses.
Further details of this comparison are
discussed in Appendix~\ref{app:directindirect}.

\FloatBarrier
\subsubsection{Baryon-charged interaction}
\label{ssub:spin1res2}
 
In the context of the \vbc\ model, the results from the
\monoHgg\ and \monoHbb\ analyses
are interpreted in the plane formed by the \bvec\ and DM
masses, due to the characteristic signature of this model involving
Higgs-strahlung from the \bvec\ mediator.
The \monoHbb\ interpretation was developed subsequently to the original publication~\cite{EXOT-2016-25}.
The results are shown in Figure~\ref{fig:SummaryZprimeB} in the
$(m_{\bvec}, m_{\chi})$ plane for $g_{q} = 1/3$, $g_{\chi} = 1$ and
$\sin \theta = 0.3 $. The dashed lines indicate the expected exclusion
contours from the two separate channels and their combination (based on best
expected limits), while the black solid line shows the observed exclusion,
presented only for the combined result.  The band around the
expected combined contour shows the effect of one-standard-deviation variation
of the total systematic uncertainties. The \monoHbb\ analysis sets the
strongest bounds in this model, excluding mediator masses up to
$1.9\;\TeV$ for all DM mass hypotheses for which the mediator
invisible decay is kinematically allowed. Due to the lower branching
ratio, the \monoHgg\ is less sensitive to this model for high mediator masses,
but it is competitive for $m_{\bvec} < 50\;\GeV$ thanks to the higher acceptance of the analysis which can trigger on the photons instead of the \met\ and
its smaller systematic uncertainties compared to the \monoHbb\ analysis.

\begin{figure}[htbp]
\centering
\includegraphics[width=0.6\linewidth]{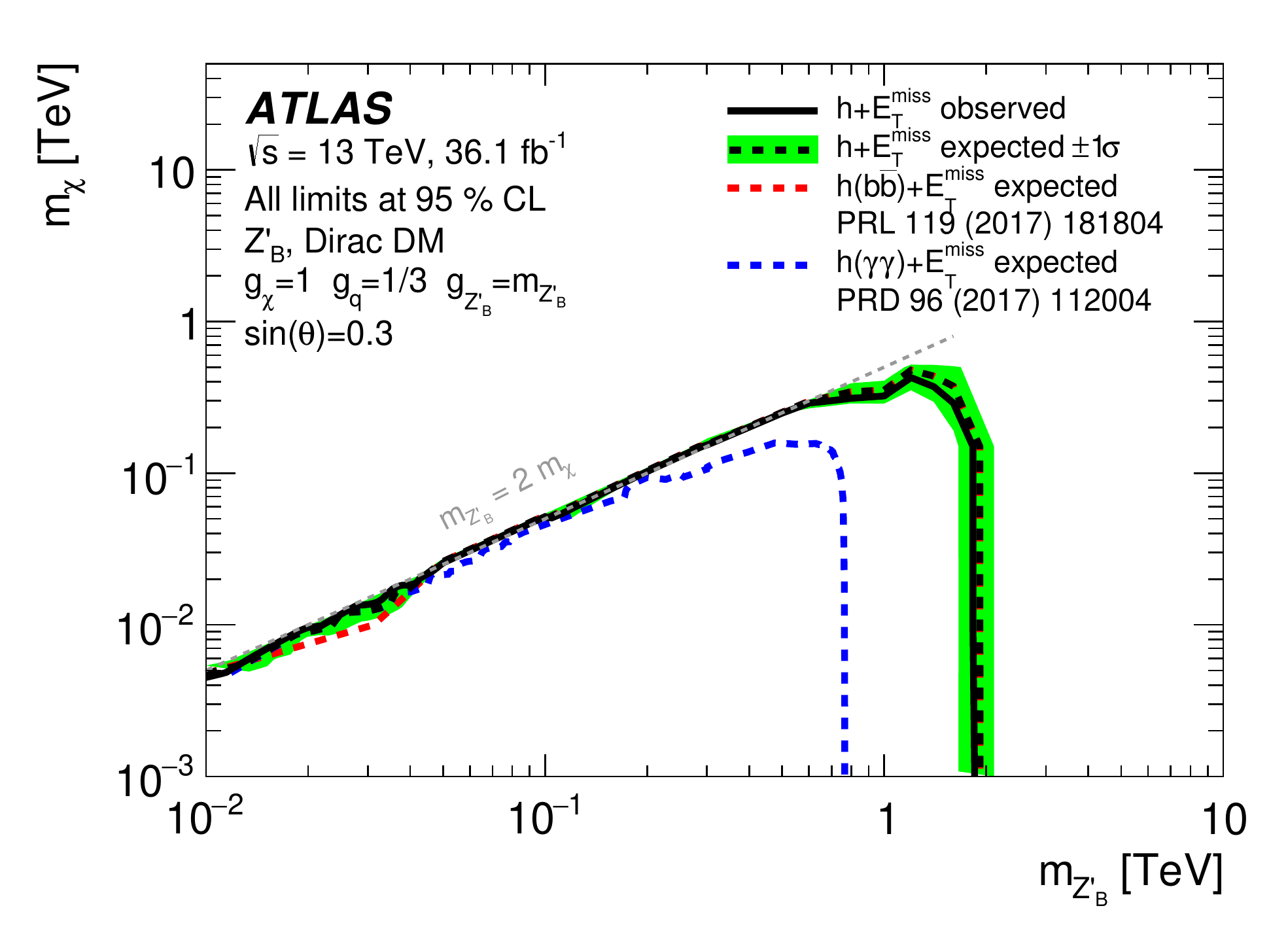}
\caption{Exclusion contours for the \vbc\ model in the
$(m_{\bvec}, m_{\chi})$ plane for $g_{q} = 1/3$, $g_{\chi} = 1$ and
$\sin \theta = 0.3 $. The dashed lines indicate the expected exclusion
contours from the two separate channels and their
combination, while the solid line shows the observed exclusion,
presented only for the combined result.
The band around the
expected combined contour shows the effect of a one-standard-deviation variation
of the total systematic uncertainties. At high mediator masses, the combined exclusion fully overlaps with the exclusion
from the \monoHbb\ analysis. }
\label{fig:SummaryZprimeB}
\end{figure}
 
\FloatBarrier
\subsubsection{Neutral flavour-changing interaction}
\label{ssub:spin1res3}
 
For the \vfc\ models,
expected and observed limits from the \monotop\ and same-sign $tt$ analyses
are derived for each independent subprocess leading to the two
signatures, as described in Section~\ref{ssub:spin1theory3} and
schematically summarised in
Figures~\ref{g:Diag-AVVFCN1}--\ref{g:Diag-AVVFCN3}. These individual results are
converted into limits for the complete \vfc\ model
following the rescaling procedure described in Section~\ref{sub:rescaleNFC}.
These results were obtained subsequently to the original analyses publications.
 
The sensitivity of the experimental analyses to this model is
explored in three scenarios that investigate different interpretation variables (where $\mathcal{B}$ is the invisible branching ratio of the mediator):
\begin{description}
\item[Scenario 1] $( m_{\nvec}, \gSM)$ interpretation plane, assuming $\gDM = 0.5$ or $\gDM =1.0$.
\item[Scenario 2] $(\mathcal{B}(\chi\bar{\chi}), \gSM)$ interpretation plane,  assuming $m_{\nvec} = 1 \;\TeV$.
\item[Scenario 3] $(\gDM, \gSM)$ interpretation plane, assuming $m_{\nvec} = 1 \;\TeV$.
\end{description}

The first scenario, presented in Figure~\ref{fig:SummaryFCNC_gSMmV},
directly identifies the constraints on the main
parameters of the model. The two different $\gDM$ coupling values
highlight the different contributions of the invisible final state
(\monotop), which probes $\gSM$ values down to $0.7$ for $1\;\TeV$ mediators and can exclude
couplings down to $0.13$ for $1.5\;\TeV$ mediators when the DM coupling is set to unity.
In this scenario the visible final state ($tt$) constrains couplings down to $0.3$ for mediator masses up to $3\;\TeV$ and it is relatively independent of $\gDM$.
This is due to the fact that  the sensitivity of this final state is dominated by  the $t$-channel exchange of the mediator
and therefore it is sensitive to $\gDM$ only through the total width of the mediator.
In this result, only mediator masses above $1\;\TeV$ are under study.
However, mediator masses down to $100\;\GeV$ are excluded by previous publications~\cite{TOPQ-2013-11}
for a coupling assumption of $g_q \sim 3\cdot 10^{-2}$.

Limits are quite similar for the two DM coupling values for the
same-sign tt analysis. The sensitivity of this final state is
dominated by  the $t$-channel exchange of \nvec\ especially for the
values of \gSM\ which are probed. This process is sensitive to
$\gDM$ only through the total with of the mediator involved in the
propagator, and therefore does not change drastically when \gDM\
is varied.

The second and third scenarios, presented in Figure~\ref{fig:SummaryFCNC_gSM},
further highlight the complementarity between the visible and invisible final states
as a function of the couplings and the invisible branching ratio of the mediator.
Couplings of the \nvec\ mediator to
SM fermions can be excluded down to $0.14$ for any value of $\gDM$ or
$\mathcal{B}(\chi\bar{\chi})$.
 
\begin{figure}[h]
\centering
\subfloat[]{\includegraphics[width=0.49\linewidth]{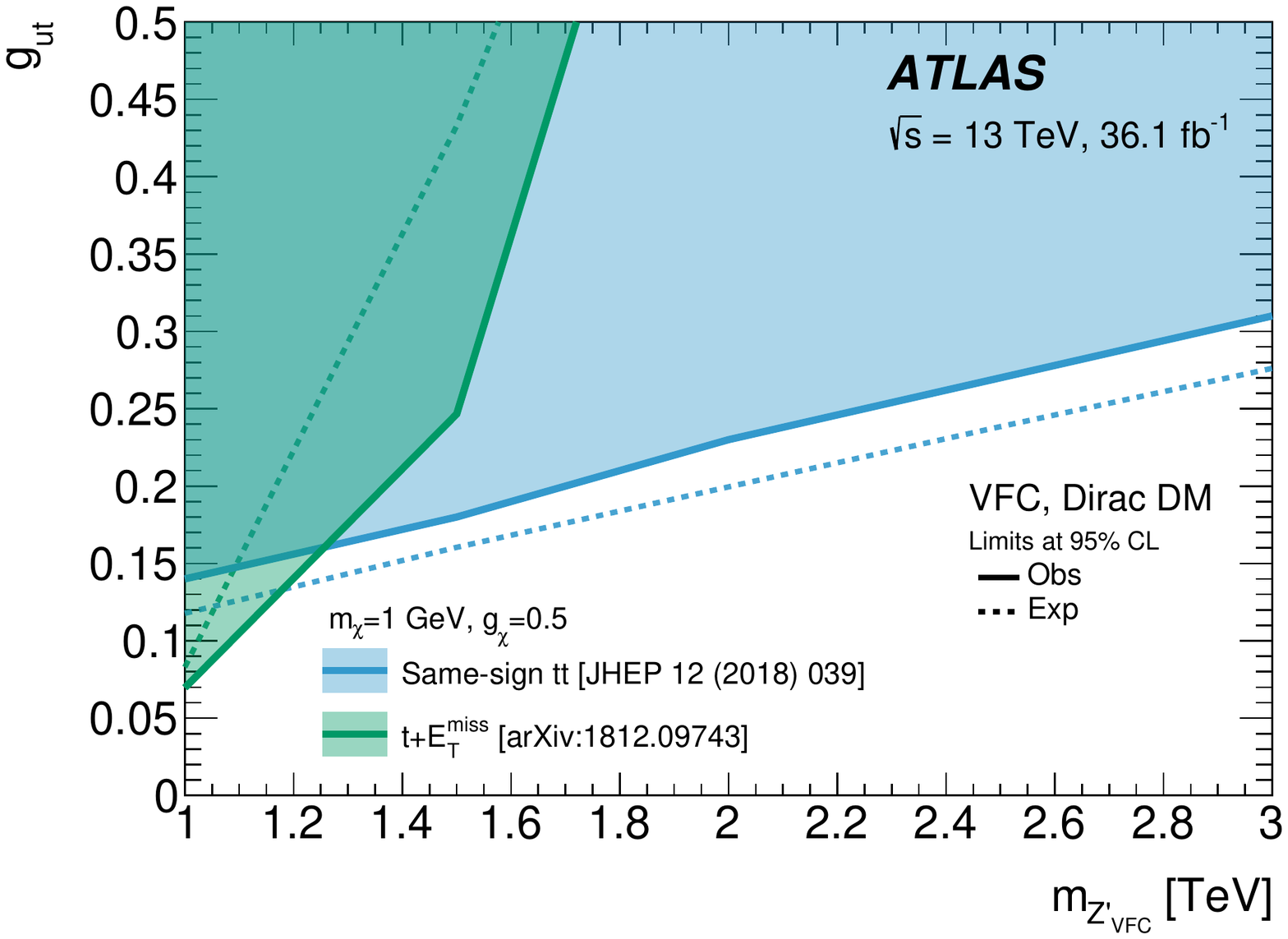}
}
\subfloat[]{\includegraphics[width=0.49\linewidth]{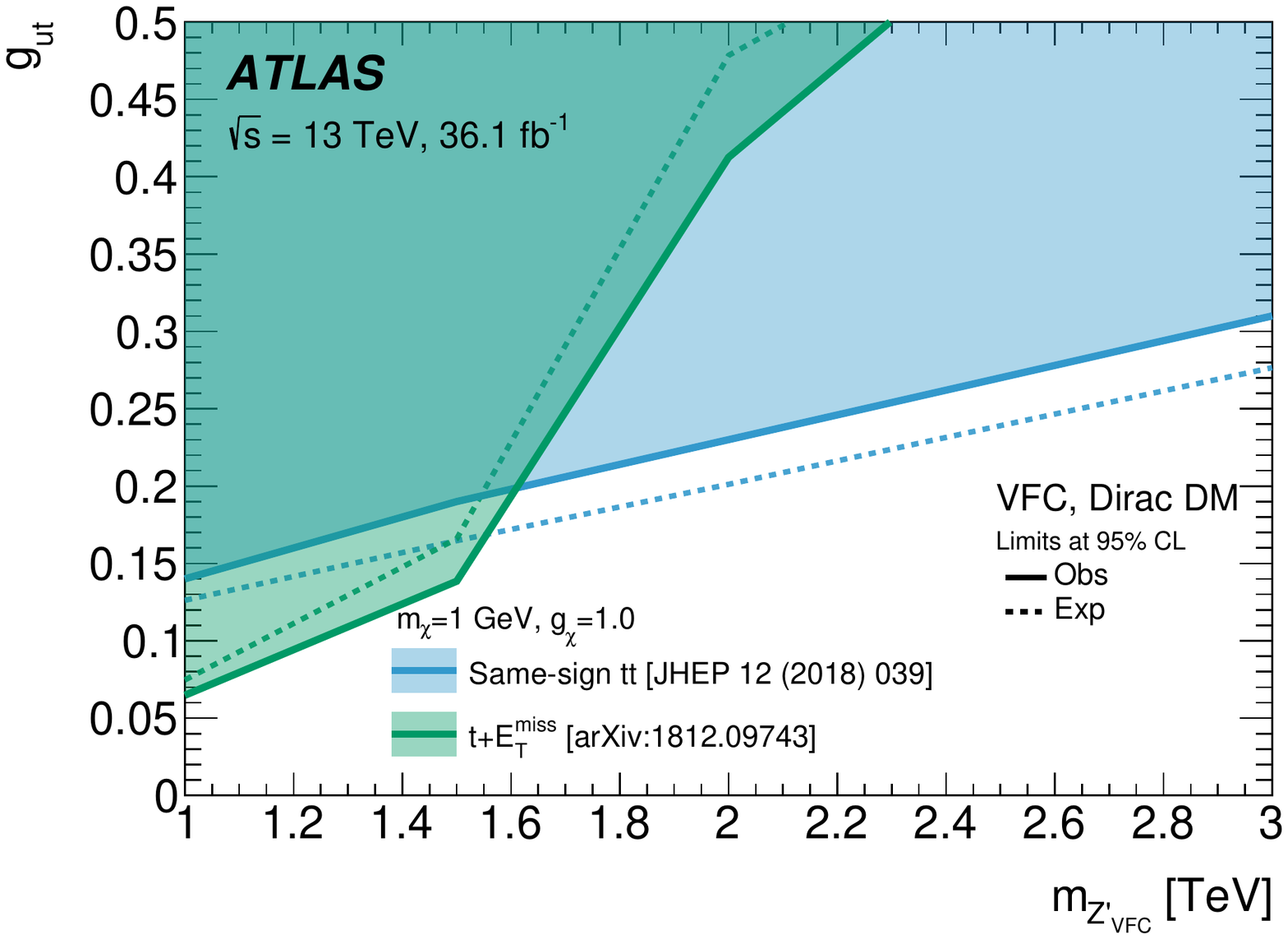}
}
\caption{Exclusion regions in the $(m_{\nvec}, \gSM)$ plane from the
\monotop\ and same-sign $tt$ analyses for the \vfc\ model. The observed exclusion is indicated
for each of the two analyses by the filled area. The mass of the DM particle is set to $1\;\GeV$ and the DM coupling, $g_\chi$, is set to (a) 0.5 or (b) 1.}
 
\label{fig:SummaryFCNC_gSMmV}
\end{figure}
\begin{figure}[h]
\centering
\subfloat[]{\includegraphics[width=0.49\linewidth]{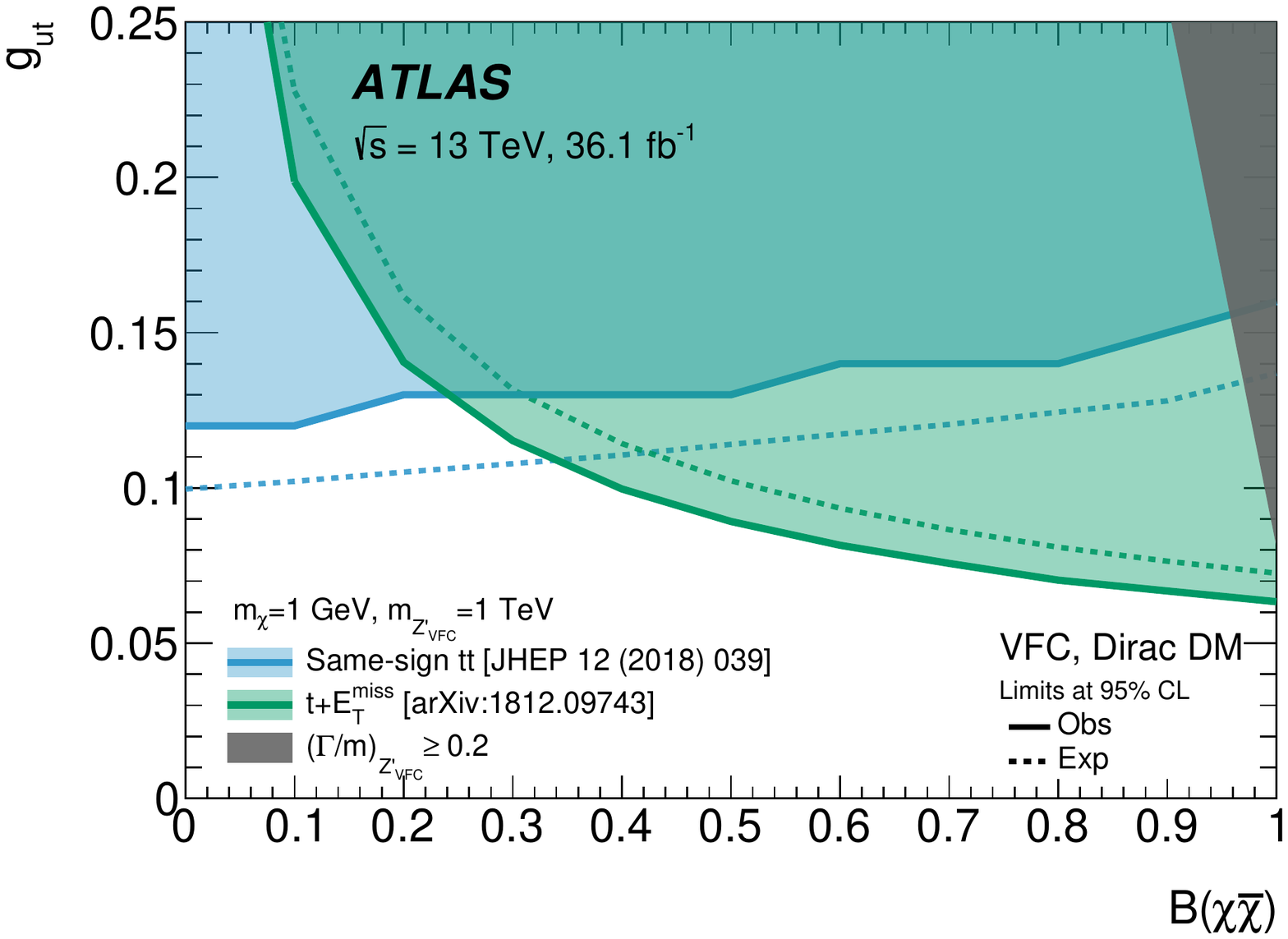}
}
\subfloat[]{\includegraphics[width=0.49\linewidth]{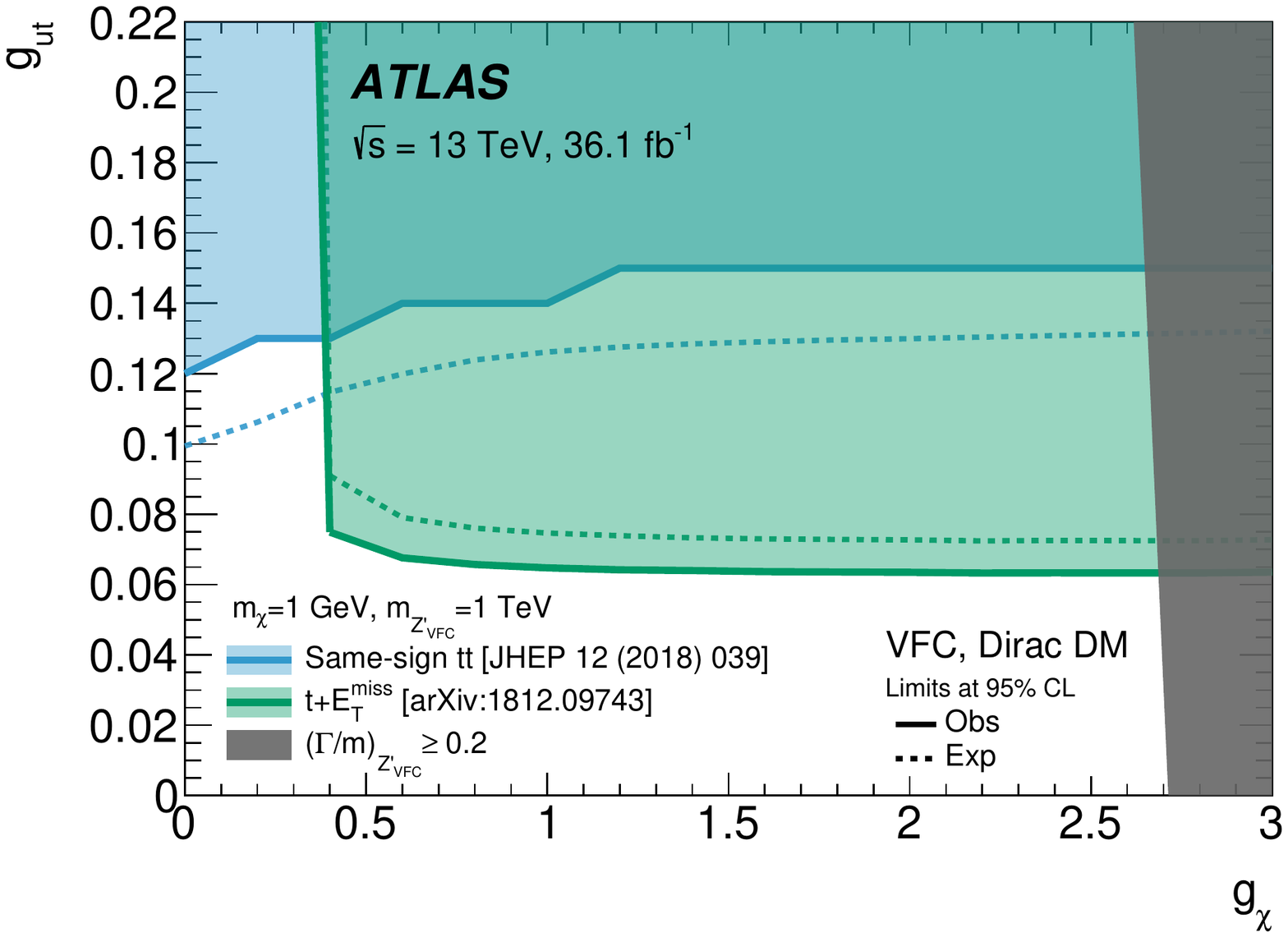}
}
\caption{Exclusion limits from the \monotop\ and same-sign $tt$ analyses for the \vfc\ model as a function
of the SM coupling $\gSM$ and (a) the DM branching ratio or (b) the DM coupling $g_\chi$. The observed exclusion is indicated
for each of the two analyses by the filled area. The mass of the DM particle is set to $1\;\GeV$ and the mass of the \nvec\ boson is set to $1\;\TeV$. The dark shaded area corresponds to an invisible partial width of the mediator above 20\%.}
\label{fig:SummaryFCNC_gSM}
\end{figure}

\FloatBarrier
\subsection{Scalar or pseudo-scalar dark matter models}
 
\subsubsection{Colour-neutral interaction}
\label{ssub:spin0res1}

The most stringent limits on S/PS models are obtained from \dmtt\
final states, which are studied in three channels assuming
fully-hadronic, semileptonic and fully-leptonic top
pair decays, respectively. The fully leptonic channel
excludes scalar-mediator models with unitary couplings $g_\chi = g_q = g = 1$ up to
mediator masses of $45\;\GeV$, setting in this mass range the
strongest upper limits on the ratio of the signal production
cross-section to the nominal cross-section (signal strength or $\sigma/\sigma(g=1.0)$), as
shown in Figure~\ref{fig:S}.
In the case of pseudo-scalar mediator models (Figure~\ref{fig:PS}),
similar sensitivity is obtained by all channels and mediator masses
in the range $15$--$25\;\GeV$ are excluded.
In all cases, a DM mass of $1\;\GeV$ is assumed, but the results are
valid for all DM mass choices for which the mediator's
decay into a pair of DM particles is kinematically allowed ($m_{\phi / a} >
2\mchi$). Pseudo-scalar mediator models can also be constrained
by \monojet\ final states, where the mediator is produced through
loop-induced gluon fusion. Although the limits obtained by this
signature are not competitive with the \dmtt\ final state, except in
the mass range above $300\;\GeV$, they
provide a complementary constraint, which would become particularly
important in case of a discovery. For the scalar model, the \monojet\ final-states
cross-section is instead too small to be probed.
Ditop resonance searches in final states with two or four tops can
also constrain this parameter space for $m_{\phi /a} > 2 m_t$.
However, \ttbar\ production through a spin-0 resonance presents a
strong interference pattern with SM top pair
production \cite{EXOT-2016-04},
which needs to be treated with care. On the other hand,
four-top final states are characterised by relatively low event yields
with the currently available integrated luminosity. For these reasons the
study of these final states is not considered here.
Finally, \dmbb\ final states are also used to set constraints on these
simplified models, resulting in upper limits on the signal strength
between $200$ and $300$ for mediator masses below $100\;\GeV$.
These results quantify the sensitivity to these models if up-type couplings
are suppressed.

\begin{figure}[htbp]
\centering
\subfloat[]{\includegraphics[width=0.7\linewidth]{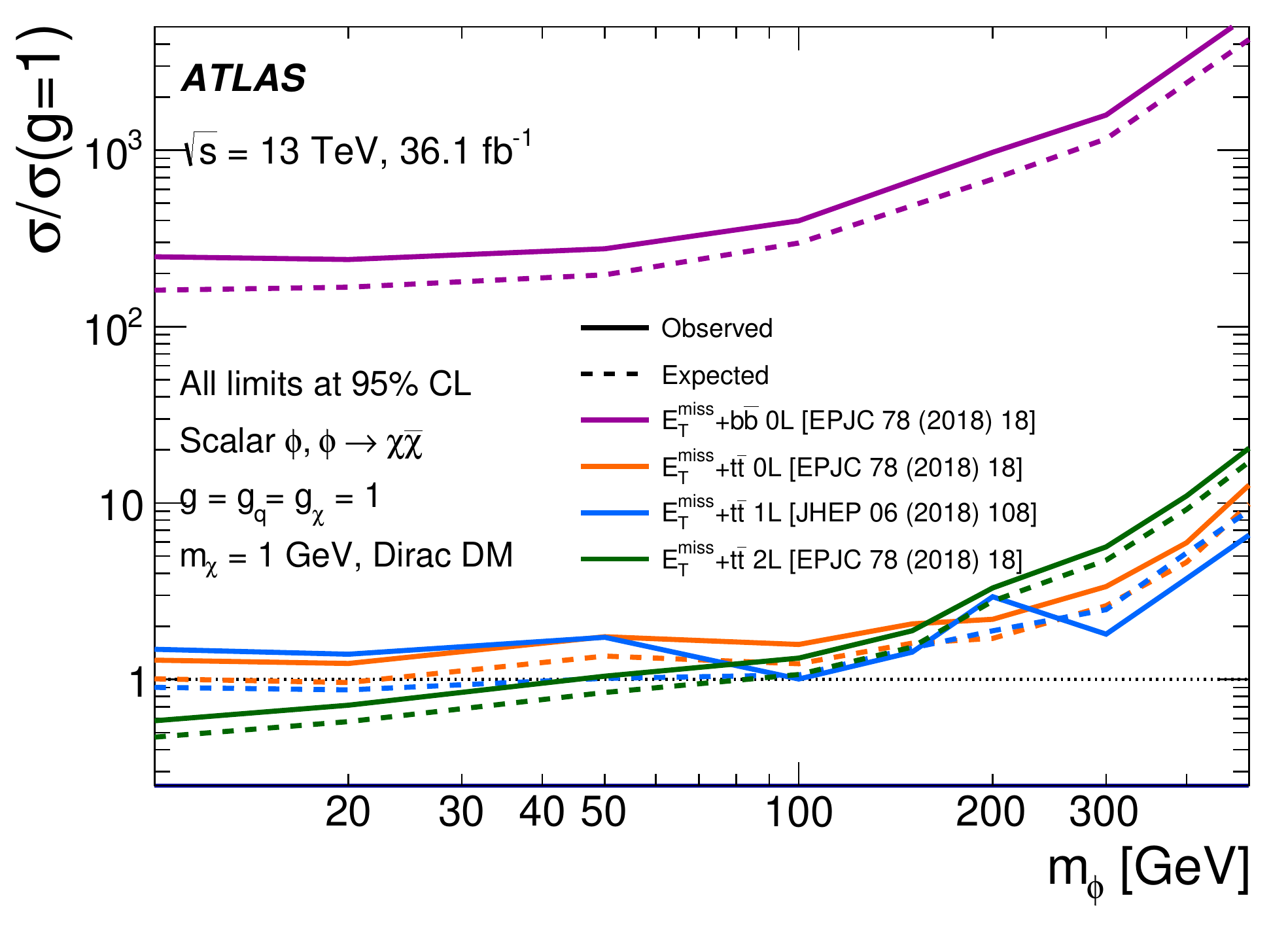}
\label{fig:S}
}
 
\subfloat[]{\includegraphics[width=0.7\linewidth]{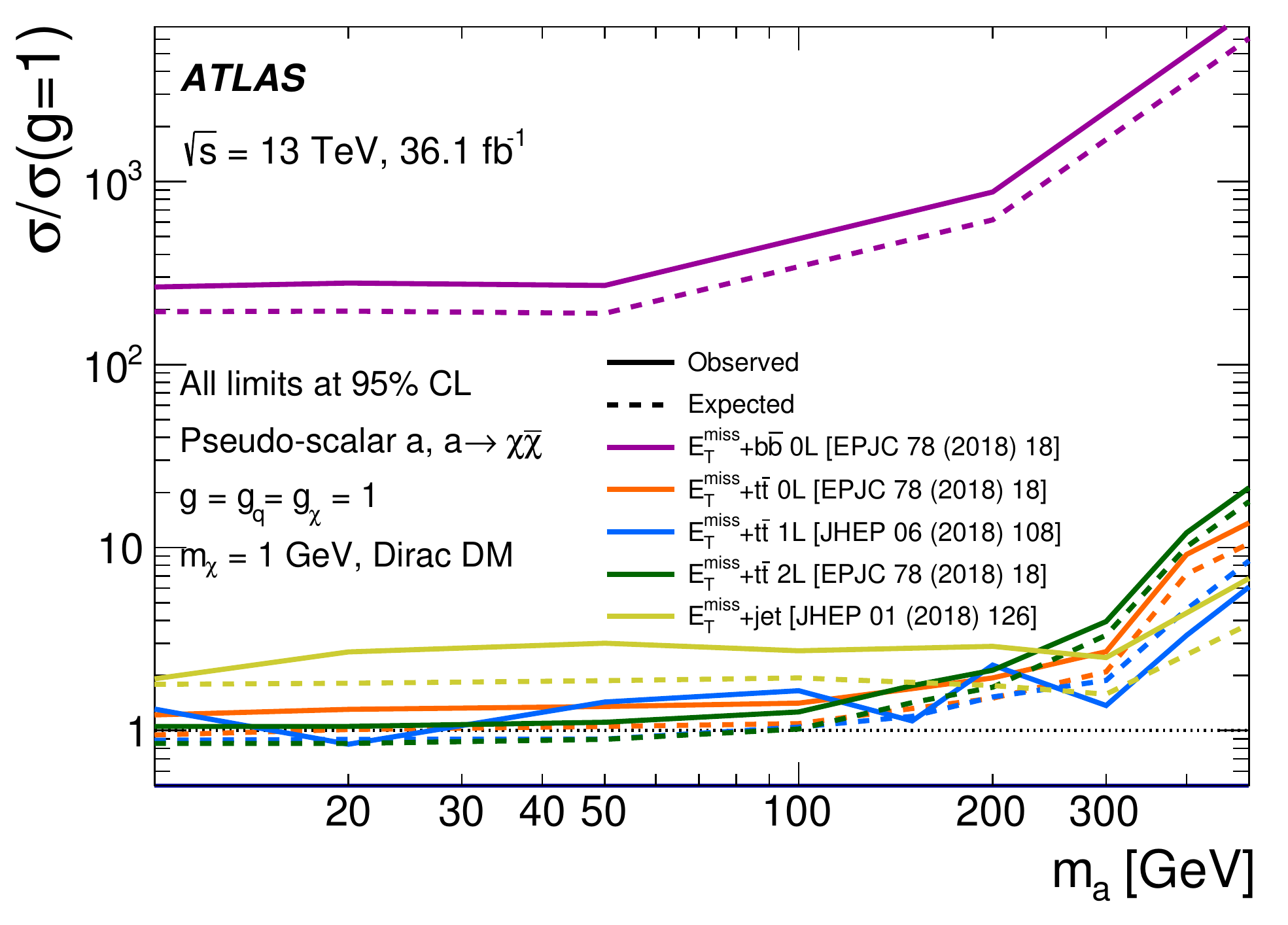}
\label{fig:PS}
}
\caption{Exclusion limits for  (a)  colour-neutral scalar or
(b) pseudo-scalar  mediator models as a function of the mediator mass for a
DM mass of $1\; \GeV$. The limits are calculated at 95\% CL and are
expressed in terms of the ratio of
the excluded cross-section to the
nominal cross-section for a coupling
assumption of $g = g_q = g_\chi =
1$. The solid (dashed) lines show the
observed (expected) exclusion limits
for each channel.}
\end{figure}
 
\FloatBarrier
\subsubsection{Colour-charged interaction}
\label{ssub:spin0res2}
 
The strongest exclusion limits on colour-charged mediators \cscal\ that couple to
first- and second-generations left-handed quarks are set by
the \monojet\ analysis. Assuming a unitary coupling, \cscal\
mediator masses up to $1.7\;\TeV$ are excluded for $m_\chi =
50\;\GeV$. Furthermore, \cscal\ mediator masses below $600\;\GeV$ are excluded
for all DM masses such that the decay $\cscal \rightarrow q \chi$ is
kinematically allowed.
The strongest exclusion limits on colour-charged mediators \cscalb\ that couple to
third-generation right-handed $b$-quarks are set by
the \monob\ analysis. Assuming a coupling set to
the value that yields a relic density value consistent with
astrophysical observations,
masses up to $1.4\;\TeV$ are excluded for $m_\chi =
1\;\GeV$.
Finally, \monotop\ final states are used to constrain the colour-charged
mediator's \cscalt\ coupling to right-handed top quarks. Mediator
masses up to $3.4\;\TeV$ are excluded, assuming a $10\;\GeV$ DM
particle mass and setting the coupling strengths of this model to: $\lambda_t = 0.4$ and $g_{s} = 0.2$.

\FloatBarrier
\subsection{Extended Higgs sector dark matter models}
\subsubsection{Two-Higgs-doublet models with a vector mediator}
\label{sub:2HDMspin1res}
 
The \thdmZ\ model is constrained by the \monoHbb\ and \monoHgg\
analyses. The results are interpreted in terms of exclusion limits in
the ($m_A$, $m_{\vvec}$) plane shown in
Figure~\ref{fig:Summary2HDMZprime}. The statistical
combination of the two analyses is also presented.
Masses of the pseudo-scalar $A$
in the range $200$--$600\;\GeV$ are excluded for
$m_{\vvec} = 1.5\;\TeV$. The limit in sensitivity is driven by the fact that the
$A\rightarrow \chi\bar\chi$ branching ratio decreases with increasing $m_A$ due to
decay channels involving top quarks or other heavy bosons of the extended Higgs sector becoming accessible (\ttbar, $HZ$ and $W^\pm H^\mp$).
At higher $m_{\vvec}$ the loss in branching ratio is combined with the
smaller production cross-section so that the reach of the analysis is limited to smaller pseudo-scalar masses.
For $m_A < 2m_{t}$ and $m_A > 2 m_\chi$, there are no  more competing decay channels
and the reach of the analysis does not depend on $m_A$ any longer.
This creates the turnover in the exclusion contour for $m_{\vvec} = 2.5\;\TeV$.
 
The two \monoH\ decay signatures are highly complementary at low \vvec\ masses,
as can be observed in the enlarged inset in the figure, while the
\monoHbb\ analysis dominates the sensitivity at high \vvec\ masses.
Due to this complementarity, the gain obtained by the statistical
combination of the two signatures
is limited to the low mass region for this model.

\begin{figure}[htbp]
\centering
\includegraphics[width=0.6\linewidth]{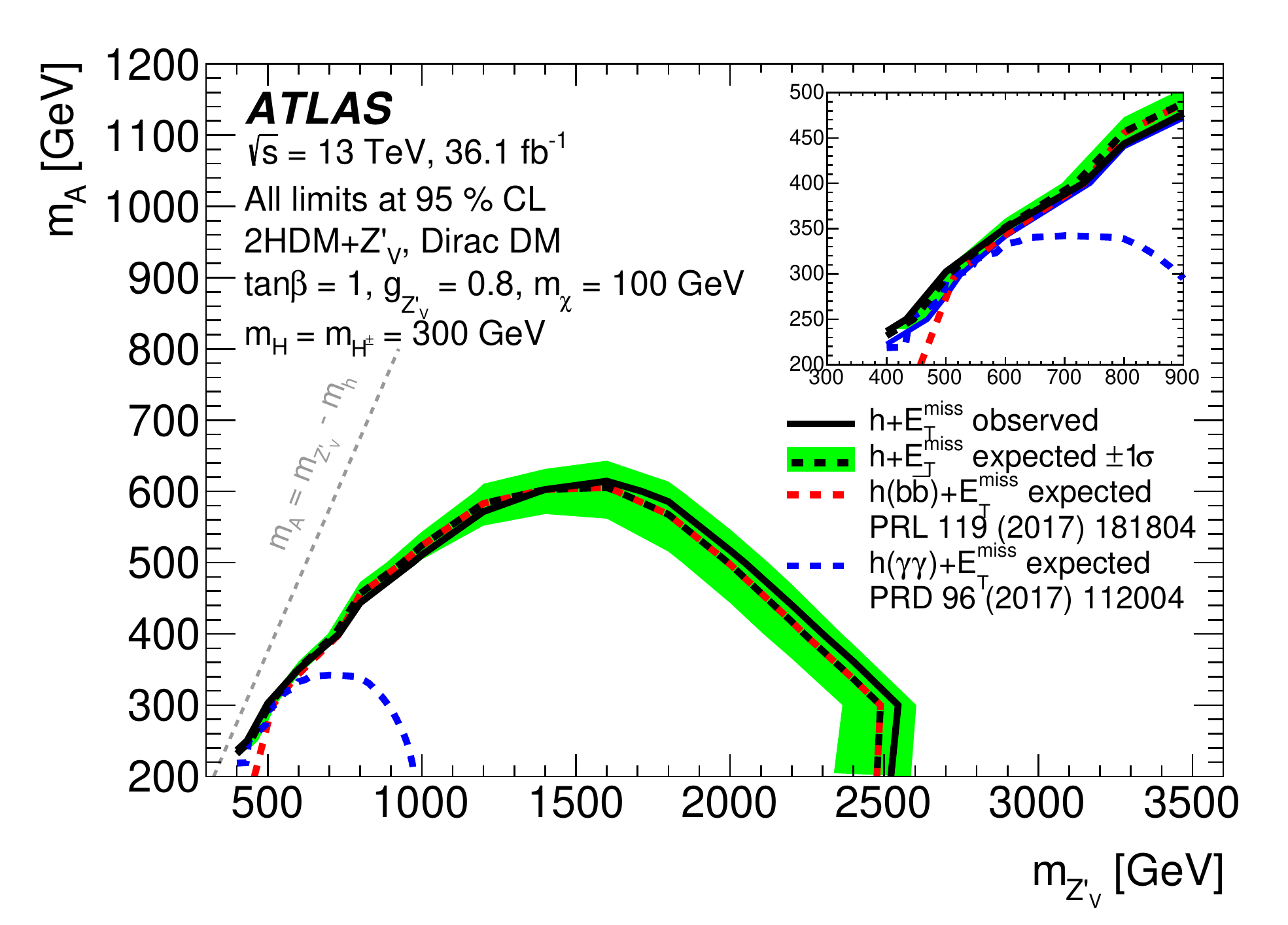}
\caption{Exclusion contours for the \thdmZ\ scenario in the
$(m_{\vvec}, m_{A})$ plane for $\tan\beta = 1$, $g_{\vvec} = 0.8$ and
$m_{\chi} = 100$~GeV. The dashed lines indicate the expected exclusion
contours from the two separate channels and their statistical
combination, while the black solid line shows the observed exclusion,
presented only for the combined result.  The band around the
expected combined contour shows the effect of a one-standard-deviation variation
of the total systematic uncertainties.
The sharp turn in the exclusion contour for $m_{\vvec} = 2.5\;\GeV$ is given by the
opening $A$ decay channels competing with the considered final state for $m_A > 2m_{t}$.
For this reason the exclusion sensitivity does not depend on $m_A$ below threshold.
The inset in the top-right side
of the panel shows a zoomed-in version of the result for low $m_{\vvec}$
masses to highlight the complementarity between the \monoHbb\ and
the \monoHgg\ analyses in this parameter region.
}
\label{fig:Summary2HDMZprime}
\end{figure}
 
\FloatBarrier
\subsubsection{Two-Higgs-doublet models with a pseudo-scalar mediator}
\label{sub:2HDMspin0res}

\begin{figure}[p]
\centering
\subfloat[]{\includegraphics[width=0.7\linewidth]{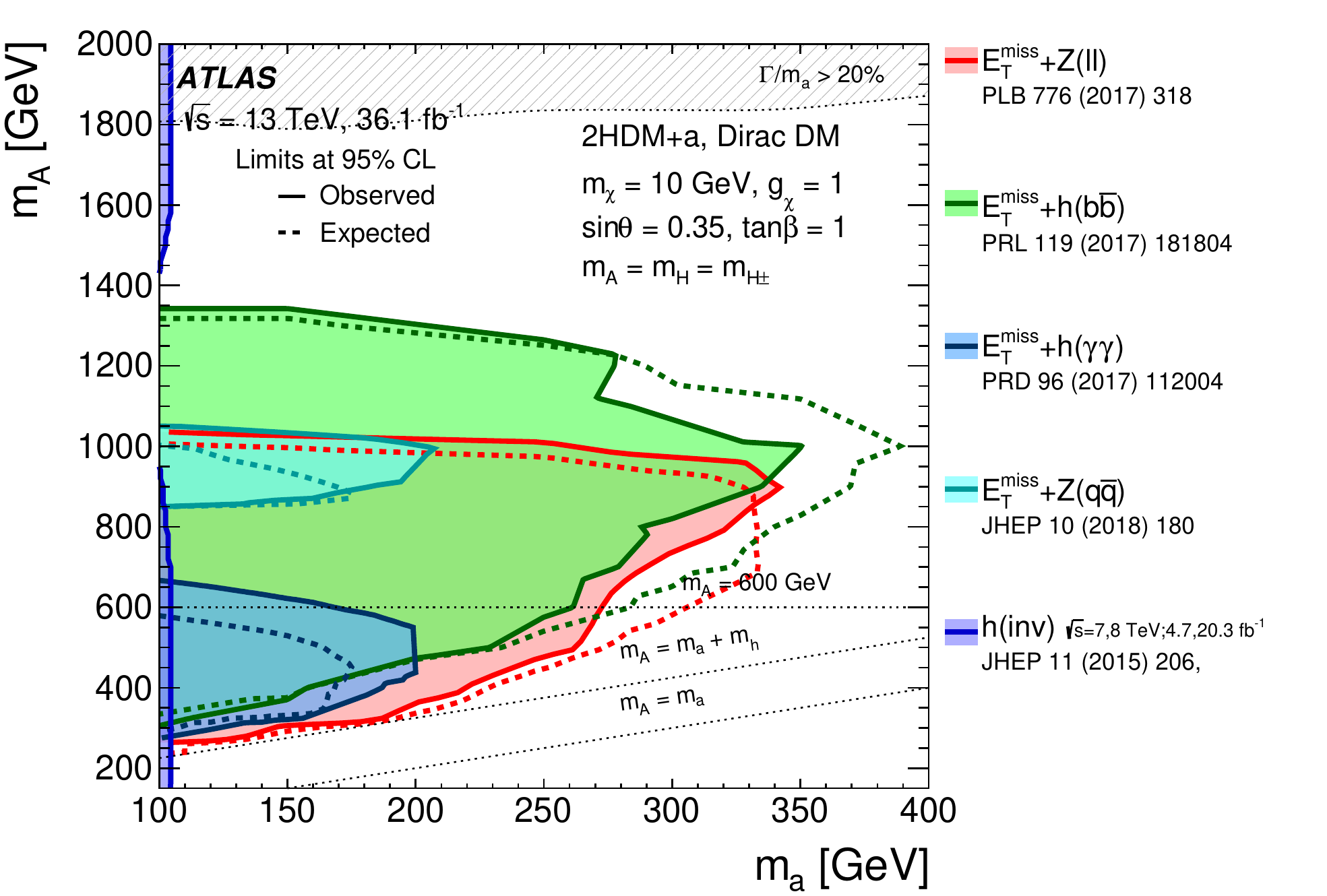}
\label{fig:Summary2HDMPS_mAma}
}  \\
\subfloat[]{\includegraphics[width=0.7\linewidth]{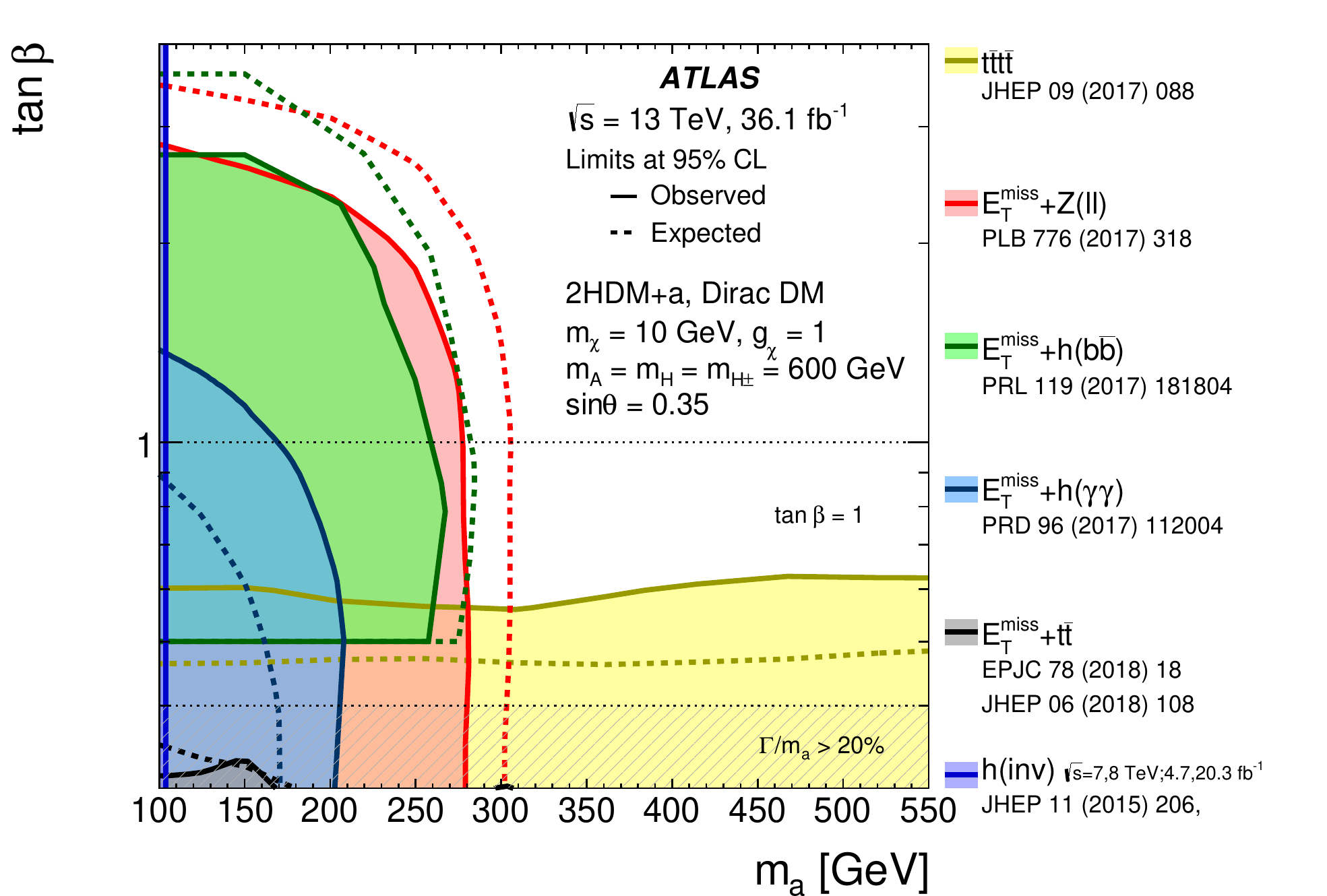}
\label{fig:Summary2HDMPS_tanbmZ}
}
\caption{Regions in the (a) $(m_{\pscal},m_A)$ and (b) $(m_{\pscal},\tan\beta)$ planes excluded by data at
95\% CL by $X+\met$ and \fourtop\ analyses, following the parameter choices of scenarios 1 and 2 of the \thdmS\ model. The dashed grey regions at
the top of (a) and the bottom of (b)
indicate the region where the width of
any of the Higgs bosons exceeds $20\%$ of its
mass.
The exclusion limits presented above conservatively neglect the contribution from \bbbar-initiated production,
which might be sizeable for $\tan\beta \geq 3$ for the \monoZ\ channel and, to a lesser extent, for the \monoH\ one.
}
\end{figure}

As highlighted in Section~\ref{sub:2HDMspin0theory}, the  \thdmS\ model is
characterised by a rich phenomenology. Constraints on this model from ATLAS searches are
presented in this paper.
Four different benchmark scenarios
are used to evaluate the sensitivity to this model achieved by the $Z/h+\met$, $\ttbar/\bbbar$+\met,
\hinv, and \fourtop\ analyses.
These four benchmark scenarios \cite{2HDMWGproxi} are consistent
with bounds from electroweak precision, flavour and Higgs observables
and are chosen to
highlight the complementarity of the various final states.
These scenarios represent two-dimensional and one-dimensional scans of
a five-dimensional parameter space, used to present the exclusion limits.
\begin{description}
\item[Scenario 1] $(m_{\pscal},m_A)$ exclusion plane assuming $\tan\beta = 1$ and $\sin\theta=0.35$;
\item[Scenario 2] $(m_{\pscal},\tan\beta)$ exclusion plane assuming  $m_A =
600\;\GeV$ and $\sin\theta=0.35$;
\item[Scenario 3] $\sin\theta$ exclusion scan assuming
\begin{enumerate}
\item[a)]  $m_A = 600\;\GeV$ , $m_{\pscal} = 200\;\GeV$ and $\tan\beta=0.5$, $1$ or $50$;
\item[b)]  $m_A = 1000\;\GeV$ , $m_{\pscal} = 350\;\GeV$ and $\tan\beta=0.5$ or $1$;
\end{enumerate}
\item[Scenario 4] $\mchi$ exclusion scan assuming $m_{A}=600$ GeV,
$m_{\pscal}=250$ GeV, $\tan\beta=1$, $\sin\theta=0.35$.
\end{description}
 
In all cases, the masses of the heavy pseudo-scalar, heavy scalar, and charged bosons are kept equal ($m_A = m_H = m_{H^\pm}$).
As visible in the results
presented in Figure~\ref{fig:Summary2HDMPS_mAma}, the exclusion
sensitivity is vastly dominated by the \monoHbb\ and \monoZll\
analyses in the first scenario. These analyses are mostly sensitive to
the production diagrams of
Figures~\ref{fig:THDMpa_feynd}~and~\ref{fig:THDMpa_feyne} and their
sensitivity depends on both pseudo-scalar mediator masses.
The maximum reach is obtained for
light pseudo-scalar $m_{\pscal}$ up to $340\;\GeV$, if the $A$ boson mass is
set to $1\;\TeV$, while for $m_{\pscal} = 150\;\GeV$ $A$ boson masses between
$280\;\GeV$ and $1.35\;\TeV$ are excluded.
The combined contours of the \monoHbb\
and \monoZll\ analyses include the \monoHgg\ and \monoZqq\ exclusion areas, although
the \monoHgg\
analysis still complements \monoHbb\ at low $(m_{\pscal},m_A)$ values.
Finally, the \hinv\ branching ratio limit constrains
very low values of $m_{\pscal}$ for $m_A$ mass below $900\;\GeV$ and above $1.4\;\TeV$,
being sensitive only to the $\pscal$
boson production cross-section.

In the context of 2HDM models, it is customary to investigate the
sensitivity in terms of the $\tan\beta$ parameter. This is achieved in
the second scenario presented in Figure~\ref{fig:Summary2HDMPS_tanbmZ}.
Although the exclusion reach is dominated also in this case by
the \monoHbb\ and \monoZll\ analyses, two additional signatures
contribute at $\tan\beta \sim 0.5$: the \dmtt\ and the \fourtop\
analyses. The sensitivity of the former analysis is driven by the
production cross-section of the $\pscal$ mediator in association with a top-quark pair and it
decreases when the decay of the light pseudo-scalar into a
top-quark pair is kinematically allowed and competes with
$\pscal\rightarrow \chi\bar\chi$. On the other hand, the sensitivity of the
latter analysis is fairly independent of $m_{\pscal}$ due to the contribution
to the total four-top production cross-section from
the heavy bosons $H/A\rightarrow \ttbar$, both of which have masses fixed to $600\;\GeV$ in this
scenario.
In the case of the \monoHbb\ analysis, the exclusion was not investigated below $\tan\beta = 0.5$. Given the non-trivial dependency of the width on $\tan\beta$ in this channel, it is not possible to extrapolate beyond the area explored.

\begin{figure}[p]
\centering
 
\subfloat[]{\includegraphics[width=0.7\linewidth]{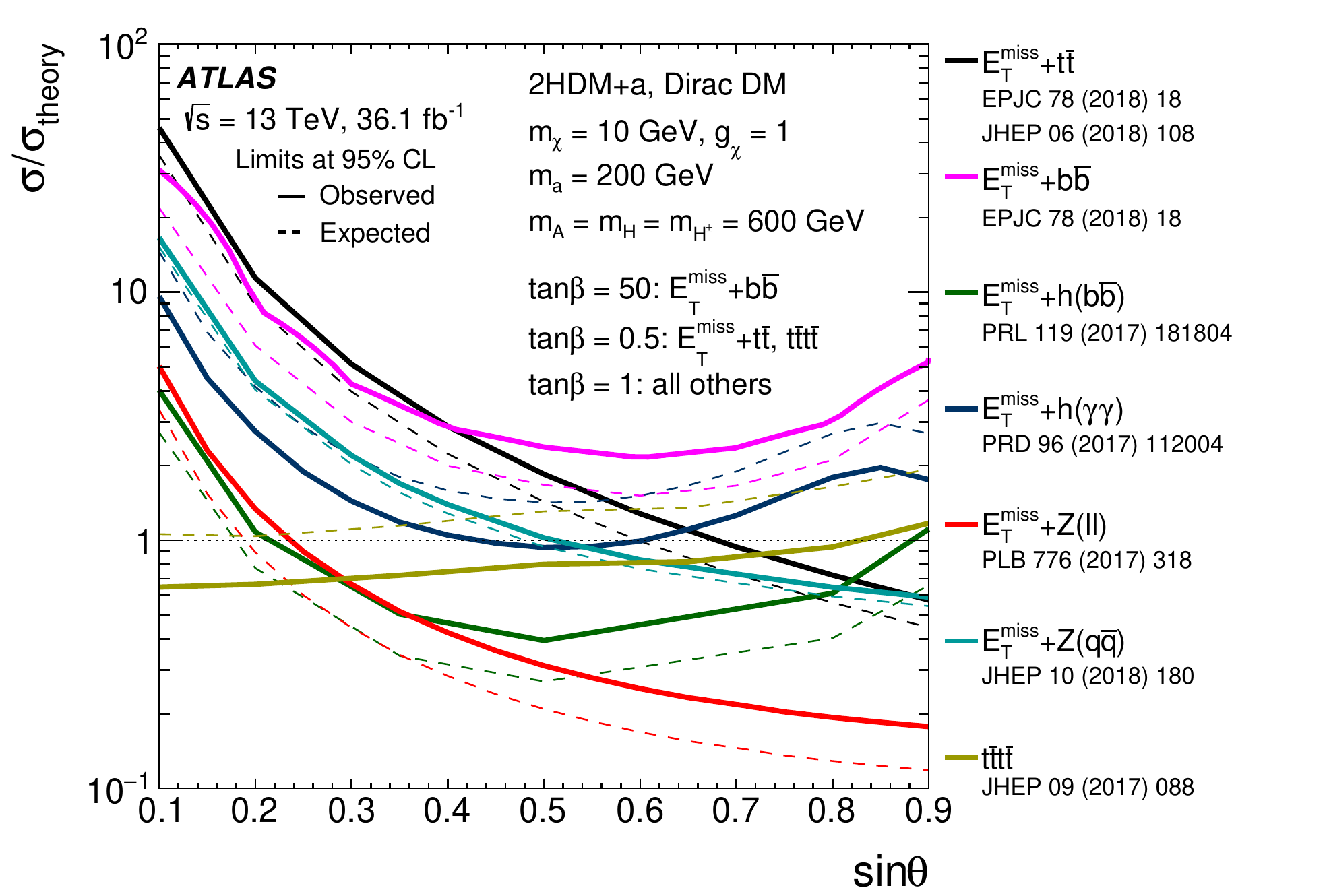}
\label{fig:Summary2HDMPS_sintheta}
} \\
\subfloat[]{\includegraphics[width=0.7\linewidth]{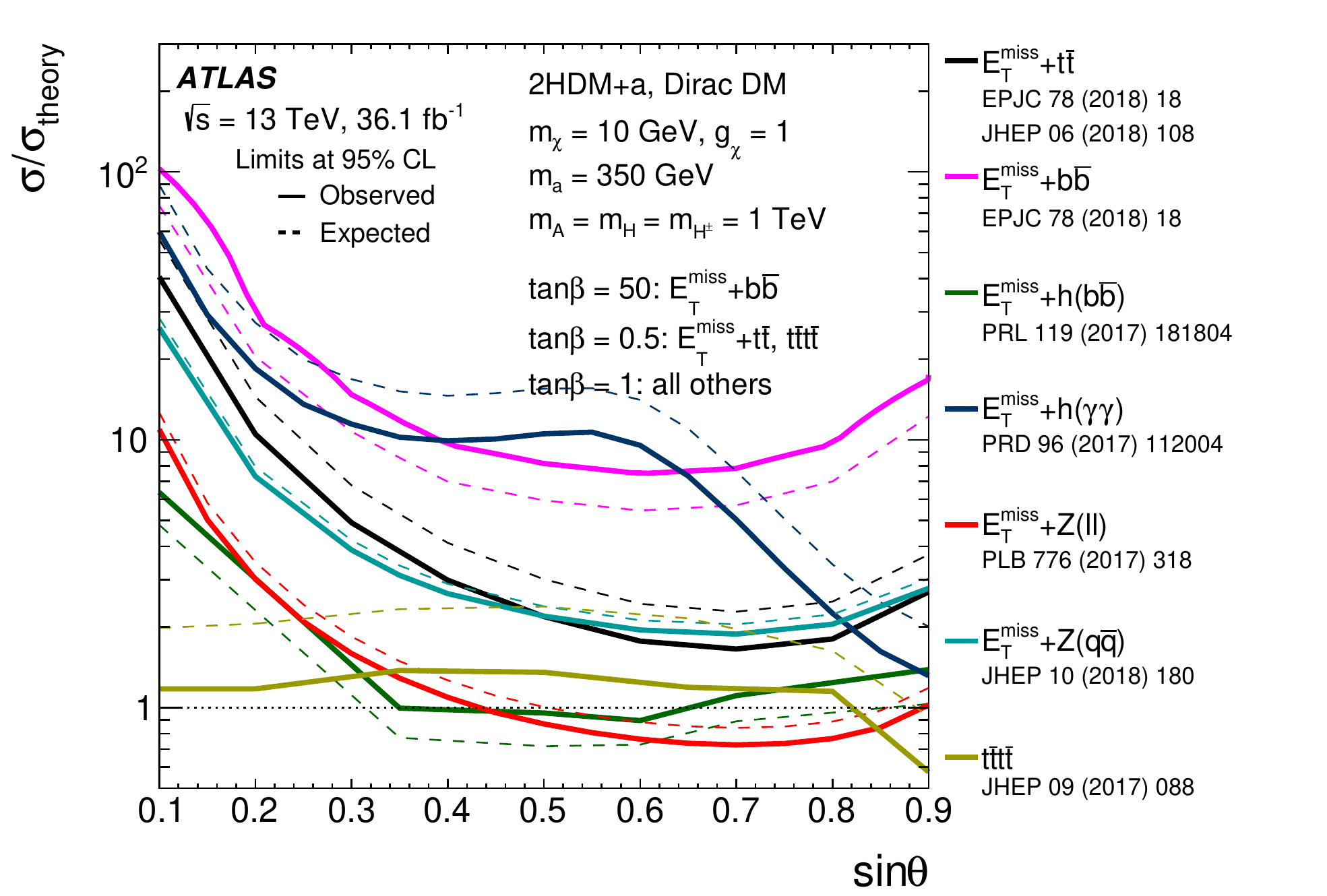}
\label{fig:Summary2HDMPS_sinthetaH}
}
\caption{Observed exclusion limits for the \thdmS\ model as a function of
$\sin\theta$, following the two parameter
choices of scenario 3, (a) low-mass and (b) high-mass  $\pscal$ hypotheses. The limits are
calculated at 95\% CL and are expressed in terms of the ratio of the
excluded cross-section to the nominal cross-section of the model.
}
\end{figure}

Figures~\ref{fig:Summary2HDMPS_sintheta}~and~\ref{fig:Summary2HDMPS_sinthetaH}
present the exclusion limits dependence on the mixing angle, $\sin\theta$,
for a low-mass and high-mass $\pscal$ hypothesis, as evaluated in the third scenario. The limits
are expressed in terms of the ratio of the
excluded cross-section to the nominal cross-section of the model.
For scenario 3a (Figure~\ref{fig:Summary2HDMPS_sintheta}), the lowest
cross-section values are excluded by the \monoZll\ and \monoHbb\
analyses. The sensitivity of both \monoZ\ analyses monotonically
improves as a function of $\sin\theta$, as the cross-section of
the non-resonant and resonant production diagrams,  in Figures~\ref{fig:THDMpa_feynd}~and~\ref{fig:THDMpa_feyne} respectively,    increases with  $\sin\theta$.
Conversely, the same production diagrams for the \monoH\ signatures
have very different dependence on the mixing angle \cite{Bauer:2017ota} in the two $m_{\pscal}$ regimes explored here.
The contribution of each diagram is also affected by the different \monoHbb\ and \monoHgg\ analysis selections.
For this scenario, both analyses
have maximum of sensitivity around $\sin\theta \sim
0.5$.
The three heavy-flavour signatures, \dmbb, \dmtt\, and \fourtop,
are presented for different $\tan\beta$ assumptions. A value of
$\tan\beta = 50$ is studied for \dmbb, with the aim of probing the
parameter space where the coupling of the $\pscal$ mediator to
down-type quarks is enhanced.
However, the \dmtt\ and \fourtop\ signatures are presented for
$\tan\beta = 0.5$ as they are not yet able to probe $\tan\beta$
values near unity. The \fourtop\ signature, in particular, shows a $\sin\theta$
dependence complementary to the other signatures
due to the combined contribution of all neutral bosons decaying into top-quark pairs and is
particularly sensitive at very small mixing angles.
Scenario 3b, presented in Figure~\ref{fig:Summary2HDMPS_sinthetaH},
sets the mass of the light pseudo-scalar so that the
$\pscal\rightarrow \ttbar$ decay is kinematically allowed, which
introduces an additional $\sin\theta$ dependence to the $X+\met$ analyses
interpreted in this scenario. For this reason, the highest sensitivity
for each analysis is found to be broadly around (or slightly below) the maximal mixing
condition ($\theta = \pi / 4$), except for the \fourtop\
and \monoH\ signatures. The \fourtop\ signature shows a constant sensitivity as a
function of $\sin\theta$ (with an increase for very high values) due to
the mass assumptions of this scenario ($m_{\pscal} = 350\;\GeV$ and $m_{A/H} =
1\;\TeV$) which cause the \fourtop\ production cross-section to
be completely dominated by the $\ttbar + \pscal(tt)$ process.
The \monoH\ signatures have a complex dependence on the mixing angle.
This is due to the different contributions of resonant and
non-resonant processes to the final selection in the two analyses.
In this case it is possible to
observe that the \monoHbb\ analysis presents
a maximum in sensitivity around the maximal mixing condition. The \monoHgg\ analysis
instead shows a local sensitivity minimum around $\sin\theta \sim 0.55$.

\begin{figure}[tb]
\centering
 
\includegraphics[width=0.7\linewidth]{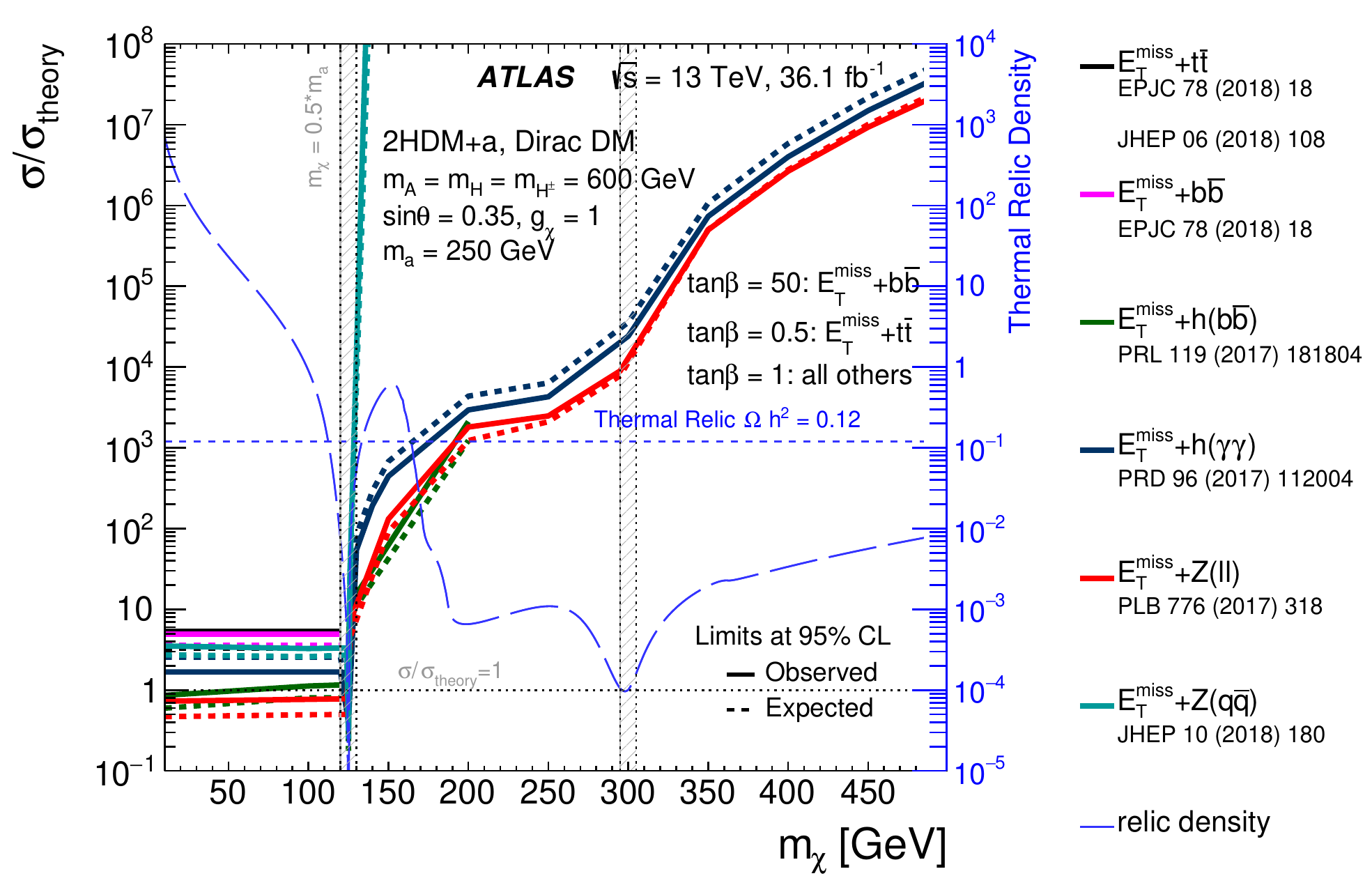}
 
\caption{Observed exclusion limits for the \thdmS\ model as a function of
$m_\chi$, following the parameter
choices of scenario 4. The limits are
calculated at 95\% CL and are expressed in terms of the ratio of the
excluded cross-section to the nominal cross-section of the model.
The relic density for each $m_{\chi}$ assumption is superimposed in the plot
(long-dashed line) and described by the right vertical axis. For DM mass values where the
relic density line is below $\Omega h^2 =
0.12$, the model depletes the relic density to below the thermal
value.
The two valleys at $m_\chi = 125\;\GeV$ and $m_\chi = 300\;\GeV$ determine the
two $\pscal$-funnel and $A$-funnel regions~\cite{Djouadi:2005dz,Bagnaschi:2015eha,2HDMWGproxi} where the predicted relic density is depleted
by the resonant enhancement of the processes $\chi\bar\chi \rightarrow A/\pscal \rightarrow \mathrm{SM}$.}
\label{fig:Summary2HDMPS_mDM2}
 
\end{figure}

Finally, Figure~\ref{fig:Summary2HDMPS_mDM2} presents the
reach of the various experimental searches in a cosmological
perspective, following the prescription of the fourth benchmark
scenario. In this case, the observed exclusion limits in terms of the
ratio of the excluded cross-section to the nominal cross-section of the model are
investigated as a function of the DM mass, which is the parameter with the strongest impact on the
relic density predicted by the \thdmS\ model. The region beyond $m_\chi = 200\;\GeV$ was not explored by the \monoHbb\ analysis,
thus the exclusion is not shown. For the same reason, the \dmbb\ exclusion
is not shown beyond $m_\chi = 125\;\GeV$.
The long-dashed line indicates the
computed relic density for the \thdmS\ model as a function of the DM mass.
The two valleys at $m_\chi = 125\;\GeV$ and $m_\chi = 300\;\GeV$ determine the
two $\pscal$-funnel and $A$-funnel regions~\cite{Djouadi:2005dz,Bagnaschi:2015eha,2HDMWGproxi} where the predicted relic density is depleted
by the resonant enhancement of the processes $\chi\bar\chi \rightarrow A/\pscal \rightarrow \mathrm{SM}$.
The plateau around and above $m_\chi \sim 200\;\GeV$ is determined by the increase in annihilation cross-section of the
DM particles close to threshold for $\chi\bar\chi \rightarrow ha \rightarrow \mathrm{SM}$ and
$\chi\bar\chi \rightarrow t\bar{t}$.
For DM masses
around  $m_{\pscal}/2$ or
$m_\chi > 170\;\GeV$ the model predicts a relic density which is equal to or below the thermal
value, $\Omega h^2 = 0.12$.
As the DM mass increases further, annihilation via single $s$-channel diagrams is more and more suppressed
and the observed DM relic density can again be reproduced. At low values of $m_{\pscal}$
this happens around $m_\chi \sim 10 \;\TeV$ and is outside the range in Figure~\ref{fig:Summary2HDMPS_mDM2}.
For all $X+\met$ signatures considered, the sensitivity is
independent of the DM mass as long as the lightest pseudo-scalar
mediator, whose mass is fixed at $250\;\GeV$ in this scenario, is allowed to decay
into a $\chi\bar\chi$ pair. The \monoZll\ analysis excludes this
parameter space. For higher DM masses, the sensitivity of
all analyses quickly decreases and no exclusion is observed. For
$m_\chi >  m_{\pscal}/2$ all parameter choices that fulfil or deplete the
relic density value are still unconstrained.

\FloatBarrier
\subsection{Scalar dark energy model}
\label{sub:DEres}
 
The results of the \monojet\ and \dmtt\ analyses are interpreted in
terms of limits on the two Lagrangian effective operators $\mathcal{L}_1$
and  $\mathcal{L}_2$ for a scalar DE model, introduced in Section~\ref{sub:DEtheory}.
The results are derived as a function of the suppression scale, $M_1$ and $M_2$, for each operator and
the effective coupling associated with the UV completion of the EFT,
$g_*$, and are shown in Figure~\ref{fig:SummaryDE}.
The EFT operators are only valid in the regime where the momentum transfer is $Q_\textrm{tr} \ll M$.
For the limits shown in Figure~\ref{fig:SummaryDE}, it is assumed that the EFT approximation is valid for events
where $Q_\textrm{tr} < g_* M$.
For events failing this requirement, the iterative rescaling procedure detailed in Ref.~\cite{Abercrombie:2015wmb} is applied.
The \dmtt\ analysis yields the most stringent constraints on the
$\mathcal{L}_1$ operator (Figure~\ref{fig:SummaryDE_L1}), as expected
from the fact that the interaction described by $\mathcal{L}_1$ is proportional to
the masses of the SM fermions to which the DE scalar couples.
The limits are obtained from the search channel (fully hadronic, semileptonic or fully leptonic top pair decays) that provides the smallest expected CL$_\mathrm{s}$ value.
The fully-hadronic and semileptonic channels contribute the
most and similarly to the final sensitivity of the analysis,
which excludes a suppression scale of about $200\;\GeV$ for $g_* \gtrsim
\pi^2$.
The \dmtt\ search is not yet sensitive to weakly coupled models, due to the high momentum transfers involved in the production of the top quarks, which are close to the exclusion limit.
 
The \monojet\ analysis yields the most stringent constraints on the
$\mathcal{L}_2$ operator (Figure~\ref{fig:SummaryDE_L2}), due to the
fact that this interaction is proportional to the momenta of the
particles involved, excluding up to $M \simeq 1.2\; \TeV$ for
$g_* \gtrsim \pi$.
Due to the absence of heavy particles in the final state, the region of EFT validity for the \monojet\ search is larger, with the constraints extending to lower values of the effective coupling.
 
These results improve upon the constraints on the disformal operator from astrophysical probes and non-collider experiments by several orders of magnitude \cite{Brax:2014vva} and also represent a significant improvement on the limits obtained by a similar reinterpretation of ATLAS and CMS results that made use of a smaller dataset at $\sqrt{s}=8$ TeV \cite{Brax:2016did}.
 
\begin{figure}[htbp]
\centering
\subfloat[]{\includegraphics[width=0.49\linewidth]{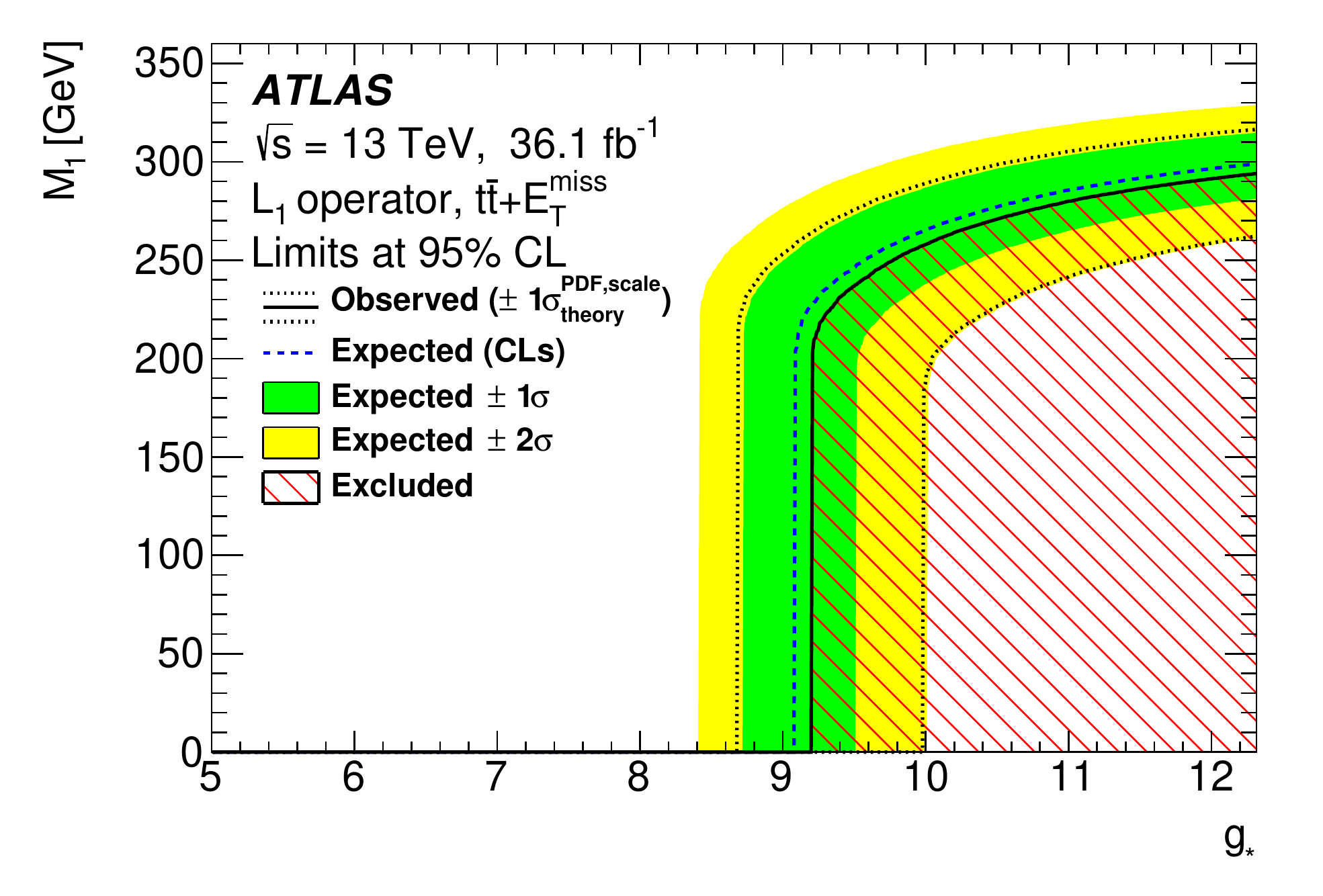}
\label{fig:SummaryDE_L1}
}
\subfloat[]{\includegraphics[width=0.49\linewidth]{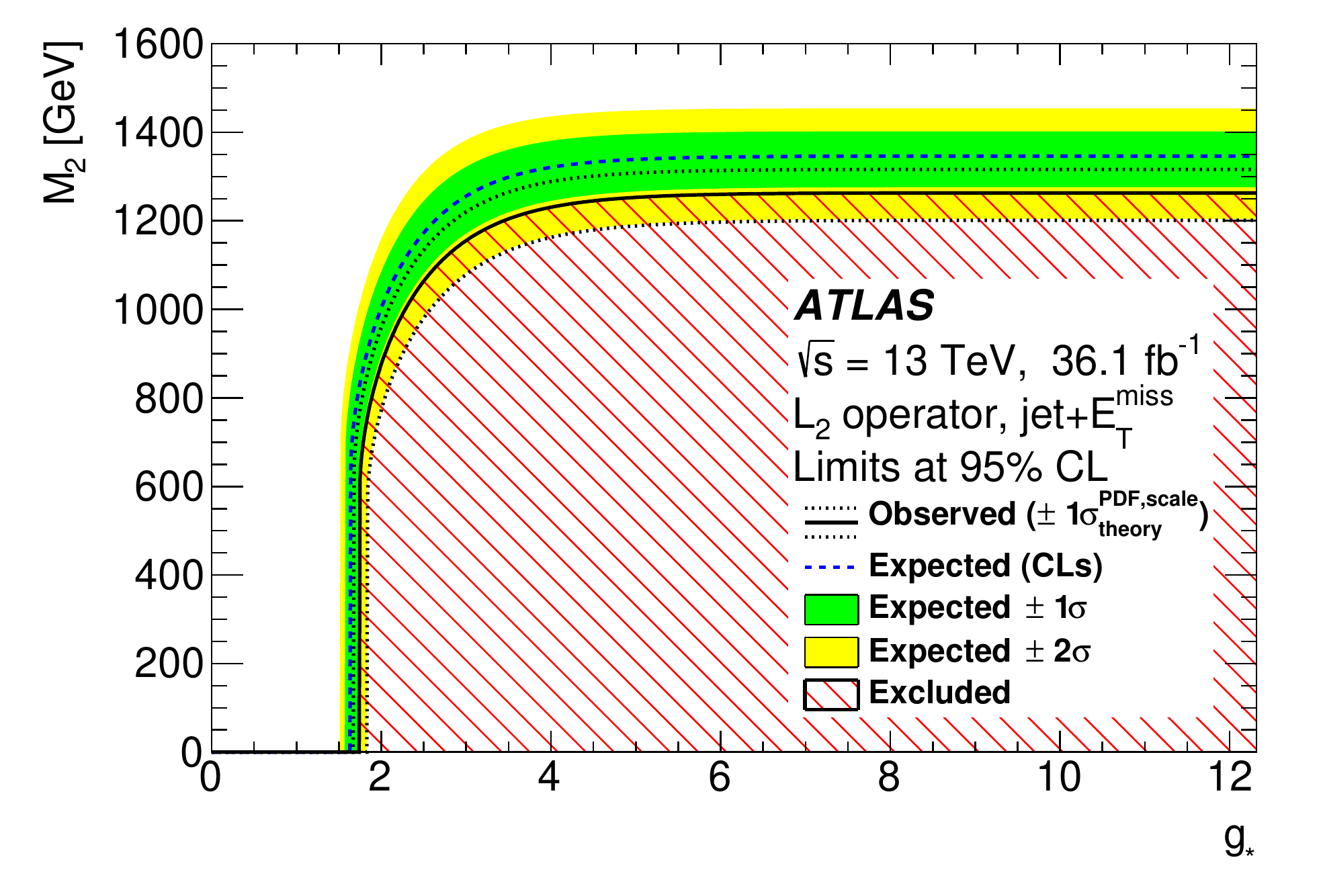}
\label{fig:SummaryDE_L2}
} \\
\caption{Exclusion plots for (a) $\mathcal{L}_1$ and (b) $\mathcal{L}_2$ on the $(g_{*},M)$ plane, after rescaling to take into account the EFT validity criterion~\cite{Abercrombie:2015wmb}.
}
\label{fig:SummaryDE}
\end{figure}

 
\FloatBarrier

\section{Conclusions}
\label{sec:conclusion}
This paper summarises the lively experimental programme	of searches for
mediator-based particle dark matter and scalar dark energy performed by the ATLAS Collaboration.
The analyses presented are based on up to 37~\ifb\ of proton--proton
collisions data at a centre-of-mass energy of $\sqrt{s} = 13\;\TeV$
collected by the ATLAS detector at the LHC in 2015 and 2016. The \hinv\ analysis
considers up to 4.7~\ifb\ at  centre-of-mass energy of $\sqrt{s} =
7\;\TeV$ and 20.3~\ifb\ at  centre-of-mass energy of $\sqrt{s} =
8\;\TeV$.
The results from the searches presented are in agreement with the
Standard Model  predictions, thus results are translated into exclusion limits on mediator-based dark matter and dark energy models.
 
Results on simplified models with the exchange of a vector or axial-vector mediator in the $s$-channel with Dirac fermions as dark matter candidates are compared across different visible and invisible final states. In particular, additional interpretations for these models are presented for \monoZll, \monoVqq\ and all resonance searches except for boosted dijet + ISR. Masses of leptophobic (leptophilic) vector and axial-vector mediators between $200\;\GeV$ and $2.5\;\TeV$ ($3.5\;\TeV$), for coupling values $g_q = 0.25$ and $g_\chi = 1$, and $m_\chi =1\;\TeV$, are excluded at 95\% CL.
Results from \monoHbb\ and \monoHgg\ final states are compared in this paper in the context of a baryon-charged interaction; masses of the $Z'_B$ boson are excluded up to $1.9\;\TeV$ for $m_\chi = 1\;\GeV$ and coupling values of $g_\chi = 1$ and $g_q = 1/3$.
Strong limits on a flavour-changing mediator model are set thank to two complementary search approaches targeting visible and invisible decays of the mediator,
\nvec. Masses up to $1.85\;\TeV$ for coupling values of $g_q = 0.35$ and $g_\chi=1.0$ are excluded
for invisible decays, while
$g_q$ coupling values between 0.14 and 0.35 for mediator masses between $1\;\TeV$ and $3\;\TeV$ are excluded for visible decays.
 
Exclusion limits for simplified model of dark matter production including a
colour-neutral scalar (pseudo-scalar) mediator are compared for \dmtt, \dmbb\ and \monojet\ final states.
Mediator masses below $45\;\GeV$ (in the range $15$--$25\;\GeV$) are excluded for dark matter particles with $m_\chi =
1\;\GeV$ and $g_\chi = 1$.
Masses for colour-charged mediators, coupling to first- and second-generation left-handed quarks, are excluded up to $1.7\;\TeV$, for $m_\chi = 50\;\GeV$ for $g_\chi = 1$.  Colour-charged  mediators that couple to right-handed $b$-quarks ($t$-quarks) are excluded for masses up to $1.4\;\TeV$ ($3.4\;\TeV$) for low dark matter masses.
 
A first interpretation of an extended two-Higgs-doublet model with an additional pseudo-scalar,
$\pscal$, which couples the dark matter particles to the Standard
Model is used to study the broad phenomenology with diverse
final-state signatures predicted by this type of model.
Masses of the
pseudo-scalar mediator, $\pscal$, are excluded up to $350\;\GeV$ for
$m_A = m_H = m_{H^{\pm}} = 1\;\TeV$, $\sin\theta = 0.35$ and
$\tan\beta = 1.0$. The \monoZll\ and \monoHbb\ searches are the most
sensitive analyses in this high mediator-mass region. Previously
published limits on a two-Higgs-doublet model with an additional
vector mediator are improved upon by the statistical combination of the two
decay channels studied: \monoHbb\ and \monoHgg. Mediator masses
between $400\;\GeV$ and $2.5\;\TeV$ are excluded for dark matter
masses of $100\;\GeV$.
 
Finally, a Horndeski model for dark energy is studied in the context of ATLAS searches.
This model introduces a dark energy scalar which couples to gravity.
Limits on the two Lagrangian effective operators, $\mathcal{L}_1$ and $\mathcal{L}_2$, are set by the \dmtt\ and \monojet\ analyses, respectively. The suppression scale is excluded up to $200\;\GeV$ for $g_* = \pi^2$ for the $\mathcal{L}_1$ operator. For the $\mathcal{L}_2$ operator, suppression scales up to $1.2\;\TeV$ for $g_* =\pi$ are excluded.
These results are the first interpretation of a dark energy model by a collider experiment.
 
In this paper, many interpretations in the context of DM and DE models were added with respect to previous publications.
This allowed to restrict very significantly the available parameter space of spin-0, spin-1 and 2HDM-based mediator-DM models as well as EFT DE models.


\clearpage
\appendix
\part*{Appendix}
\addcontentsline{toc}{part}{Appendix}
 
\section{Signal models generation details}
\label{app:rescaling}
 
 
The model implementations, settings and parameter scans used in this paper follow the
prescriptions of the DM Forum/LHC DM Working Group
\cite{Abercrombie:2015wmb,Boveia:2016mrp,Albert:2017onk,2HDMWGproxi}.
and all generation settings used for signal models in this paper are
summarised in Tables~\ref{tab:MCvectorsignal},~\ref{tab:MCscalarsignal}~and~\ref{tab:MCdesignal}.

\subsection{V/AV models}
\label{sub:rescaleVAV}
For all V/AV models, reconstructed samples were produced only
for a specific reference scenario (either a vector or an axial-vector leptophobic mediator model).
Rescaling factors for the acceptance ($w_\mathcal{A}$) and the cross-section ($w_\sigma$) were
calculated to match the acceptance and cross-section
of each of the other scenarios to the
reference. The acceptance weights were calculated for each
$(m_{Z'},m_\chi)$ mass hypothesis as the ratio of
the particle-level acceptance for each of the NLO benchmark models considered ($\mathcal{A}_{\mathrm{truth}}^{\mathrm{NLO}}$)
to
the particle-level acceptance of the analysis  for the reference NLO scenario in a fiducial region ($\mathcal{A}_{\mathrm{truth}}^{\mathrm{ref}}$):
\begin{equation*}
w_\mathcal{A}(m_{Z'},m_\chi) =
\frac{\mathcal{A}_{\mathrm{truth}}^{\mathrm{NLO}}(m_{Z'},m_\chi)}
{\mathcal{A}_{\mathrm{truth}}^{\mathrm{ref}}(m_{Z'},m_\chi)}.
\end{equation*}
The
cross-section weights were calculated for each $(m_{Z'},m_\chi)$
mass hypothesis in a similar way, as the ratio of the reference cross-section at NLO to each
cross-section of the four NLO benchmark models.   The acceptance
rescaling weights were found to be consistent with unity for the
$Z'(\chi\bar{\chi})+j$ and  $Z'(\chi\bar{\chi})+\gamma$ signatures.
 
A few specific
exceptions apply to this treatment. In case of the $Z'(\chi\bar{\chi})+j$
signature, the cross-section rescaling factors were calculated from LO
samples (DMSimp \cite{Backovic:2015soa,Afik:2018rxl} generated with \AMCATNLO 2.4.3
(LO)~\cite{Alwall:2014hca}) and applied to the
samples described in Table~\ref{tab:MCvectorsignal}. In the specific case of the
$Z'(\chi\bar{\chi})+V$ signature, the baseline samples were generated at LO
and rescaled at particle level to match  the NLO samples
described in the table. Finally, the exclusions from the resonance searches (dijet, dilepton, \dibjet)
as a function of the $(m_{Z'},m_\chi)$ interpretations are derived from the limits calculated for Gaussian-shape resonances
\cite{EXOT-2014-15}, and the samples in Table~\ref{tab:MCvectorsignal} are
only used to derive the cross-section normalisation for the final results and the
limits for the leptophobic \avec\ mediator models as a function of the
universal coupling strength.
The $Z'(t\bar t)$ samples were obtained from the
topcolour-assisted technicolour samples of \cite{EXOT-2015-04} rescaled at
particle level to match the DMSimp models described in Table~\ref{tab:MCvectorsignal}.
The correction weights between the two samples were calculated from
the bin-by-bin ratio of the invariant mass distributions of
the \ttbar\ system for the two samples at particle level.
An additional uncertainty is assigned to account for this procedure
as described in Section~\ref{sec:syst}.
 
\begin{table}
\caption{Details of the generation setup and Universal FeynRules Output (UFO)
model used for the spin-1
mediator simplified models, for each signature considered in this
paper. }
\label{tab:MCvectorsignal}
\scalebox{0.75}{
\begin{tabular}{m{38mm}m{28mm}m{74mm}m{15mm}m{45mm}}
\toprule
Model and Final State & UFO & Generator and Parton Shower & Cross-section & Additional details \\
\toprule
$Z'(\chi\bar{\chi})+j$       & DMV \cite{EXOT-2016-27,urldmv} &\textsc{powheg-box} v2~\cite{Alioli:2010xd} +  \PYTHIA 8.205~\cite{Sjostrand:2007gs} & NLO & Particle-level rescaling of leptophobic \avec\ scenario of Ref.~\cite{EXOT-2016-27} (see Appendix~\ref{sub:rescaleVAV})  \\\hline
$Z'(\chi\bar{\chi})+\gamma$  & DMSimp \cite{Backovic:2015soa,urldmsimp} & \AMCATNLO 2.4.3 (NLO)~\cite{Alwall:2014hca} +  \PYTHIA 8.212 & NLO & Leptophobic \avec\ scenario simulated, other scenarios obtained by cross-section rescaling  (see Appendix~\ref{sub:rescaleVAV})\\\hline
$Z'(\chi\bar{\chi})+V$       & DMSimp   & \AMCATNLO 2.5.3 (NLO) +  \PYTHIA 8.212                       & NLO & Particle-level rescaling of LO samples of Ref.~\cite{EXOT-2016-23} to each of the four NLO scenarios (see Appendix~\ref{sub:rescaleVAV})\\\hline
$Z'(qq)$ or $Z'(qq)$+ISR & DMSimp          & \AMCATNLO 2.2.3 (NLO) +  \PYTHIA 8.210                       & NLO &Leptophobic \avec\ scenario simulated, other scenario obtained by Gaussian resonance limits and cross-section rescaling~\cite{EXOT-2014-15} \\\hline
$Z'(\bbbar)$ & DMSimp          & \AMCATNLO 2.2.3 (NLO) +  \PYTHIA 8.210                       & NLO &Leptophobic \avec\ scenario simulated, other scenario obtained by Gaussian resonance limits and cross-section rescaling~\cite{EXOT-2014-15} \\\hline
$Z'(\ell\ell)$ & DMSimp          & \AMCATNLO 2.2.3 (NLO)                       & NLO & Gaussian resonance limits and cross-section rescaling~\cite{EXOT-2014-15} \\\hline
$Z'(t\bar t)$            & DMSimp  & \AMCATNLO 2.4.3 (LO) +  \PYTHIA 8.186 & LO & Particle-level rescaling of the topcolour-assisted technicolour samples of Ref.~\cite{EXOT-2015-04} (see Appendix~\ref{sub:rescaleVAV}) \\\hline
\toprule
$\nvec(\chi\bar{\chi})/\nvec(\bar{u}t)$ & MonotopDMF~\cite{urlmonotop}& \AMCATNLO 2.2.3 (LO) +\PYTHIA 8.210& LO  & Ref.~\cite{EXOT-2017-16} and Appendix~\ref{sub:rescaleNFC}\\\hline
$\bvec(\chi\bar{\chi})+h$ & Higgs\_scalar~\cite{Carpenter:2013xra,urlzbar}& \AMCATNLO 2.2.3 (LO) +\PYTHIA 8.186& LO   & Ref.~\cite{HIGG-2016-18}, simulated for \monoHbb \\\hline
\thdmZ & Zp2HDM~\cite{Berlin:2014cfa,url2hdmz}& \AMCATNLO 2.2.3 (LO) +\PYTHIA 8.186& LO & Ref.~\cite{EXOT-2016-25,HIGG-2016-18} \\
\bottomrule
\end{tabular}
}
\end{table}
 
\begin{table}
\caption{Details of the generation setup and Universal FeynRules Output (UFO)
model used for the spin-0
mediator models, for each signature considered in this
paper. }
 
\label{tab:MCscalarsignal}
\scalebox{0.75}{
\begin{tabular}{m{38mm}m{31mm}m{72mm}m{15mm}m{45mm}}
\toprule
Model and Final State & UFO & Generator and Parton Shower & Cross-section & Additional details \\
\toprule
$\pscal(\chi\bar{\chi})+j$       & DMS\_tloop~\cite{Buckley:2014fba,urlscalmj} & \textsc{powheg-box} v2 +  \PYTHIA 8.205 & NLO & Ref.~\cite{EXOT-2016-27}  \\\hline
$\phi(\chi\bar{\chi})+\ttbar$    & DMScalarMed\_loop \cite{Buckley:2014fba,urldmhf} & \AMCATNLO 2.3.3 (LO) + \PYTHIA 8.186 & NLO~\cite{Backovic:2015soa} & Up to one additional parton. Ref.~\cite{SUSY-2016-18}  \\\hline
$\phi(\chi\bar{\chi})+\bbbar$    & DMScalarMed\_loop & \AMCATNLO 2.3.3 (LO) + \PYTHIA 8.186 & NLO~\cite{Afik:2018rxl} & Up to one additional parton. Ref.~\cite{SUSY-2016-18}  \\\hline
$\cscal$                   & dmS\_T~\cite{Papucci:2014iwa,urltchan}& \AMCATNLO 2.3.3 (LO) + \PYTHIA 8.186 & LO  & Ref.~\cite{EXOT-2016-27,Papucci:2014iwa}  \\\hline
$\cscalb$                  & DM\_Bflavored~\cite{Agrawal:2014una,urlbfdm}& \AMCATNLO 2.3.3 (LO) + \PYTHIA 8.186 & LO  &  Ref.~\cite{SUSY-2016-18}  \\\hline
$\cscalt$                  & MonotopDMF~\cite{EXOT-2017-16,urlmonotop}& \AMCATNLO 2.3.3 (LO) +\PYTHIA 8.210& LO  & Ref.~\cite{EXOT-2017-16}\\
\toprule
\thdmS: $\chi\bar{\chi}+\ttbar/\bbbar$ & Pseudoscalar\_2HDM \cite{Bauer:2017ota,url2hdma} & \AMCATNLO 2.3.3 (LO)  &  NLO & Cross-section based rescaling from simplified model samples of Ref.~\cite{SUSY-2016-18}  \\\hline
\thdmS: $\chi\bar{\chi}+Z$ & Pseudoscalar\_2HDM & \AMCATNLO 2.4.3 (LO) + \PYTHIA 8.212 & LO & Only gluon-initiated production considered~\cite{Bauer:2017ota}\\\hline
\thdmS: $\chi\bar{\chi}+h$ & Pseudoscalar\_2HDM & \AMCATNLO 2.4.3 (LO) + \PYTHIA 8.212 & LO & $b$-quark-initiated production considered only for $\tan\beta \geq 10$\\\hline
\thdmS: $4t$         & Pseudoscalar\_2HDM & \AMCATNLO 2.4.3 (LO) + \PYTHIA 8.212 & LO & Ref.~\cite{Bauer:2017ota}  \\
\bottomrule
\end{tabular}
}
\end{table}
 
\begin{table}
\caption{Details of the generation setup and Universal FeynRules Output (UFO)
model used for the dark energy
model.}
\label{tab:MCdesignal}
\scalebox{0.75}{
\begin{tabular}{llll}
\toprule
Model & UFO & Generator and Parton Shower & Cross-section \\
\toprule
Dark Energy         & Standard\_Model\_cosmo\_no\_c10~\cite{Brax:2016did} & \AMCATNLO 2.6.1 (LO) + \PYTHIA 8.212 & LO \\
\bottomrule
\end{tabular}
}
\end{table}

\subsection{VFC model}
\label{sub:rescaleNFC}
 
The \vfc\ model is studied in two final states: $pp\rightarrow
t\nvec \rightarrow t\chi\bar{\chi}$ and $pp\rightarrow tt(j)$ (via $\nvec$).
 
A complete set of models with the full ATLAS detector simulation was
generated as a function of $m(\nvec)$ and assuming minimal width and
unitary couplings, following the
generation settings summarised in Table~\ref{tab:MCvectorsignal}.
In order to assess the experimental constraints on all the the model parameters,
\gSM, \gDM\ and subsequently $\Gamma(\nvec)$,
a rescaling procedure is applied.
 
In the case of the $t\chi\bar{\chi}$ final state,
for each point in the parameter space, rescaling factors were
calculated at particle level to match the acceptance and cross-section
to those of the simulated
reference model.
 
Three different matrix-element amplitudes contribute to the same-sign top-quark
signature ($tt(j)$) relevant for this model:  (i) prompt $tt$
production in Figure~\ref{g:Diag-AVVFCN2}, (ii) on-shell mediator
in Figure~\ref{g:Diag-AVVFCN1}, (iii) off-shell mediator
in Figure~\ref{g:Diag-AVVFCN3}.
The relative contributions of the three amplitudes depend on the model
parameters, but not the kinematic properties of each process.
The three subprocesses, which were generated
separately with full detector simulation, are combined according to
the following formula to model the signal kinematics for any
choice  of parameter values ($\xi$):
\begin{equation*}
\text{d}\sigma(\boldsymbol{\xi}) \; = \; \alpha(\boldsymbol{\xi}) \, \text{d}\sigma^{\text{ref}}_{tt} \;
+\;  \beta(\boldsymbol{\xi}) \, \text{d}\sigma^{\text{ref}}_{\text{OnShell}} \; +\; \gamma(\boldsymbol{\xi}) \, \text{d}\sigma^{\text{ref}}_{\text{OffShell}}.
\end{equation*}
The functions $\alpha, \beta,\gamma$ can be computed with \textsc{MadGraph} as the ratio of the desired cross-section to the baseline cross-section:
\begin{equation*}
\alpha(\boldsymbol{\xi}) \;\equiv\; \frac{\sigma_{tt}(\boldsymbol{\xi})}{\sigma_{tt}(\boldsymbol{\xi}_{\text{ref}})} ~~~,~~~
\beta(\boldsymbol{\xi})  \;\equiv\; \frac{\sigma_{\text{OnShell}}(\boldsymbol{\xi})}{\sigma_{\text{OnShell}}(\boldsymbol{\xi}_{\text{ref}})} ~~~, \text{ and }~~~
\gamma(\boldsymbol{\xi}) \;\equiv\; \frac{\sigma_{\text{OffShell}}(\boldsymbol{\xi})}{\sigma_{\text{OffShell}}(\boldsymbol{\xi}_{\text{ref}})}.
\end{equation*}

\subsection{\thdmS\ models with heavy-flavour final states}
 
The $\chi\bar{\chi}+\ttbar/\bbbar$ signature of the \thdmS\ model can be
successfully described \cite{2HDMWGproxi} as the superposition of the associated production of two heavy-flavour quarks with
either the light or the heavy pseudo-scalar mediator, which subsequently decays into a $\chi\bar{\chi}$ pair.
When the masses of the two pseudo-scalar mediators are sufficiently different, the contributions of the two processes can be factorised, and the \thdmS\ model can be described
in terms of two sets of colour-neutral pseudo-scalar simplified models, each corresponding to the desired
choice for $m_{\pscal}$ and $m_A$.
 
The acceptance $\mathcal{A}$ of the analysis for each point of interest in the \thdmS\ parameter space is therefore derived as:
\begin{equation*}
\mathcal{A}_{\mathrm{2HDM}}(m_A,m_{\pscal})=\frac{\sigma_{\pscal} \times \mathcal{A}_{\mathrm{simp}}(m_{\pscal})+
\sigma_A \times \mathcal{A}_{\mathrm{simp}}(m_A)}{\sigma_{\pscal}+\sigma_A},
\end{equation*}
where $\mathcal{A}_{\mathrm{simp}}$ is the acceptance of the analysis for the colour-neutral pseudo-scalar simplified model
for a certain mass choice of the $A(\pscal)$-boson, and $\sigma_{\pscal}$ computed in fully reconstructed samples and $\sigma_A$ are the production cross-sections
for $pp\rightarrow \ttbar \pscal (\rightarrow\chi\bar{\chi})$ and $pp\rightarrow \ttbar A(\rightarrow\chi\bar{\chi}$), respectively.
\footnote{The procedure is also valid for $pp\rightarrow \bbbar \pscal/A$ production. However the impact of the correction
was found to be minimal~\cite{2HDMWGproxi} and is neglected in this paper.}
This rescaling is valid in the on-shell region, $m_{\pscal} (m_A) > 2m_\chi$~\cite{2HDMWGproxi}.
 
 
\section{Comparison with direct and indirect searches}
\label{app:directindirect}
 
 
Searches for  weakly-interactive massive particles (WIMPs)~\cite{Steigman:1984ac} represent the
current paradigm for searches for particle dark matter (DM).
Within this paradigm, understanding the nature of DM requires a
measurement of its interaction cross-section with Standard Model particles.
This can be achieved using three complementary methods \cite{Strigari:2013iaa}, schematically
depicted in Figure~\ref{app:DMdetection} and briefly outlined in the following.

\paragraph{Direct Searches} These searches aim to measure the elastic scattering of DM with nuclei in low background
underground detectors such as CRESST-III \cite{Angloher:2015ewa},
LUX \cite{Akerib:2016vxi}, PICO \cite{Amole:2017dex}, DEAP \cite{Amaudruz:2017ekt}, PandaX \cite{Cui:2017nnn,},
XENON \cite{Aprile:2018dbl,Aprile:2017iyp} and SuperCDMS \cite{Agnese:2017jvy,Agnese:2017njq,}.
These direct detection experiments
ultimately measure the strength of the interactions between WIMPs and the partons composing protons and neutrons and are
sensitive to the properties of the DM halo around Earth.
 
\paragraph{Indirect Searches} These searches aim to measure the annihilations or decays of DM particles in astrophysical systems,
by means of neutrino detectors such as SuperKamiokande \cite{Choi:2015ara} or IceCube \cite{Aartsen:2018mxl} or by means
of either ground or space telescopes, for example the
H.E.S.S. Cherenkov telescope \cite{hess2,hess1}, AMS~\cite{Aguilar:2019owu} and Fermi-LAT \cite{TheFermi-LAT:2017vmf}.
This measurement closely relates to the processes that determine the abundance of DM in the early universe
and the interpretation of the results depends on the DM distribution in the universe as well as the SM
particles into which the DM preferentially annihilates or decays.
 
\paragraph{Collider Searches} These searches aim to discover DM particles and to measure the DM production cross-section through collisions of high-energy particles.
The most stringent results to date on WIMPs are provided by the ATLAS~\cite{EXOT-2016-23,EXOT-2016-32,HIGG-2016-18,EXOT-2016-25,HIGG-2016-28,SUSY-2016-18,EXOT-2016-27,SUSY-2016-16,SUSY-2016-15,HIGG-2015-03,EXOT-2017-16},
CMS~\cite{CMS-EXO-16-037,CMS-EXO-16-012,CMS-EXO-16-005,CMS-EXO-16-039,Sirunyan:2017leh} and LHCb \cite{Aaij:2016isa,Aaij:2017rft} experiments at the LHC.
Sub-$\GeV$ DM candidates are also constrained by the MiniBooNE experiment at Fermilab~\cite{Aguilar-Arevalo:2018wea}.
The interpretation of these results closely depends on the underlying mechanisms that couple DM to SM particles. In the simplified model framework considered in this paper this underlying mechanism is assumed to be the production of
new mediator(s) state(s) which subsequently decay into DM.
 
\begin{figure}[bh]
\centering
\includegraphics[width=.38\textwidth]{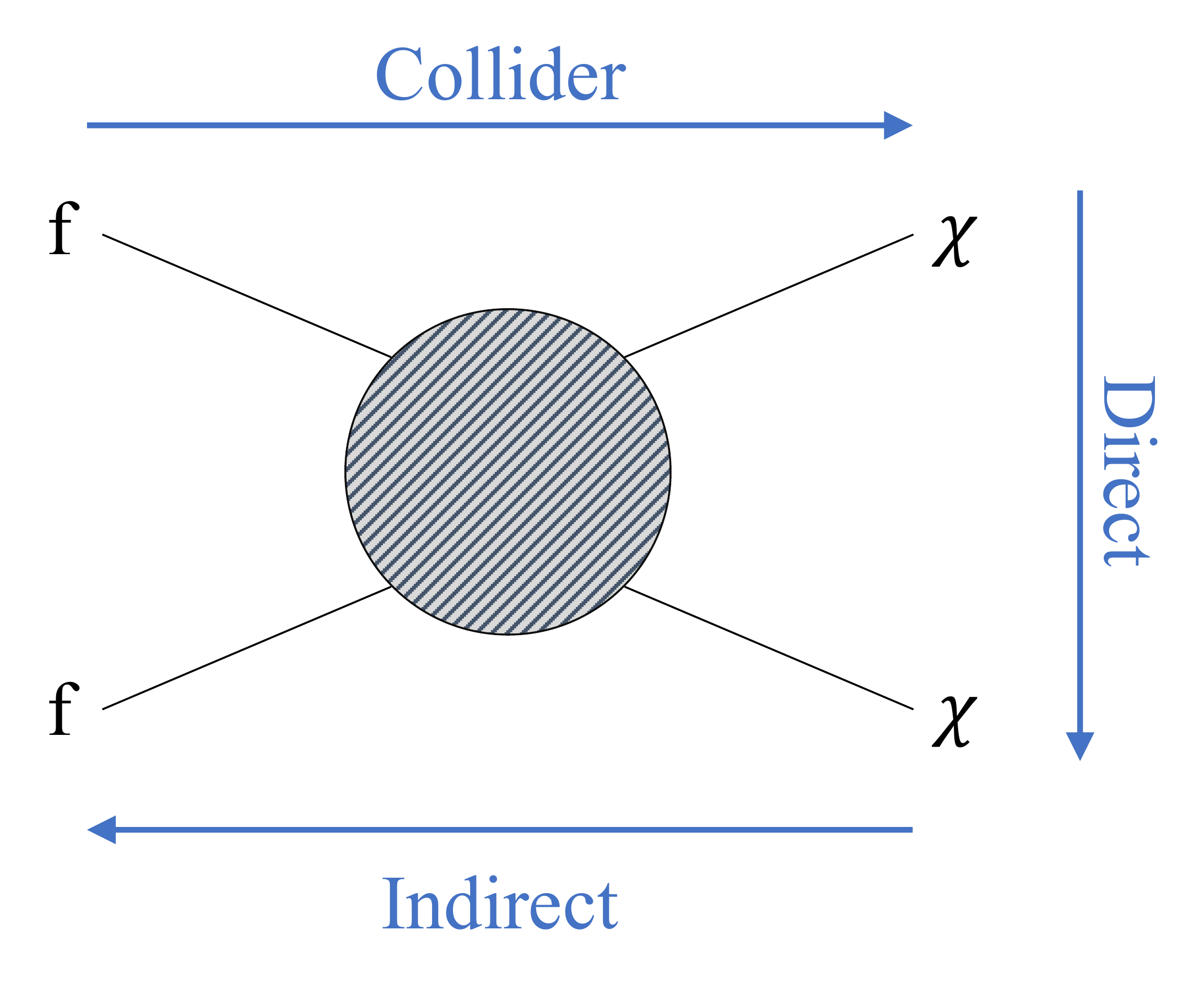}
\caption{Schematic summary of the complementary approaches used in searches for WIMP DM. }
\label{app:DMdetection}
\end{figure}
 
The present understanding of the DM puzzle is encompassed in the summary and comparison of the experimental results of these three approaches.
Likewise, the discovery of DM as an elementary particle will require determination of its interaction cross-section with SM particles via each of these methods.
It is convenient and customary to compare these different approaches
in terms of spin-dependent (spin-independent) $\chi$--nucleon
scattering cross-sections as a function of the DM mass.
In this paper, the ATLAS exclusion limits are converted into bounds on
the $\chi$--nucleon
scattering cross-sections for the following
models:
\begin{itemize}
\item Vector and axial-vector neutral (V/AV) mediator models (two of the benchmark coupling
scenarios, see Section~\ref{ssub:spin1res1} for details).
\item Vector baryon-charged (VBC) mediator model.
\item Scalar colour-neutral  mediator model.
\end{itemize}
For each model, the translation procedure to convert
and compare these limits is well defined and described in
Ref.~\cite{Boveia:2016mrp}.
The interpretation in the
spin-dependent (SD) and spin-independent (SI) DM--nucleon cross-sections,
$\sigma_{SD}$ and $\sigma_{SI}$, respectively, depends on the mediator
mass and the couplings assumptions. Each comparison is valid solely
in the context of the specific model and coupling assumptions.
The ATLAS limits are always shown at 95\% confidence level (CL)
and the direct detection limits at 90\% CL.
 
ATLAS exclusion limits for pseudo-scalar colour-neutral mediator
models should be compared with indirect search experiments in terms
of the DM annihilation
cross-section $\left<\sigma v_{\mathrm{rel}}\right>$, as
the rate in direct searches experiments is suppressed by additional
velocity-dependent terms entering the cross-section.
However, the observed exclusion limits for pseudo-scalar mediator
models with a unitary coupling assumption are limited to a very narrow
mass range, due to a small data excess in the analysis
(Figure~\ref{fig:PS}). Therefore, this comparison is deferred to the
results with the full Run-2 dataset.

The observed limits for the V/AV, VBC and scalar mediator models  are compared
with limits from direct search experiments in
Figures~\ref{fig:AVDD1}--\ref{fig:AVDD3}.
The excluded regions are indicated by shaded areas inside the
contours. Each combined contour summarises the ATLAS results for each
considered model, obtained by using the best expected limit for each parameter point in the figure.
When the contour does not close inside the
plotted area, the exclusion of smaller
scattering cross-sections does not imply that
larger scattering cross-sections (beyond the vertical axis range) are also
excluded.

The spin-dependent WIMP--neutron (WIMP--proton)
scattering cross-section in the context of the leptophobic
\avec\ mediator model is shown in the upper (lower) panel of
Figure~\ref{fig:AVDD1}. The difference between the WIMP--proton and WIMP--neutron cross-sections
is negligible. The ATLAS exclusion curves are therefore identical in
the two panels. The collider searches for this specific leptophobic
axial-vector model complement the reach of the direct searches and extend beyond it,
being particularly sensitive in the low-DM-mass parameter space,  where the
LUX and PICO experiments have less sensitivity due
to the very low-energy recoils that such low-mass dark matter
particles would induce. As in the case of the interpretation of the
results in terms of mediator and DM masses (Section~\ref{ssub:spin1res1}),
if the values chosen for the couplings are reduced,
the relative interplay
between direct and collider searches changes. This is exemplified
by the change of lepton and quark couplings in the leptophilic
\avec\ mediator model shown in Figure~\ref{fig:AVDD2}, where the reach
of resonant dijet final states is greatly reduced in favour of
dilepton searches (differently for the two scenarios) and limited to mediator masses above $200\;\GeV$.
The sensitivity of the $\met+X$ searches is the same for the two models in
Figures~\ref{fig:AVDD1}~and~\ref{fig:AVDD2b}. This is a
coincidental result of two opposite effects \cite{Boveia:2016mrp}:
the fact that the scattering cross-section limit is inversely
proportional to the mediator mass reach (raised to the fourth power), which is higher in the
leptophobic mediator model (Figure~\ref{fig:SummaryAV_A1}), and the
fact that the $\sigma_{SD}$ limit is proportional to $g_q^2$, which
is lower in the leptophilic mediator model.

The spin-independent WIMP--nucleon scattering cross-section results for
leptophobic, leptophilic, or baryon-charged vector mediator $Z'$ and scalar
colour-neutral mediator \scal\ are compared with the most stringent
direct detection limits to date from the LUX, CRESST-II, XENON1T,
SuperCDMS and
PandaX experiments in Figure~\ref{fig:AVDD3}.
One contour for each model is presented in the figure and it includes
the combination, based on the best expected limit for each parameter
point,  of every analysis considered for each model and presented in
Section~\ref{sec:result}. The excluded regions are indicated by shaded
areas inside the contour. As before, when the contour does not close inside the
plotted area, the exclusion of smaller
scattering cross-sections does not imply that
larger scattering cross-sections (beyond the vertical axis range) are also
excluded. The collider searches in this case complement the reach of the direct
searches for $m_\chi \lesssim 5\;\GeV$.
By comparing the exclusion reach of the ATLAS searches for each of the
four models considered in Figure~\ref{fig:AVDD3}, it is possible to
gauge the importance of the production mechanism assumptions for
the collider limits, which represent a complementary
and not exclusive approach to DM searches with respect to direct and
indirect searches.

\begin{figure}[htbp]
\centering
\subfloat[]{\includegraphics[width=0.75\linewidth]{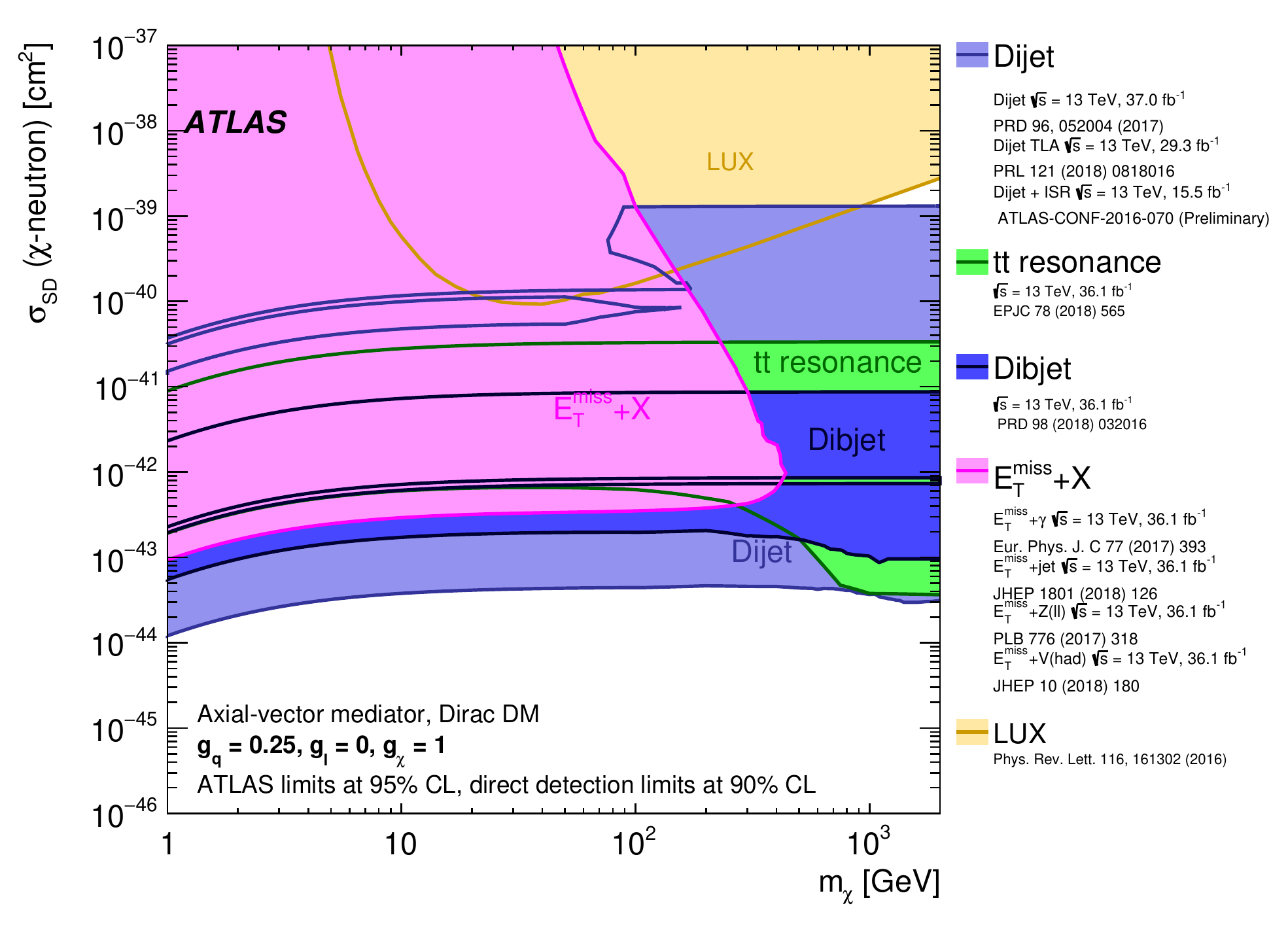}
}\\
\subfloat[]{\includegraphics[width=0.75\linewidth]{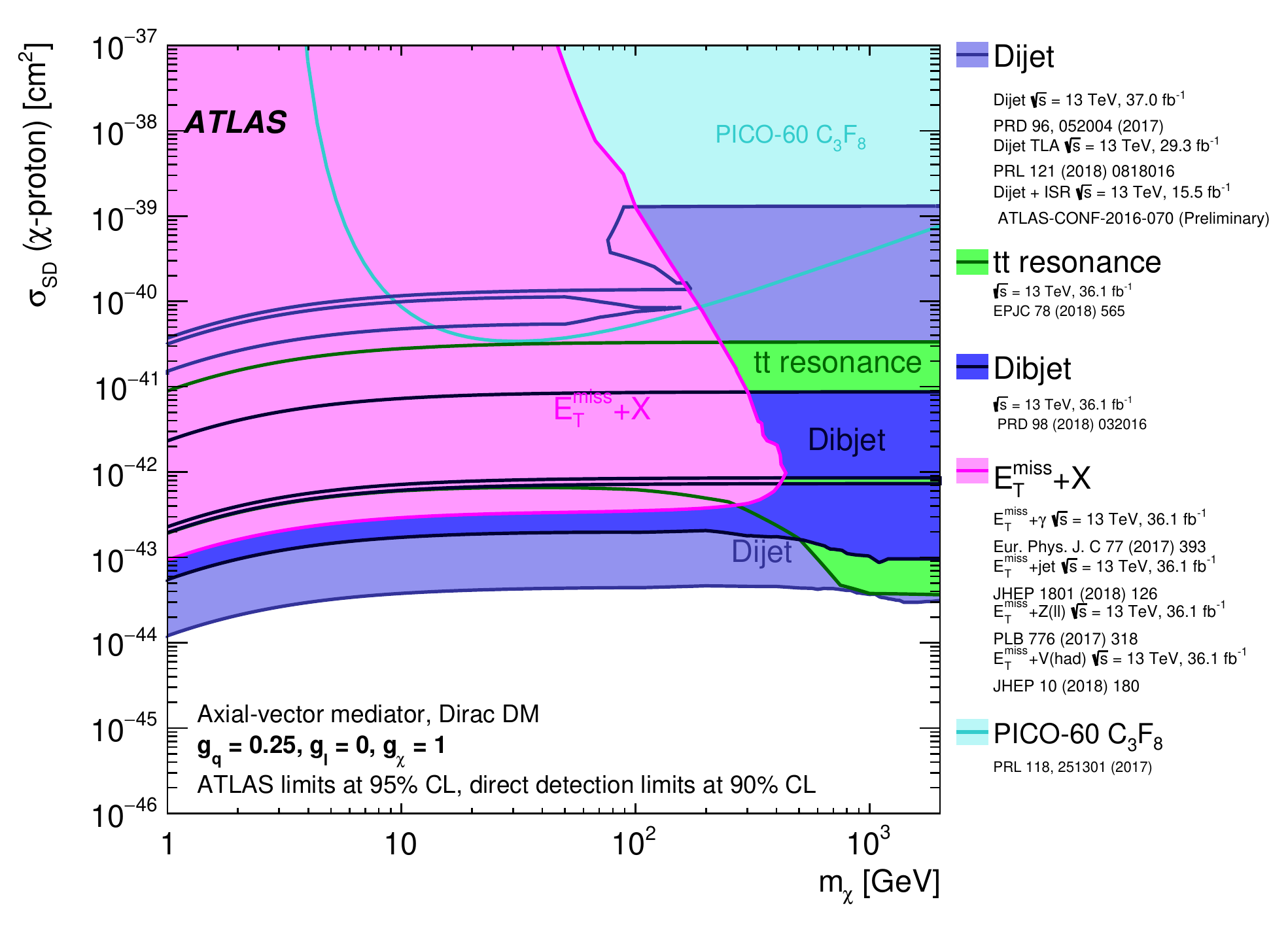}
}
\caption{A comparison of the inferred limits with the constraints from
direct detection experiments on  (a) the spin-dependent WIMP--neutron or (b) WIMP--proton
scattering cross-section in the context of the $Z'$-like simplified
model with axial-vector couplings. The results from this analysis,
excluding the region to the left of the contour, are compared with
limits from direct detection experiments. LHC limits are shown at 95\% CL and
direct detection limits at 90\% CL. The comparison is valid solely
in the context of this model, assuming a mediator width fixed by the
dark matter mass and coupling values $g_q= 0.1, g_\ell= 0.1$, and $g_{\chi} =
1$. LHC searches and direct detection experiments exclude the shaded
areas. Exclusions of smaller scattering cross-sections do not imply
that larger scattering cross-sections are also excluded. The
resonance and $\met$+X exclusion region represents the union of
exclusions from all analyses of that type.}
\label{fig:AVDD1}
\end{figure}

\begin{figure}[htbp]
\centering
\subfloat[]{\includegraphics[width=0.75\linewidth]{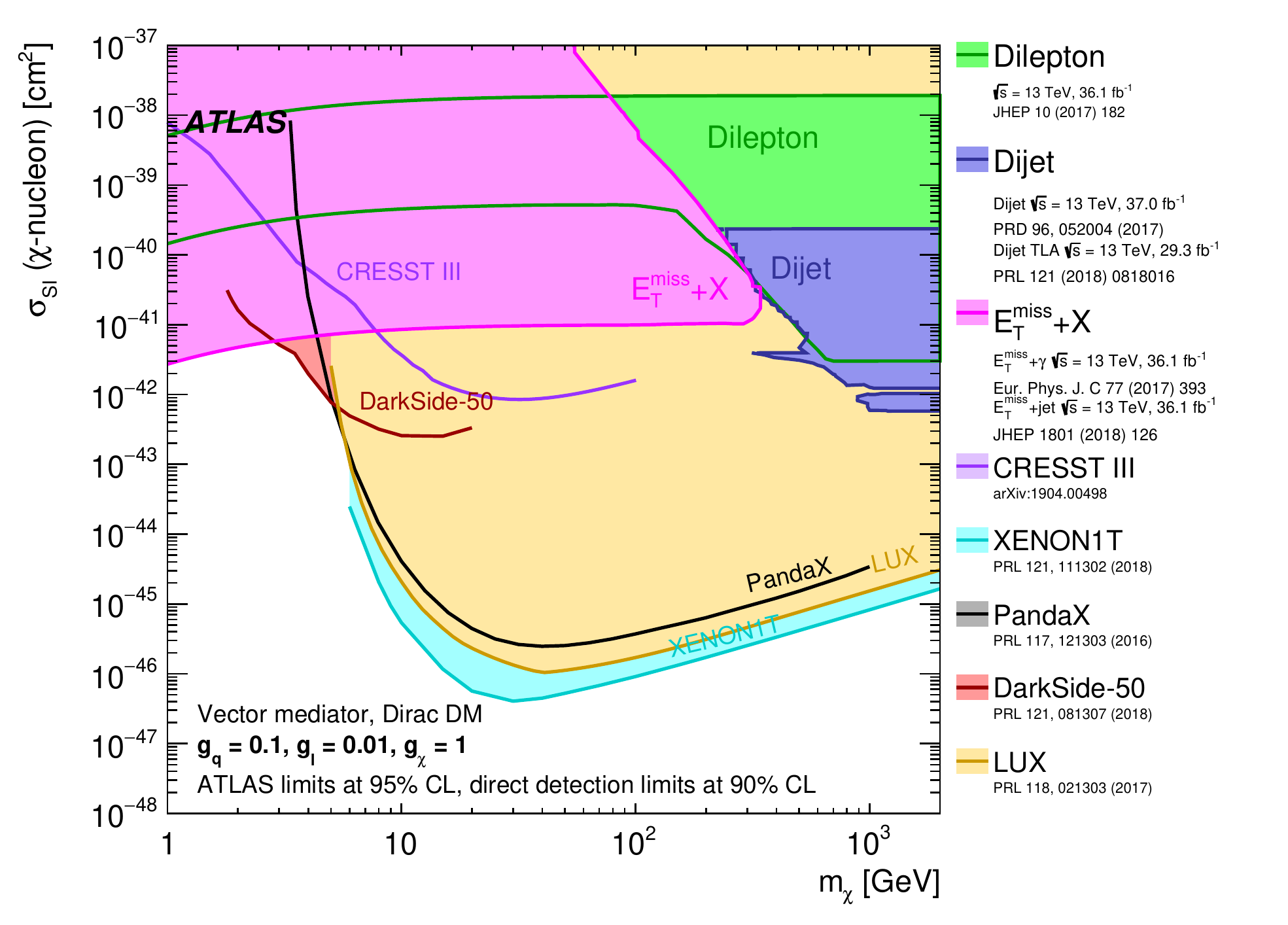}
}\\
\subfloat[]{\includegraphics[width=0.75\linewidth]{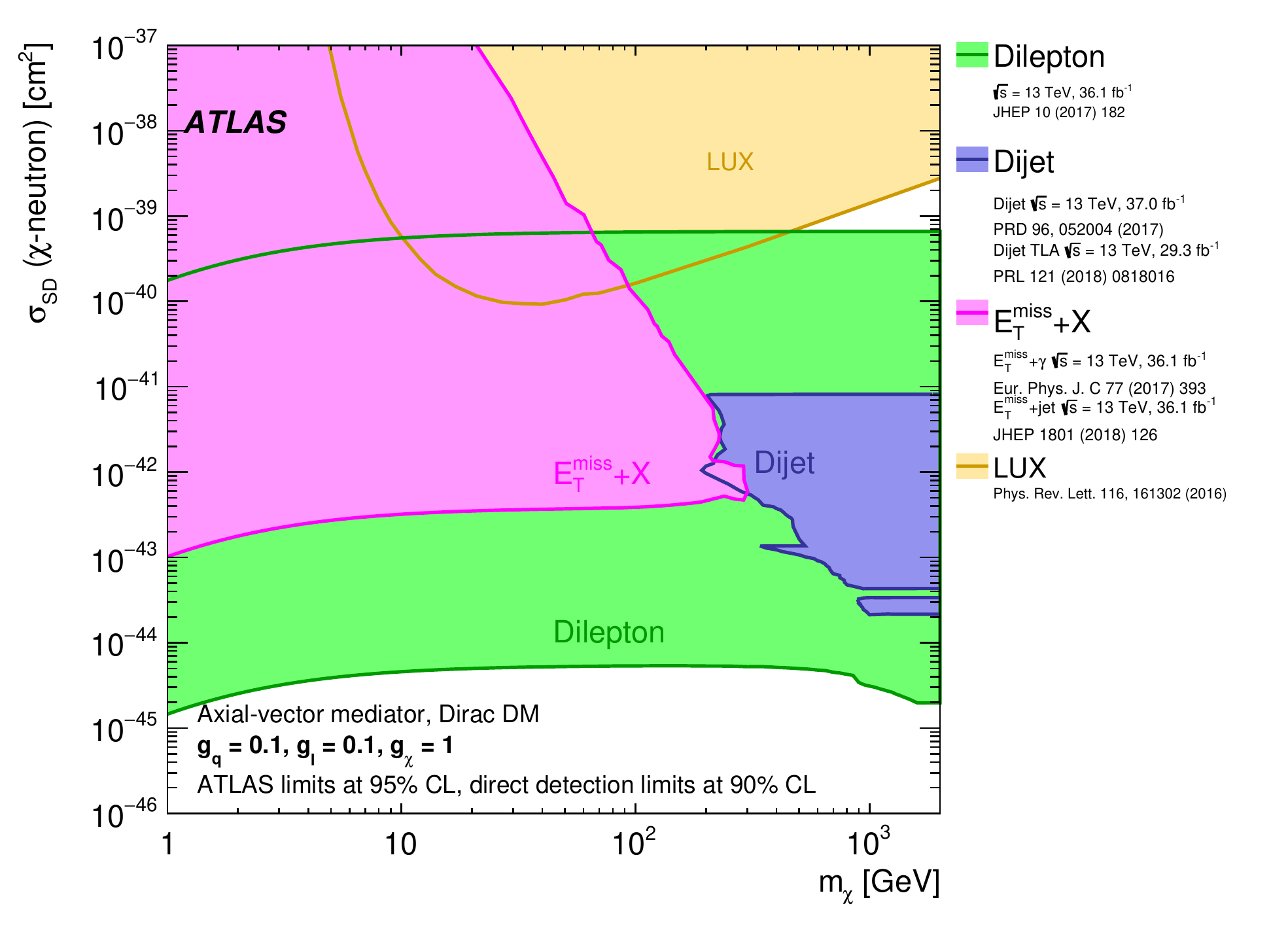}
\label{fig:AVDD2b}
}
\caption{A comparison of the inferred limits with the constraints from
direct detection experiments on the spin-independent WIMP--nucleon (spin-dependent WIMP--neutron)
scattering cross-section in the context of  (a) the $Z'$-like simplified
model with leptophilic vector or  (b) axial-vector couplings. The results from this analysis,
excluding the region to the left of the contour, are compared with
limits from the direct detection experiments. LHC limits are shown at 95\% CL and
direct detection limits at 90\% CL. The comparison is valid solely
in the context of this model, assuming a mediator width fixed by the
dark matter mass and the coupling values highlighted in each figure.
LHC searches and direct detection experiments exclude the shaded
areas. Exclusions of smaller scattering cross-sections do not imply
that larger scattering cross-sections are also excluded. The
resonance and $\met$+X exclusion region represents the union of
exclusions from all analyses of that type.}
\label{fig:AVDD2}
 
\end{figure}

\begin{figure}[htbp]
\centering
\includegraphics[width=0.75\linewidth]{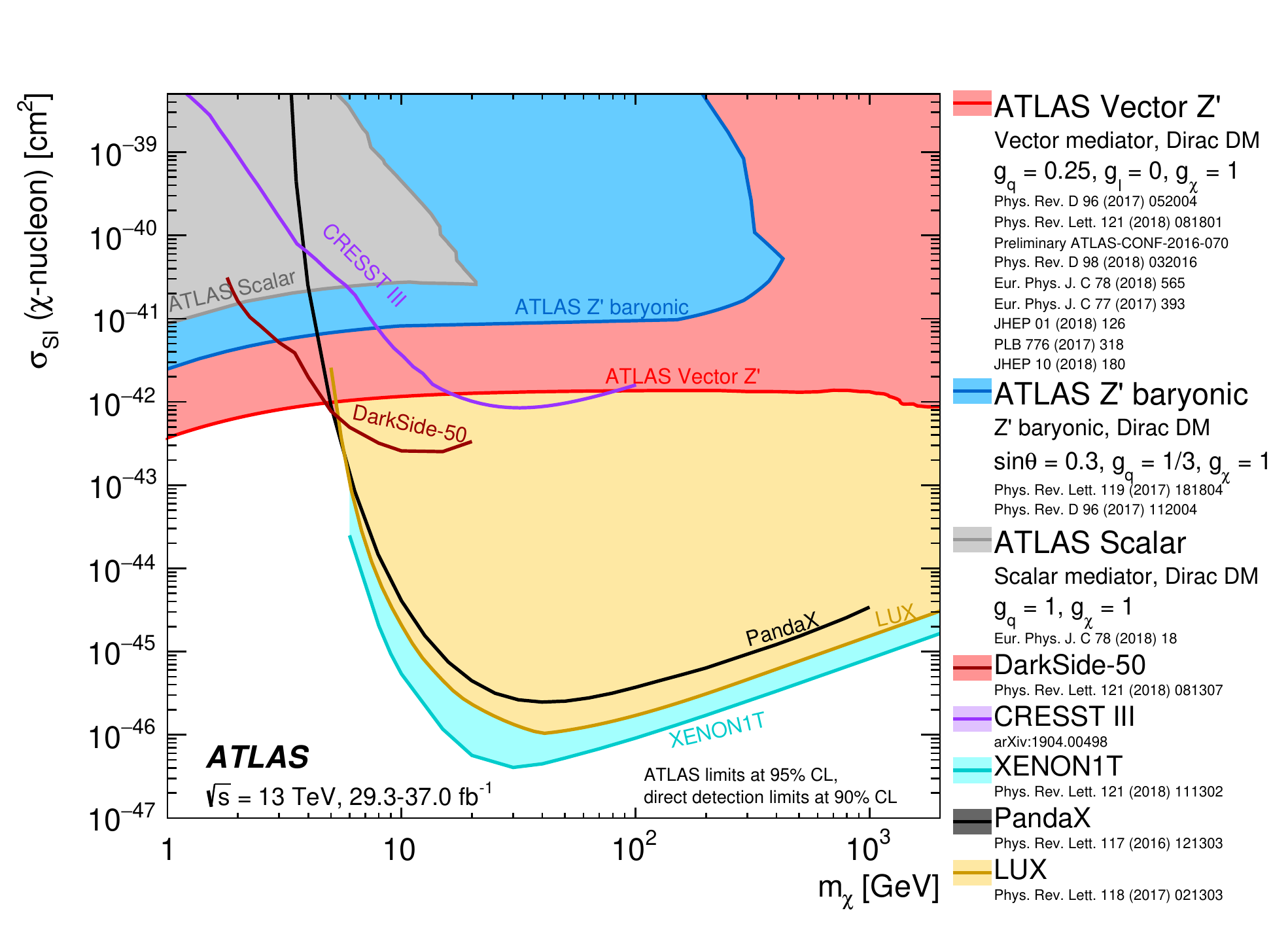}
\caption{
A comparison of the inferred limits with the constraints from
direct detection experiments on the spin-independent WIMP--nucleon
scattering cross-section.
The results from ATLAS analyses, excluding the shaded regions,
are compared with
limits from direct detection experiments. LHC limits are shown at
95\% CL and direct detection limits at 90\% CL. The comparison is
valid solely in the context of this model, assuming a mediator width
fixed by the dark matter mass and coupling values $g_q= 0.25$, $g_\ell
= 0$ or $g_q = 0.1$, $g_\ell = 0.01$ for the neutral-mediator
model
and coupling $g_q = 0.33$ for the baryon-charged mediator. The coupling
to the DM particle $g_\chi$, is set to unity in all cases.
LHC searches and direct detection experiments exclude the shaded
areas. Exclusions of smaller scattering cross-sections do not imply
that larger scattering cross-sections are also excluded. The single
dijet and \met+$X$ exclusion region represents the union of
exclusions from all analyses of that type.}
\label{fig:AVDD3}
 
\end{figure}

 
\FloatBarrier
\section*{Acknowledgements}
 
 
We thank CERN for the very successful operation of the LHC, as well as the
support staff from our institutions without whom ATLAS could not be
operated efficiently.
 
We acknowledge the support of ANPCyT, Argentina; YerPhI, Armenia; ARC, Australia; BMWFW and FWF, Austria; ANAS, Azerbaijan; SSTC, Belarus; CNPq and FAPESP, Brazil; NSERC, NRC and CFI, Canada; CERN; CONICYT, Chile; CAS, MOST and NSFC, China; COLCIENCIAS, Colombia; MSMT CR, MPO CR and VSC CR, Czech Republic; DNRF and DNSRC, Denmark; IN2P3-CNRS, CEA-DRF/IRFU, France; SRNSFG, Georgia; BMBF, HGF, and MPG, Germany; GSRT, Greece; RGC, Hong Kong SAR, China; ISF and Benoziyo Center, Israel; INFN, Italy; MEXT and JSPS, Japan; CNRST, Morocco; NWO, Netherlands; RCN, Norway; MNiSW and NCN, Poland; FCT, Portugal; MNE/IFA, Romania; MES of Russia and NRC KI, Russian Federation; JINR; MESTD, Serbia; MSSR, Slovakia; ARRS and MIZ\v{S}, Slovenia; DST/NRF, South Africa; MINECO, Spain; SRC and Wallenberg Foundation, Sweden; SERI, SNSF and Cantons of Bern and Geneva, Switzerland; MOST, Taiwan; TAEK, Turkey; STFC, United Kingdom; DOE and NSF, United States of America. In addition, individual groups and members have received support from BCKDF, CANARIE, CRC and Compute Canada, Canada; COST, ERC, ERDF, Horizon 2020, and Marie Sk{\l}odowska-Curie Actions, European Union; Investissements d' Avenir Labex and Idex, ANR, France; DFG and AvH Foundation, Germany; Herakleitos, Thales and Aristeia programmes co-financed by EU-ESF and the Greek NSRF, Greece; BSF-NSF and GIF, Israel; CERCA Programme Generalitat de Catalunya, Spain; The Royal Society and Leverhulme Trust, United Kingdom.
 
The crucial computing support from all WLCG partners is acknowledged gratefully, in particular from CERN, the ATLAS Tier-1 facilities at TRIUMF (Canada), NDGF (Denmark, Norway, Sweden), CC-IN2P3 (France), KIT/GridKA (Germany), INFN-CNAF (Italy), NL-T1 (Netherlands), PIC (Spain), ASGC (Taiwan), RAL (UK) and BNL (USA), the Tier-2 facilities worldwide and large non-WLCG resource providers. Major contributors of computing resources are listed in Ref.~\cite{ATL-GEN-PUB-2016-002}.
 

\clearpage
\printbibliography

\clearpage 
 
\begin{flushleft}
{\Large The ATLAS Collaboration}

\bigskip

M.~Aaboud$^\textrm{\scriptsize 35d}$,    
G.~Aad$^\textrm{\scriptsize 101}$,    
B.~Abbott$^\textrm{\scriptsize 128}$,    
D.C.~Abbott$^\textrm{\scriptsize 102}$,    
O.~Abdinov$^\textrm{\scriptsize 13,*}$,    
D.K.~Abhayasinghe$^\textrm{\scriptsize 93}$,    
S.H.~Abidi$^\textrm{\scriptsize 167}$,    
O.S.~AbouZeid$^\textrm{\scriptsize 40}$,    
N.L.~Abraham$^\textrm{\scriptsize 156}$,    
H.~Abramowicz$^\textrm{\scriptsize 161}$
% ...
\bigskip
\\

% ...
$^{13}$Institute of Physics, Azerbaijan Academy of Sciences, Baku; Azerbaijan.\\
% ...
$^{35}$$^{(a)}$Facult\'e des Sciences Ain Chock, R\'eseau Universitaire de Physique des Hautes Energies - Universit\'e Hassan II, Casablanca;$^{(b)}$Facult\'{e} des Sciences, Universit\'{e} Ibn-Tofail, K\'{e}nitra;$^{(c)}$Facult\'e des Sciences Semlalia, Universit\'e Cadi Ayyad, LPHEA-Marrakech;$^{(d)}$Facult\'e des Sciences, Universit\'e Mohamed Premier and LPTPM, Oujda;$^{(e)}$Facult\'e des sciences, Universit\'e Mohammed V, Rabat; Morocco.\\
% ...
$^{40}$Niels Bohr Institute, University of Copenhagen, Copenhagen; Denmark.\\
% ...
$^{93}$Department of Physics, Royal Holloway University of London, Egham; United Kingdom.\\
% ...
$^{101}$CPPM, Aix-Marseille Universit\'e, CNRS/IN2P3, Marseille; France.\\
$^{102}$Department of Physics, University of Massachusetts, Amherst MA; United States of America.\\
% ...
$^{128}$Homer L. Dodge Department of Physics and Astronomy, University of Oklahoma, Norman OK; United States of America.\\
% ...
$^{156}$Department of Physics and Astronomy, University of Sussex, Brighton; United Kingdom.\\
% ...
$^{161}$Raymond and Beverly Sackler School of Physics and Astronomy, Tel Aviv University, Tel Aviv; Israel.\\
% ...
$^{167}$Department of Physics, University of Toronto, Toronto ON; Canada.\\
% ...
$^{a}$ Also at Borough of Manhattan Community College, City University of New York, New York NY; United States of America.\\
$^{b}$ Also at Centre for High Performance Computing, CSIR Campus, Rosebank, Cape Town; South Africa.\\
$^{c}$ Also at CERN, Geneva; Switzerland.\\
$^{d}$ Also at CPPM, Aix-Marseille Universit\'e, CNRS/IN2P3, Marseille; France.\\
$^{e}$ Also at D\'epartement de Physique Nucl\'eaire et Corpusculaire, Universit\'e de Gen\`eve, Gen\`eve; Switzerland.\\
% ...
$^{*}$ Deceased

\end{flushleft}


\end{document}